\documentclass[3p,times]{elsarticle}

\usepackage{ecrcadapted}

\volume{175}

\firstpage{1}

\journalname{Published by Elsevier: Physica E}

\runauth{G. Trambly de Laissardière et al.}

\jid{procs}

\jnltitlelogo{Physica E}

\biboptions{sort&compress}
\usepackage[breaklinks,colorlinks,citecolor=blue,linkcolor=blue,urlcolor=blue]{hyperref}

\usepackage{amssymb}

\usepackage{ulem}

\usepackage[figuresright]{rotating}

\begin{document}

\newcommand{\spd}{$sp$-$d$}
\newcommand{\spde}{$sp$-$d$ }

\newcommand{\ef}{$E_{F}$}
\newcommand{\efe}{$E_{F}$ }

\newcommand{\dd}{{d}}
\newcommand{\du}{{d}}
\newcommand{\Ange}{{\rm \AA} }
\newcommand{\Ang}{{\rm \AA}}

\newcommand{\B}{\mathrm{B}}
\newcommand{\NB}{\mathrm{NB}}
\newcommand{\RRe}{\mathrm{Re}}
\newcommand{\IIm}{\mathrm{Im}}
\newcommand{\donc}{\Longrightarrow}
\newcommand{\ds}{\displaystyle}


\begin{frontmatter}

\dochead{}

\title{
Electronic structure and transport in materials with flat bands:\\ 2D materials and quasicrystals}

\author[label1]{Guy Trambly de Laissardi\`ere}
\ead{guy.trambly@cyu.fr}
\author[label1]{Somepalli Venkateswarlu}
\author[label2,label1]{Ahmed Missaoui}
\author[label2]{\\Ghassen Jema\"i}
\author[label2]{Khouloud Chika}
\author[label4]{Javad Vahedi}
\author[label3,label5]{Omid Faizy Namarvar}
\author[label3]{\\Jean-Pierre Julien}
\author[label1]{Andreas Honecker}
\author[label3]{Laurence Magaud}
\author[label2]{Jouda Jemaa Khabthani}
\author[label3]{Didier Mayou}

\address[label1]{Laboratoire de Physique Th\'eorique et Mod\'elisation, CY Cergy Paris Universit\'e, CNRS, 95302 Cergy-Pontoise, France.}

\address[label2]{Laboratoire de Physique de la Mati\`{e}re Condens\'{e}e, D\'epartement de Physique, Facult\'e des Sciences de Tunis, University El Manar, \\ Tunis 1060, Tunisia.}

\address[label4]{Institute of Condensed Matter Theory and Optics, Friedrich-Schiller-University Jena, Max-Wien-Platz 1, 07743 Jena, Germany.}

\address[label3]{Institut NEEL, CNRS and Universit\'e Grenoble Alpes, 38042 Grenoble, France.}

\address[label5]{Laboratoire de Chimie de la Matière Condensée de Paris, CNRS, Sorbonne Université, 75252  Paris, France.}

\begin{abstract}
In this review, we present recent works on materials whose common point is the presence of electronic bands of very low dispersion, called ``flat bands'', which are 
due to specific atomic order effects without electron interactions.
These states are always indicative of some form of confinement and have significant consequences on the electronic structure, transport properties and magnetism of these materials.
A first part is devoted to the cases where this confinement is due to the long-range geometry of the defect-free structure. We have thus studied periodic approximant structures of quasiperiodic Penrose and octagonal tilings, and twisted bilayers of graphene or transition metal dichalcogenides (TMDs) whose rotation angle between the two layers assumes a special value, called ``magic angle''. 
In these materials, the flat bands correspond to electronic states distributed over a very large number of atoms (several hundreds or even thousands of atoms) and are very sensitive to small structural distortions such as ``heterostrain''. We have shown that their electronic transport properties cannot be described by usual Bloch-Boltzmann theories, because the interband terms of the velocity operator dominate  the intraband terms as far as quantum diffusion is concerned. 
In the case of twisted bilayer graphene, flat bands can induce a magnetic state and other electron-electron correlation effects. 
The second part focuses on two-dimensional nanomaterials in the presence of local point defects that cause
resonant
electronic states (vacancies, adsorbed atoms or molecules). 
We present studies on monolayer graphene, twisted or Bernal bilayer graphene, carbon nanotubes,
monolayer and multilayer black phosphorene, 
and monolayer TMDs.
A recent result is the discovery that the selective functionalization of a Bernal bilayer graphene sublattice leads to a metallic or insulating behavior depending on the functionalized sublattice type. This result, which seems to be confirmed by very recent experimental measurements, suggests that functionalization can be a key parameter to control the electronic properties of two-dimensional materials.

\vspace{1 cm}

\end{abstract}


\end{frontmatter}



\vspace{0.5cm}
\noindent

\tableofcontents

\section{Introduction}
\label{Sec_intro}

For around two decades, the physics of 2D materials has been steadily progressing, both in terms of the fundamental insights it provides and its potential applications\,\cite{Geim13,Lin2023,Lei2022,KUMBHAKAR2023106671}. 
As far as their electronic properties are concerned, a central aspect is the existence of electronic states with very low energy dispersion, known as ``flat-band states'' or ``almost flat-band states''. 
The origin of these states is diverse. 
On the one hand, they can be caused by surfaces, edges or local defects in the crystallographic structure, such as adsorbates.
On the other hand, they may also be the consequence of electronic confinement mechanisms by complex geometries, such as in  moir\'e
patterns in the twisted bilayer 2D materials. 
Another important class of materials with flat bands which have been much studied both theoretically and experimentally consists of quasicrystals and their so-called crystalline approximants. In their perfect structure these systems generate flat bands at all energies that are related to the existence of critical states. These materials provide an important example for systems with flat bands, diffferent from, but connected to moir\'e patterns in 2D materials. 
In this review,  we present some of the flat-band states of these two types and discuss aspects of their electronic properties that we consider essential and we have been working on. 
Let us emphasize that our approach uses realistic tight-binding Hamiltonian models so that we can treat aspects which are relevant to actual materials. 

We note that beyond the 2D materials and quasicrystals which are considered in this review there are many other partially overlapping notions of flat bands in condensed
matter physics, see, e.g., Refs.~\cite{Tasaki1998,Richter2004,Derzhko2007,Leykam18,DanieliAndreanovLeykamFlach2024,Liu20,Liu22,Mitscherling22,Regnault2022,Bernevig24,Calugaru22,Thumin23,Thumin25,Jiang25},
including flat-band ferromagnetism in the Hubbard model \cite{Tasaki1998}
and localized eigenstates arising from flat bands in quantum spin systems \cite{Richter2004,Derzhko2007}
and photonic lattices \cite{Leykam18,DanieliAndreanovLeykamFlach2024}.
As discussed in the literature\,\cite{Heikkil16,Song19,Liu21_topo,Regnault2022,Calugaru22,Song22,Bernevig24,Thumin23,Thumin25,NunezRegueiro24,Jiang25,Kwan25,Iugov25,Thompson25,Mesple25,Duncan20,Manna24},
these  flat-band states and the exotic properties such as superconductivity may result from topological effects.
We will not enter into any further  details about these systems here, but instead refer to the aforementioned review articles.

\begin{figure}[t!] 
\begin{center}
\includegraphics[width=12cm]{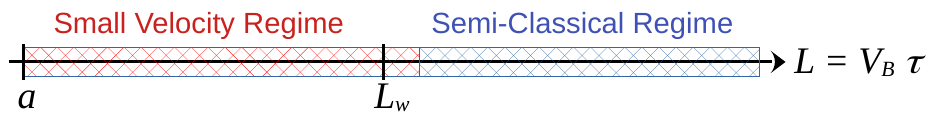}
\end{center}
\caption{ \label{Fig_SVR_Lw}
Schematic transport regime diagram versus the elastic or inelastic  mean free path $L$ due to defects. $a$ is the interatomic distance and $L_w$ the extent of wave packets of charge carriers.
}
\end{figure}

Flat bands are associated to confinement which is expected  to occur on some typical length scale. This length should be important for electronic structure (in the paramagnetic or magnetic state)  and also for transport.  The usual theories to understand the electronic transport properties of materials for which  electronic interactions are not strong, such as metals or semiconductors, are based on the semi-classical approximation and Bloch-Boltzmann theories \cite{Ashcroft76}.
These approaches  
use 
the concept of wave packets which travel at a given velocity (equal to the band velocity) between two scattering events. As we show in this review, the minimal extent $L_w$ of a wave packet depends on the properties of the perfect structure and may play a prominent role, especially when the band velocity becomes small. 
The semi-classical model applies well provided that the spatial extent $L_w$ of charge carrier wave packets is small compared to the mean free path $L$, i.e.,\ the average distance between two collisions on static scatterers (point defects) or dynamic scatterers (electron/phonon interactions or effect of an external magnetic field). 
In a crystal, electronic energy bands $E_n( \mathbf{k})$ can be defined and the semi-classical condition is thus written,
\begin{equation}
L_w \ll L = V_{\B}\, \tau {\rm ~~~with~} V_{\B} = \frac{1}{\hbar} \, 
\left\langle \frac{\partial E_n( \mathbf{k})}{\partial k_i} \right\rangle_{E_n=E_F} , 
\label{Eq_conditionSC}
\end{equation}
where $V_{\B}$ is the Boltzmann (intraband) velocity in the $i$-direction and $\tau$ the elastic or inelastic scattering time. 
As we will see in the following, $L_w$ results from a complex effect of the structure. It depends on $E_F$ but may also depend on $\tau$.
When specific mechanisms localize or confine charge carriers, the velocity $V_{\B}$ can be greatly reduced. 
If the localization mechanism confines the states to large sizes, typically much larger than the interatomic distance, then
$L_w$ is large and the condition (\ref{Eq_conditionSC}) breaks down, rendering the semi-classical approximation invalid (Fig.\,\ref{Fig_SVR_Lw}).
This unusual localization mechanism, called ``Small Velocity Regime'' (SVR), has been highlighted in the approximants of aluminum-based quasicrystals to  qualiltatively understand their anomalous conductivity, which, while being gap-free, behave like semiconductors \cite{Trambly06,Mayou08}. 
More recently, this regime has been also demonstrated in the electronic states confined by magic-angle twisted bilayer graphene (TBG) \cite{Trambly16}.
SVR, which is not due to defects in the crystallographic structure, is the first source of the electronic flat bands presented here. 
The second origin, which is better known but also very interesting, both from a theoretical point of view and for its potential applications, is the existence of localized states due to local defects, in particular resonant defects such as atoms or molecules adsorbed onto the surface of the 2D material (functionalization).

The paper is organized as follows.
The section\,\ref{Sec_flatBand_geo} presents  cases where the flat bands are due to the particular geometry of the systems without defects: quasiperiodic tilings (Sec.\,\ref{Sec_QC}), twisted bilayer graphene (Sec.\,\ref{Sec_QD-TBG}) and twisted bilayer transition metal dichalcogenides (TMDs) (Sec.\,\ref{Sec_TB_MoS2}).
Section\,\ref{Sec_FlatBands_Defects} focuses on  quantum transport studies in systems where flat bands arise from static defects:
monolayer and bilayer graphene (Bernal AB bilayer and twisted Bilayer Graphene (TBG)) with defects randomly distributed throughout the structure (Sec.\,\ref{Sec_biL_rand}) or selectively in specific atomic sublattices (Sec.\,\ref{Sec_SelectiveDefects}),
metallic Single Wall Carbon Nanotubes (SWCNTs) (Sec.\,\ref{Sec_Transp_nanotubes}),
monolayer TMDs (Sec.\,\ref{Sec_Transp_TMD}) and monolayer and multilayer black phosphorene (Sec.\,\ref{Sec_phosphorene}).   
Our aim is to describe the properties of real materials as much as possible. 
Therefore, we pay particular attention to the tight-binding (TB)  model of the studied systems.  
In the framework of the linear response theory and the Kubo-Greenwood approach\,\cite{Kubo57,Greenwood58,Kubo91},
to study quantum diffusion, we use numerical methods in reciprocal space for systems without defects and whose unit cell is not too large, and a real-space method for very large systems. 
These numerical methods to calculate quantum transport by recursion\,\cite{Mayou88,Mayou95,Roche97,Roche99,Triozon02,Trambly13} or by diagonalization~\cite{Trambly06,Trambly16} are presented in the \ref{Sec_NumMethods}.
These methods include all quantum effects for non-interacting charge carriers, including multiple scattering effects.\\

\noindent
{\it Quasicrystals, related complex intermetallic alloys and quasiperiodic tilings.}\\
Immediately after the discovery by 
D.\ Shechtman, I.\ Blech, D.\ Gratias, and J.~W.\ Cahn  \cite{shechtman84} of quasiperiodic intermetallics, one major question was raised about the physical properties of  phases with this new type of order, non-periodic but with long-range order. 
In particular, one expected that  electronic and thermal properties of quasicrystals could be deeply affected \cite{Poon92,Berger94}. 
The description of electrons or phonons in periodic phases rests on the Bloch theorem which cannot be applied to a quasiperiodic structure. Within several decades a series of new quasiperiodic phases, periodic approximants and related complex intermetallic alloys were discovered and intensively studied. These investigations have taught  us that indeed 
electron and phonon properties could be strongly
affected by this new type of order with respect to usual intermetallic alloys (crystals or amorphous alloys).

The first quasiperiodic alloys, found in Al-Mn systems, were metastable and contained many structural defects \cite{shechtman84}. As a consequence they had conduction properties similar to those of amorphous metals with resistivities in the range 100--500\,$\mu\Omega$cm. 
In 1986, the first stable icosahedral phase was discovered in AlLiCu systems. This phase was still defective and although  its resistivity was higher (800\,$\rm \mu\Omega$cm), it was still comparable to that of amorphous metals. 
The real breakthrough came with the discovery of the stable AlCuFe icosahedral phase, having a high structural order. The resistivity of these very well ordered systems were very high, of the order of $\rm 10^4\,{\rm \mu\Omega cm}$ \cite{Klein91,Mayou93}, which gave a considerable interest in their conduction properties.  Within a few years several important electronic characteristics of these phases  were experimentally demonstrated.  The density of states in AlCuFe was smaller than in Al, about one third of that of pure Al, but still largely metallic. The conductivity presented a set of characteristics that were either  those of semiconductors or those of normal metals. In particular  weak-localization effects were observed which are typical of amorphous metals. 
Yet the conductivity  increases with the number of defects just as in semiconductors.  
Optical measurements showed that the Drude peak, characteristic of normal metals,  was absent. 
In 1993 another breakthrough was the discovery of  AlPdRe  which had resistivities in the range of  $\rm 10^6\,{\rm \mu\Omega cm}$ \cite{Pierce93,Berger93,Akiyama93}. 
This system gave the possibility of studying a metal-insulator transition in a quasiperiodic phase.  There are still many questions concerning electronic transport in AlPdRe phases \cite{Delahaye15}. 

Since the discovery of quasicrystals, the view of the role of quasiperiodic order has evolved. 
On  one hand, the long-range quasiperiodic order
can induce electronic states which are neither localized nor
extended and called 
\textit{``critical states''} 
(see Refs.\,\cite{Kohmoto86,Fujiwara89_Fibo,Sire90,Passaro92,Sire94,Mayou94,Guarneri94,Piechon96,SchulzBaldes98,Zhong98,Roche97,Roche99,Mayou00,Triozon02,Jagannathan07,Trambly11_QC,Trambly14_QC,Trambly17,Collins17,Chiaracane21,Grimm21,Jagannathan23,Jagannathan24,Bellissard24}
and Refs.\ therein). 
On the other hand,
for the electronic properties of most known alloys, it appears that the medium-range order, spanning one or a few nanometers, is the actual length scale that determines these properties. 
This observation has led the scientific  community to adopt a larger point of view and consider quasicrystals as an example of a larger class. This  class of complex metallic alloys contains  quasicrystals, approximants and alloys with large and complex unit cells with possibly hundreds of atoms in the unit cell. \\

\newpage
\noindent
{\it Graphene and related two-dimensional (2D) nanomaterials.}\\
Graphene is a 2D carbon material which takes the form of
a planar  lattice of $sp^{2}$ bonded atoms. 
Since it could be isolated in 2004 \cite{Novoselov04,Berger04,Hashimoto04,Novoselov2005,Zhang05},
it is of great interest both theoretically and experimentally.
Its honeycomb lattice consists of two sublattices which give a specific  property to the wave function, 
the so-called chirality. 
The linear dispersion relation close to the charge neutrality point implies that electron states are well described by an effective massless Dirac equation at low energy (Dirac cone)  
\cite{Wallace47,Novoselov2005,Zhang05,Berger06,Sadowski06}. 
The properties of  electrons in
graphene, obtained  from the corresponding Dirac equation, are fundamentally different from those deriving from  the  Schr\"odinger equation.  
In particular, the quantum Hall effect is quantized in integer and half-integer multiples of the fundamental conductance quantum \cite{Novoselov2005,Zhang05}, and can even be observed  at room temperature \cite{Novoselov2007}. Another example of the unique behavior of
Dirac electrons is the so-called Klein paradox  which is  intimately related to the chirality of their wavefunction \cite{Katsnelson2006}. The Klein paradox is the fact that, in a one-dimensional configuration, a potential barrier is perfectly transparent for electrons. 
As a result, it is challenging to localize Dirac electrons using an electrostatic potential. 
However, achieving such confinement could be highly beneficial, particularly for the development of elementary devices.

Graphene is a zero gap material,  which may limit its potential applications. 
However, electronic transport in graphene is sensitive to
static defects such as frozen ripples, screened charged impurities, or local defects like vacancies 
or adsorbates 
\cite{Peres06,Pereira06,Pereira08a,
Lherbier08,Lherbier08b,Leconte10,Leconte11,
Skrypnyk10,Trambly13}.
One way to create and tune a gap in graphene is the selective functionalization  which has been used, e.g., with hydrogen adsorption on a moir\'e pattern of Graphene-Ir(111) \cite{Jorgensen16}.

The study of graphene has also opened up the way to the study of the broader families of 2D materials 
such as hexagonal boron nitrite (h-BN), transition metal dichalcogenide (TMDs), phosphorene, thin oxides... In fact it opens up the way to ``flatland'' \cite{Novoselov11}, a new paradigm for electronics and optoelectronics. 2D materials emerge as the cutting edge of physical and chemical science and engineering.
Stacking layered materials is also a very powerful method to tailor their electronic properties \cite{Geim13}. 
The properties not only depend on the choice of materials to be stacked but also on the details of the relative arrangement of the layers,
e.g., the rotation angle $\theta$ between the two layers in twisted bilayers of graphene (TBG)
\cite{LopesdosSantos07}.
It has thus been shown theoretically \cite{LopesdosSantos07,Trambly10,Bistritzer10,SuarezMorell10,Bistritzer11,Trambly12,Santos12} and experimentally \cite{Li10,Brihuega12} that TBGs, forming a moir\'e pattern \cite{Campanera07,Mele10,Gratias23,Gratias20} at ``magic angles'' \cite{Bistritzer10} 
confine electrons in a tunable way in terms of the angle of rotation $\theta$ of one layer with respect to the other. 
In 2018, it has been experimentally proven that this localization by geometry induces strong electronic correlations and a superconducting state for specific angles \cite{Cao18a,Cao18b,Choi2019,Jiang2019,Xie19}. 
The rotation angle $\theta$ is a key parameter to fix the geometry and the properties of the flat bands due to the moir\'e pattern, but it has also been shown that other effects are  very important in particular, the ``heterostrain'' (axial strain in one of the two layers) \cite{Huder18,Bi19,Mesple21} 
(see also \cite{Kerelsky2019,Gao21,Pantaleon21,Mannai21,Parker21,XIONG2023129048,Escudero24,Yu2025})
and the atomic relaxation
\cite{Uchida14,Wijk15,Dai16,Nam17,Gargiulo17,Choi18,Angeli18,Lin18,Lin18b,Lin18c,Koshino18,Lucignano19,Yoo19,Guinea19,Liu21,Nguyen21,Yananose21,Kazmierczak21,Leconte22,Ezzi24,Ceferino24}.
Despite numerous studies  on these new systems
(see, e.g., Refs.\,\cite{Trambly16,Chung18,Andelkovic18,Omid20,Ciepielewski22,Ciepielewski24,Sinha24,guerrero24_tBLG_disorder_Kubo,Rai24,Kim24,GobboKhun25,Guerrero25}), the consequences of this electronic localization by a moir\'e pattern on electrical transport properties are still  poorly known. 
Moreover in 2018, ``quasiperiodic Dirac states'' have been observed by diffraction in bilayer graphene with a rotation angle 
$\theta = 30^\circ$ \cite{Ahn18,Yao18,Yu19,Yan19}, thus showing the great richness and variety of possible phases in these systems.
For an overview of the electronic properties of TBGs and their potential applications, see, e.g., Refs.\,\cite{Rozhkov16,Catarina19,Wang19,andrei2020graphene,Carr20,Mogera20,Ledwith21,Sun24_revue,Bernevig24}.
Since the discovery of TBG, many other homo- or hetero-twisted 2D materials have been discovered, in particular based on semiconductor materials such as TMDs and h-BN\,
\cite{Wang15_reviewTMD,Liu15_reviewTMD,
Duong17_review,Latil23},
In the latter, depending on the rotation angle, flat bands and quasiperiodic states may also exist\,\cite{Xian19,Ochoa20,Walet21,Sponza24}
(see also Refs.\ in Sec.\,\ref{Sec_TB_MoS2}).

\section{Flat bands due to complex crystal geometry without defects}
\label{Sec_flatBand_geo}

In condensed matter physics, electronic localization is often associated with a short-range confinement effect. This is particularly the case when local static defects or edge effects produce localized states and thus flat bands. 
We will study examples of these flat bands and their consequences on electrical conduction in  
section\,\ref{Sec_FlatBands_Defects}.
However, it is now well established that complex structures without defects can also give rise to confined states with very low energy dispersion, i.e.,\ with large lifetimes. 
This is thus the case in quasicrystals, their periodic approximants, and complex structures close to quasicrystals from the point of view of atomic order at medium and/or long distances. 
Since the 1990's, it has been experimentally proven  that this particular crystalline order can produce intermetallic alloys, such as AlPdRe, AlPdMn and AlCuFe, which are almost insulators \cite{Poon92,Berger93,Berger94,Delahaye15}.
A characteristic of this new electronic localization due to quasiperiodic order is that the confined states are localized on a large number of atoms (several hundreds or thousands) in real space. 
Twisted bilayer 2D materials constitute another example of such electronic confinement by the crystalline structure. 
It has indeed been shown, first in TBG \cite{Trambly10,Bistritzer10,SuarezMorell10,Bistritzer11}, 
and then in other twisted bilayer 2D materials, e.g., Refs.\,\cite{Pan18,Venky20,Zhan20,Ghiotto21,Conti23,Guo25}, 
that for some angles the lowest-energy states have a very low energy dispersion and therefore a very low velocity. 
As for quasicrystals and related phases, these states are spread over a very large number of atoms in the moir\'e cell resulting from the rotation angle between the two layers. 
In 2018, it was experimentally proven that this electronic localization by the moir\'e  pattern can induce strong electronic correlations and a superconducting state for some fillings of the four flat bands \cite{Cao18a,Cao18b}.

In this 
section, we present our work on quasiperiodic tilings (Sec. \ref{Sec_QC}) and twisted bilayer 2D materials (Secs.\,\ref{Sec_QD-TBG} and \ref{Sec_TB_MoS2}) with emphasis on the properties of these confined states by defect-free crystallographic structures. 
In order to understand the analysis which has led to these results, it is important to present first (Sec. \ref{Sec_SVR}) the ``Small Velocity Regime'' of electronic states localized on a large number of atoms.

\subsection{The Small Velocity Regime (SVR)}
\label{Sec_SVR}

The concept of SVR was developed in 2006 to understand the anomalous conductivity of quasicrystals\,\cite{Trambly06,Mayou08,Trambly08,Trambly14_QC} and complex intermetallic alloys \cite{Trambly14_CRAS} (Sec.~\ref{Sec_QC}). 
More recently we have shown that this concept also applies  well to the quantum diffusion of charge carriers in the moir\'e pattern of twisted bilayer graphene near the magic angles (Sec. \ref{Sec_QD-TBG}) \cite{Trambly16}, and to the universal conductivity of the midgap states in monolayer and bilayer graphene (Sec.~\ref{Sec_QL_selective_BLG}) \cite{Bouzerar21,Jouda21}. 

\begin{figure}[t!]
\includegraphics[width=16.4cm]{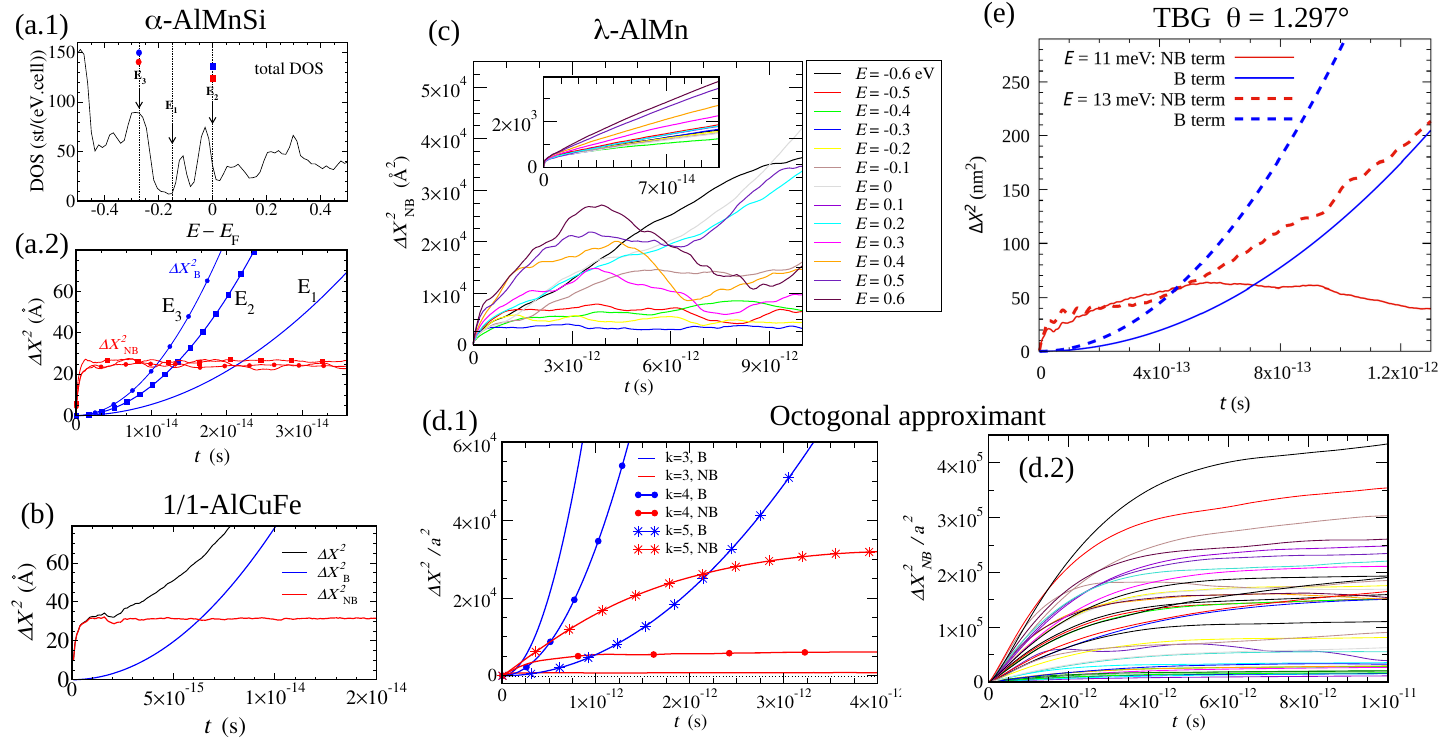}
\caption{\label{Fig_DX2-NB}
Examples of Boltzmann and non-Boltzmann contributions to the average square spreading, 
$\Delta X^2 = \Delta X_{\B}^2+ \Delta X_{\NB}^2$, versus times $t$, for various energies $E$, in different periodic systems we have studied. 
(a) Realistic $\alpha$-AlMnSi approximant of quasicrystals:
(a.1) zoom of total DOS, (a.2) $\Delta X^2$ for three energies close to Fermi energy $E_F$ (see (a.1))  \cite{Trambly06,Trambly08}.
(b) Realistic 1/1-AlCuFe approximant at $E_F$ \cite{Trambly08}.
(c) $\lambda$-AlMn complex intermetallic alloy \cite{Trambly14_CRAS} that is not an approximant, but that has local and medium orders similar to an AlMn quasicrystal. 
The DOS of $\alpha$-AlMnSi and $\lambda$-AlMn are shown Fig.\,\ref{Fig_Conduc_logT_sept13}.
(d) Octogonal tiling approximants:
(d.1) at the same energy $E$, for approximants $k=3$, 4, 5, with increasingly large cell size (see Fig.\,\ref{Fig_octogonal}); (d.2) approximant $k=5$ for many energies.
TB calculations are done for a pure hopping potential\,\cite{Trambly11_QC}.
Note that similar results have been obtained for the Penrose tiling \cite{Trambly14_QC}.
(e) (25,26) Twisted bilayer graphene  ($\theta =1.297^o$) close to the magic angle for two energies in the flat bands  (see Sec.\,\ref{Sec_flatBands_in_tBLG}) \cite{Trambly16}.
$\Delta X_{\NB}^2(E,t)$ is calculated from Eq.\,(\ref{Eq_DeltaX2_NB}).
For (a), (b) and (c) eigenstates are calculated by DFT (LMTO\,\cite{Andersen75}), whereas for (d) and (e), by TB models (see Refs.).
}
\end{figure}

The SVR can be understood by considering firstly
clean 
crystals at zero temperature. Once the band structure is calculated the average quadratic spreading $\Delta X^2(E,t)$ along the $x$ axis,
\begin{equation}
\Delta X^2(E,t) = \left\langle \big( \hat{X}(t)-\hat{X}(t=0) \big)^{2} \right\rangle_{E}, 
\label{Eq_DX2_def}
\end{equation}
can be computed in the basis of Bloch states \cite{Trambly06} (see \ref{Sec_NumMeth_withoutDef}).
In (\ref{Eq_DX2_def}), $\langle \cdots \rangle_E$ is the average on all states at energy $E$. 
The average square spreading is the sum of two terms:
\begin{equation}
\Delta X^2(E,t) = \Delta X_{\B}^2(E,t) + \Delta X_{\NB}^2(E,t).
\label{Eq_DeltaX2-1}
\end{equation}
The first term (Boltzmann term) is the ballistic contribution
at the energy $E$,
\begin{equation}
\Delta X_{\B}^2(E,t) = V_{\B}^2(E) \, t^2
\label{Eq_DeltaX2B-1}
\end{equation}
$V_{\B}$ is the Boltzmann velocity in the $x$ direction. 
The semi-classical theory
is equivalent to taking into account only this Boltzmann term.
The second term,
$\Delta X^2_{\NB}(E,t)$,  is
a non-ballistic (non-Boltzmann) contribution.
This term can refer to a bounded contribution (over time) that is associated with the minimal size 
$L_w$ of a wave packet, as discussed in the introduction. Roughly speaking, for a system without defects, $L_w$ at energy $E$ is the quadratic time average of the non-Boltzmann square spreading,
\begin{equation}
L_w(E,t) \approx \sqrt{ \Delta X_{\NB}^2(E,t) }. 
\label{Eq_LW}
\end{equation}
When the eigenstates of a periodic system are known (by the Density Functional Theory (DFT) method or Tight Binding (TB) method), it is possible to calculate 
$\Delta X_{\B}$ and $\Delta X_{\NB}$ at any energy and any time (see \ref{Sec_NumMeth_withoutDef}).
The ballistic (Boltzmann) contribution and non-Boltzmann contribution are related to the intraband and interband matrix elements of the velocity operator, respectively. 
Note that the importance of interband terms, sometimes called ``quantum'' terms as opposed to ``classical'' intraband terms, has been already discussed in topological flat-band materials\,\cite{Mitscherling22,Regnault2022,Calugaru22,Thumin23,DanieliAndreanovLeykamFlach2024,Thumin25,Jiang25}.
Some examples for $\Delta X_{\B}(t)$ and $\Delta X_{\NB}(t)$ are shown in Fig.\,\ref{Fig_DX2-NB}.

\begin{figure}
\begin{center}
\includegraphics[width=14cm]{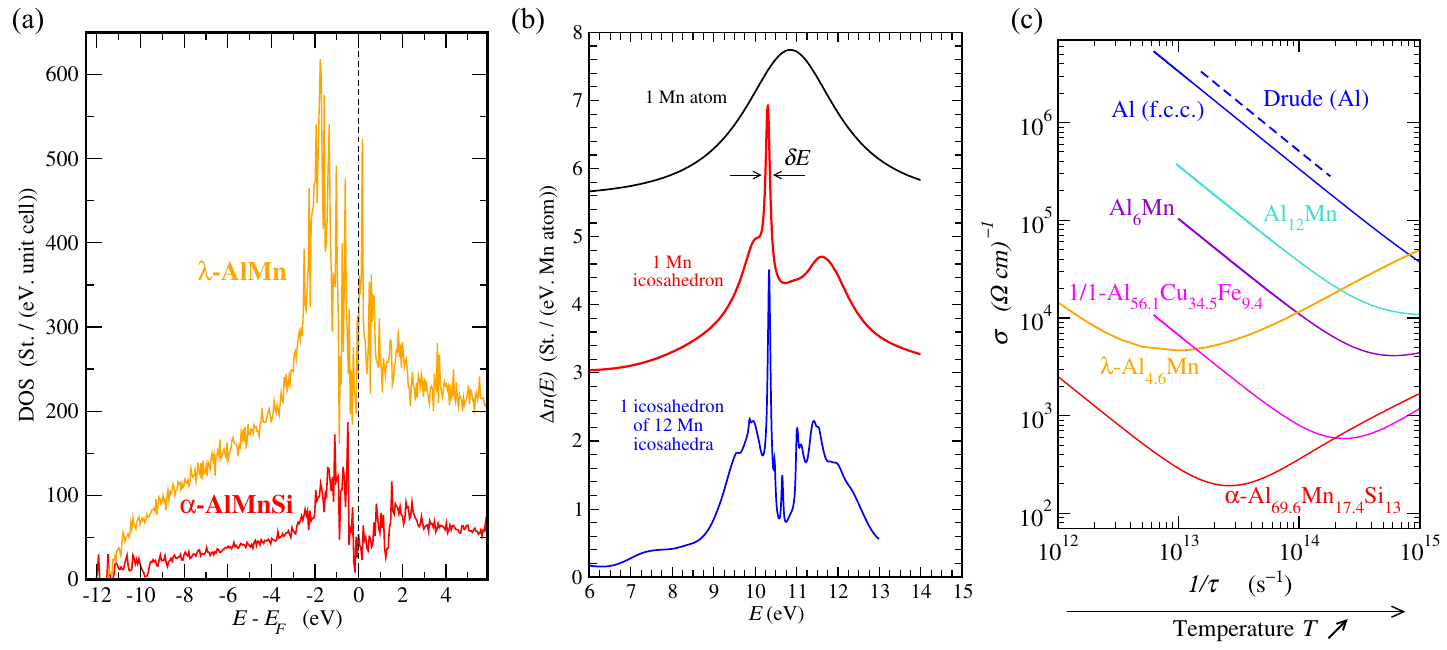}
\end{center}
\caption{\label{Fig_Conduc_logT_sept13}
Electronic properties of actual Al-based quasicrystal approximants containing transition metals. 
(a) DFT (LMTO\,\cite{Andersen75}) Density of states (DOS) of $\alpha$-AlMnSi
approximants (140 atoms/cell)\,\cite{Fujiwara89,Fujiwara93,Trambly08} 
and $\lambda$-AlMn (540 atoms/cell) complex intermetallic alloy, which is not a quasicrystal approximant, but its local atomic order is similar to that of AlMn quasicrystal\,\cite{Trambly09}.
(b) Variation of DOS due to Mn atoms in a metallic medium simulating an Al matrix, for 1 Mn atom (Friedel virtual bound state\,\cite{Friedel56,Anderson61}) and the Mn icosahedron commonly found in AlMn and AlPdMn quasicrystals and their approximants, and an icosahedron consisting of 12 Mn icosahedrons (144 atoms), built according to an inflation rule characteristic of periodic structures\,\cite{Trambly97}.
The spiky peaks are the signature of confined states by the clusters at both length scales.
Their width, $\delta E$, corresponds to a lifetime of the order of $\hbar/\delta E \approx 10^{-13}$\,s.  
(c) DC conductivity $\sigma$ versus the inverse scattering time $1/\tau$ in log–log scale. For each phase, numerical calculations are performed from DFT (LMTO) wave functions \cite{Trambly14_CRAS}.
At room temperature, realistic scattering time $\tau$ values are about $10^{-14}$\,s or  $10^{-13}$\,s \cite{Mayou93}, which corresponds to situations where non-Boltzmann terms may dominate (right part of the curves) in the approximants 1/1-AlCuFe, $\alpha$-AlMnSi, and the complex intermetallic alloys $\lambda$-AlMn.
}
\end{figure}

As already mentioned, the semi-classical theory of transport in crystals is based on the concept of a charge carrier wave packet propagating at a Boltzmann velocity $V_{\rm B}$. 
However the validity of a wave-packet concept requires that the extent $L_w$ of the wave packet is smaller than the distance $V_{\rm B} \tau$ of traveling between two scattering events, i.e.,\ $L_w(E_F,t=\tau)  \ll V_{\rm B} \tau$ (Eq.\,(\ref{Eq_conditionSC})).
In systems with a small Boltzmann velocity and a sufficiently large wave-packet extent, this condition may be no longer valid,
\begin{equation}
 L_w (E_F,t=\tau) \, \gtrsim \,  L = V_{\rm B} \, \tau 
\label{Eq_SVR}
\end{equation}
Thus, when condition (\ref{Eq_SVR}) is satisfied,
the semi-classical (Boltzmann) approach for transport is no more valid, and 
a new diffusion regime, 
the small velocity regime (SVR), is reached.
This makes it possible to draw up a simple phase diagram of the charge carrier transport regime (Fig.\,\ref{Fig_SVR_Lw}) as a function of the scattering length, $L=V_{\B} \tau$, due to elastic or inelastic defects. 
For realistic values of scattering time, $\tau \approx 10^{-14}$\,s or  $\tau \approx 10^{-13}$\,s, in quasicrystals and approximants \cite{Mayou93}, DFT calculations  have shown that the SVR is reached for many energies. 
Some results for $\alpha$-AlMnSi, $1/1$-AlCuFe and $\lambda$-AlMn are presented in Refs.\,\cite{Trambly06,Trambly08,Trambly14_CRAS}, respectively, and Figs.\,\ref{Fig_DX2-NB} and  \ref{Fig_Conduc_logT_sept13}(c).
For magic angle twisted bilayer graphene (TBG), SVR is reached for states with energies corresponding to the low-energy flat-band states\,\cite{Trambly16} (Fig.\,\ref{Fig_DX2-NB} and Sec.\,\ref{Sec_TBG_QD}).
The numerical results show different dependence of $L_w$ with the time $t$. 
Depending on the system and the energy, $L_w$ may increase with $t$ or saturate at large $t$. 
In this last case, one can define a new length,
\begin{equation}
L_w^{\infty}(E_F) = \lim_{t -> \infty} L_w(E_F,t),
\end{equation}
which is characteristic of the properties of quantum diffusion due to the crystallographic
structure at the medium and long length scale.
Numerical calculations show that the length $L_w$ may be smaller than the size of the crystal cell, but not necessarily (Fig.\,\ref{Fig_DX2-NB}).
In this case,
when condition (\ref{Eq_SVR}) is satisfied for realistic Fermi energy values, one obtains simple equations for the diffusivity which (from Eq.\,(\ref{Eq_DeltaX2-1})) is the sum of contributions from Boltzmann and non-Boltzmann terms that can be estimate in the presence of elastic or inelastic defects within the Relaxation Time Approximation (RTA, see \ref{Sec_RTA}),
\begin{equation}
\mathcal{D}(E_{F}) = V_{\rm B}^2(E_{F}) \,\tau + \frac{1}{2} \frac{(L_w^{\infty}(E_{F}))^2}{\tau},
\label{Eq_SVR_D}
\end{equation}
and the conductivity,
$\sigma(E_{F}) = e^2 n(E_{F})\,\mathcal{D}(E_{F})$,
\begin{equation}
\sigma(E_{F}) 
= e^2 n(E_{F})\,V_{\rm B}^2(E_{F})\, \tau + \frac{1}{2} e^2 n(E_{F}) 
\frac{(L_w^{\infty}(E_{F}))^2}{\tau},
\label{Eq_SVR_sigma}
\end{equation}
where the first terms are the Boltzmann terms and the second terms the non-Boltzmann terms.
The Boltzmann terms are ballistic, i.e., $\sigma$ increases linearly as $\tau$ increases, whereas the non-Boltzmann term is ``insulating like'', i.e., $\sigma$ decreases as $\tau$ increases.

It is important to remark that $L_w$ can even diverge at large time $t$, as it is the case for monolayer graphene when $E \rightarrow E_D$ \cite{Trambly16}, without the SVR condition (\ref{Eq_SVR}) being satisfied. 
However, for structures with large cells, such as approximants of quasicrystals and magic-angle TBGs, the SVR is achieved by a combined effect of a small velocity $V_{\B}$ and a large length $L_w$, due to the confinement of electrons by the medium order of the geometric structure with life time confinement of the order of magnitude of the scattering time $\tau$.

\subsection{Quantum diffusion in quasicrystals and quasiperiodic tilings}
\label{Sec_QC}

The present review is devoted to the localization mechanism that induces flat-band states.
Rather than describe the properties of all quasicrystals, we will focus on Al-based quasicrystals, which are good examples of intermetallic alloys in which the atomic structure induces very specific localization at medium and long distances. 
Of course, the notion of band does not exist in quasicrystals, which have no translation periodicity. 
However, it is possible to consider approximate periodic structures, called ``approximants'', with local and medium-range atomic orders similar to that of quasicrystals. 
It has been shown experimentally that approximants have electronic properties similar to those of quasicrystals\,\cite{Berger94,Mayou94}. 
In the approximants, the bands are well defined and the flat bands correspond to a particular confinement of charge carriers by the quasiperiodic structure found also in the quasicrystals themselves.
Experimentally, there are 3D quasicrystals, such as the icosahedral AlPdMn, AlCuFe and AlPdRe phases, and quasicrystals that are quasiperiodic in 2D and periodic in the third direction, such as the decagonal AlCuCo, AlCoNi phases. 
The localization mechanisms that are at the origin of the flat band in these complex materials are not directly related to the spatial dimension and apply to both 3D and 2D quasicrystals. 
It therefore seems appropriate to briefly review them here. 

Experimental investigations have indicated that the conduction
properties of many stable Al-based quasicrystals (AlCuFe, AlPdMn, AlPdRe, ...) are unusual and differ strongly 
from those of more simple intermetallic alloys
\cite{Klein91,Poon92,Berger93,Pierce93,Akiyama93,Berger94,Mayou93,Delahaye15,Macia23}.
In particular, their conductivity increases when the static defects density increases and when the temperature increases (inversed Mathiessen rule). 
It appears also that the medium-range order and the chemical order --over one or a few nanometers--
have a decisive influence.
In particular it has been shown that the role of transition metals in the electronic and magnetic properties of aluminum based quasicrystals and more generally aluminum based complex intermetallic alloys is essential to reinforce the effect of the quasiperiodic potential on electronic states  \cite{Fujiwara89,Fujiwara93,Trambly93,Mayou93_QC_d,Krajvci95,
Trambly95,Trambly95_Ester,Trambly97,Roche97_revue,Simonet98,Hippert99,Trambly00,Trambly03,Trambly05}.
There is now strong evidence that these non-standard properties result from the breakdown of the semi-classical Bloch-Boltzmann theory due to the SVR (Sec. \ref{Sec_SVR}).
This phenomenon is presented in the following section (Sec.~\ref{Sec_SVR_approx}) for realistic approximants and related complex intermetallic phases. 
In Sec.~\ref{Sec_SVR_tiling}, it is shown that long-range  quasiperiodic order induces also SVR in Penrose tilings and octagonal 2D tilings without defects.

\subsubsection{Quantum diffusion in aluminum-based approximants and complex intermetallic phases}
\label{Sec_SVR_approx}

The DOS of periodic approximants of aluminium-based quasicrystals
and  related complex phases
(Fig.\,\ref{Fig_Conduc_logT_sept13}(a)) 
is characterized by two essential aspects. 
On the one hand, the presence of a pseudogap of typically 0.5\,eV near the Fermi level  contributes strongly to the stability of these phases by a Hume-Rothery mechanism\,\cite{Massalski78}, i.e.,\ by a Fermi sphere lying on the Bragg planes associated with intense peaks in the diffraction images (pseudo-Brillouin zone) \cite{Poon92,Trambly05,Mizutani12}. 
On the other hand, the presence of almost flat bands at many energies suggests that the DOS is composed of many spiky peaks, as predicted by DFT calculations\,\cite{Fujiwara89,Fujiwara93,Krajvci95,Roche97_revue,NguyenManh03,Trambly09} and confirmed by surface spectroscopy experiments on quasicrystals
\,\cite{Widmer09,Mader10}. It has also been shown that these spiky peaks are the signature of states confined in the local atomic order characteristic of medium-range quasiperiodicity, in particular by clusters of transition metal atoms~\cite{Trambly97,Trambly09} (Fig.\,\ref{Fig_Conduc_logT_sept13}(b)).

Figure\,\ref{Fig_Conduc_logT_sept13}(c) compares the conductivity calculated in approximants and complex phases  
with simple phases that have a standard metallic behavior. 
From equation (\ref{Eq_SVR_D}), it is clear that the Boltzmann (non-Boltzmann) term increases (decreases) when
$\tau$ increases. The minimum of diffusivity (conductivity) is thus obtained when, \cite{Trambly14_CRAS}
\begin{equation}
\tau = \tau^* ~~{\rm with}~ \tau^* = \frac{L_w^{\infty}(E_{F})}{\sqrt{2} \, V_{\rm B}(E_{F})} .
\end{equation} 
For a scattering time $\tau$ such that $\tau > \tau^*$, Boltzmann terms dominate and the diffusivity (conductivity) increases as $\tau$ increases.  
Since $\tau$ decreases with increasing defect concentration and temperature, the behavior is {\it metallic like}: $\sigma$ decreases when defects and/or temperature increase.
By contrast, for $\tau < \tau^*$, the conductivity increases when defects and/or temperature increase, and 
the behavior is {\it insulating like}.
From DFT calculations (Fig.\,\ref{Fig_Conduc_logT_sept13}) in realistic phases, $\tau^*$ is around a few $10^{-14}$ or $\sim 10^{-13}$\,s. 
These scattering time values correspond to scattering time estimates in quasicrystals and approximants from transport measurements at low temperature (4\,K) \cite{Poon92,Berger93,Mayou93,Delahaye15}.
Therefore when the temperature increases from low values, these complex phases are insulating like, as experimentally found.
From equation (\ref{Eq_SVR_sigma}), when the Boltzmann term is negligible, i.e.,\ $\tau \ll \tau^*$, the conductivity follows the {inverse Mathiessen rule} found experimentally \cite{Berger93,Mayou93}:
\begin{equation}
\sigma(T) = \sigma_{\rm 4 K} + \Delta \sigma(T) .
\end{equation}

In $\alpha$-AlMnSi, the minimum value of the conductivity obtained from DFT calculations, $\sigma(E_{F},\tau^*)$, is about 200\,$\rm (\Omega cm)^{-1}$, which is in good agreement with measurements \cite{Berger93}.
This value is very low with respect to standard metallic alloys (Fig.\,\ref{Fig_Conduc_logT_sept13}) as 
expected in Al-based quasicrystals.
In the complex metallic alloys $\lambda$-AlMn, the minimum value of  $\sigma(E_{F},\tau^*)$ is not so low, but the insulating like regime is obtained for a larger range of $\tau$ values.  
This shows that the SVR can be observed in a great number of complex metallic alloys even if their conductivity is not very low. 
Indeed, J.\ Dolin$\rm \check{s}$ek et al. \cite{Dolinsek07,Dolinsek09} (see also the review \cite{Dolinsek12} and Refs.\ therein) successfully analyzed the experimental transport properties of several complex metallic phases using a phenomenological transport model based on the SVR.  
More recently, K.\ Kitahara and K.\ Kimura  have shown the importance of intraband terms in  thermoelectric properties of Al-Cu-Ir  quasicrystalline approximants
which confirms the importance of terms beyond the classical Boltzmann approximation\,\cite{Kitahara21}.

\subsubsection{Sub-diffusive states in 2D quasiperiodic tilings}
\label{Sec_SVR_tiling}

The specific role of long-range quasiperiodic order in electronic properties is
still an open question in spite of a large number of studies (see, e.g., Refs.\,\cite{Kohmoto86,Fujiwara89_Fibo,Sire90,Passaro92,Sire94,Mayou94,Guarneri94,Piechon96,SchulzBaldes98,Zhong98,Roche97,Roche99,Mayou00,Triozon02,Jagannathan07,Trambly11_QC,Trambly14_QC,Trambly17,Collins17,Chiaracane21,Grimm21,Jagannathan23,Jagannathan24,Bellissard24}
and Refs.\ therein). 
Numerous studies confirm the existence of critical states, which are neither extended nor localized but are characterized by a power-law temporal decay of the wave function envelope at large distances. 
The existence of such critical states has been proven only for 1D quasiperiodic chains and 2D quasiperiodic tilings at certain very specific energies. In the presence of critical states, charge carrier diffusion at sufficiently large times $t$ follows a power law,  
the average spreading, $\Delta X = \sqrt{\Delta X^2}$,  is expected to  be\,\cite{Guarneri94,Piechon96,SchulzBaldes98},
\begin{equation}
\Delta X(E,t) \propto t ^{\,\beta} ~{\rm at~large~} t,
\label{eq_Lt}
\end{equation} 
where $\beta$, $0 \le \beta \le 1$, is an exponent depending on energy $E$. 
Different wave-packet expansion regimes can be found: 
\begin{itemize}
\item In usual metallic crystals without static defects, $\beta = 1$ and the propagation is {\it ballistic}.
\item For $0.5 < \beta < 1$, states are called {\it super-diffusive}. 
\item For $\beta =0.5$ the regime is called {\it diffusive}. 
In disordered 2D systems, this regime always exists\,\cite{Lee85}, but it may be reached for a very large time.
\item For $0 < \beta < 0.5$, states are called {\it sub-diffusive}. 
\item For localized states, i.e.,\ in usual insulators, $\beta = 0$.
\end{itemize}  
When disorder is introduced in the perfect approximant or perfect quasicrystal in the form of static defects (elastic scatterers) and/or inelastic scattering (temperature, magnetic field, ...), the defects induce scattering and the propagation of the wave packet is expected to be diffusive for times $t$ larger than a scattering time $\tau$. 
The diffusivity $\mathcal{D}$ of charge carriers at energy $E$ can then be estimated by,\footnote{This approach is less accurate than the Relaxation Time Approximation (RTA)  we used to calculate conductivity in systems with defects 
(section \ref{Sec_FlatBands_Defects} and \ref{Sec_RTA}), but it gives qualitatively similar results.}
\begin{equation}
\mathcal{D}(E,\tau) \approx \frac{\Delta X^2(E,t=\tau)}{2 \tau} \propto \tau^{2 \beta(E) - 1},
\end{equation} 
and the conductivity $\sigma$ at zero frequency is given by the Einstein formula: 
\vskip -.2cm 
\begin{equation}
\sigma(E_F,\tau) = e^2 n(E_F) \, \mathcal{D}(E_F,\tau) \propto \tau^{2\beta(E_F)-1},
\end{equation} 
where $n(E)$ is the total density of states (DOS).

The case $0.5 < \beta < 1$ (super-diffusive regime), leads to transport properties similar to a metal,
since the conductivity decreases when disorder increases --i.e.,\ when $\tau$ decreases--.
Conversely, for $0 < \beta < 0.5$ (sub-diffusive regime), the conductivity increases when disorder increases as in real quasicrystals. 
Many authors consider 
\cite{Sire90,Passaro92,Mayou00,
Guarneri94,Piechon96,SchulzBaldes98} 
that critical states could lead to $\beta < 0.5$ but it has not yet been shown in 2D or 3D quasiperiodic structures 
(except for some very specific energies). 

We have studied \cite{Trambly11_QC,Trambly14_QC,Trambly17} two 2D quasiperidoc tilings, the octagonal 
--Ammann-Benenker-- tiling  \cite{Duneau89,Socolar89,Ammann92,Katz94,Jagannathan24} 
and the Penrose tiling \cite{Katz94,Duneau94}.   
Our numerical calculations require the study of periodic structures (see \ref{Sec_NumImplementation}), so we have studied crystalline approximant structure series of increasingly large size to find the effect of long-range quasiperiodicity.
The periodic approximants of the octagonal tiling were provided by C.\ Oguey according to \cite{Duneau89}, and those of the Penrose tiling by A.\ Sz\'all\'as and A.\ Jagannathan according to \cite{Duneau94}.  
Quantum diffusion in the smallest approximants, i.e.,\ those with sufficiently  small cell, has been studied from the eigenstates calculated by diagonalization 
(\ref{Method_Reciprocal}).  
This allowed us to show \cite{Trambly11_QC,Trambly14_QC} the importance of non-Boltzmann terms in these systems as we found for the realistic approximants (Sec. \ref{Sec_SVR_approx}). 
More recently, larger approximants have been studied by the recursion method in real space 
(\ref{Method_Real}). 
Here, we focus exclusively on the latter approaches to discuss some general aspects of quantum diffusion in quasiperiodic tilings, specifically in the context of results obtained for octagonal tilings \cite{Trambly17} and the Penrose tiling.

\begin{figure}[t]
\begin{center}
\includegraphics[width=0.9\textwidth]{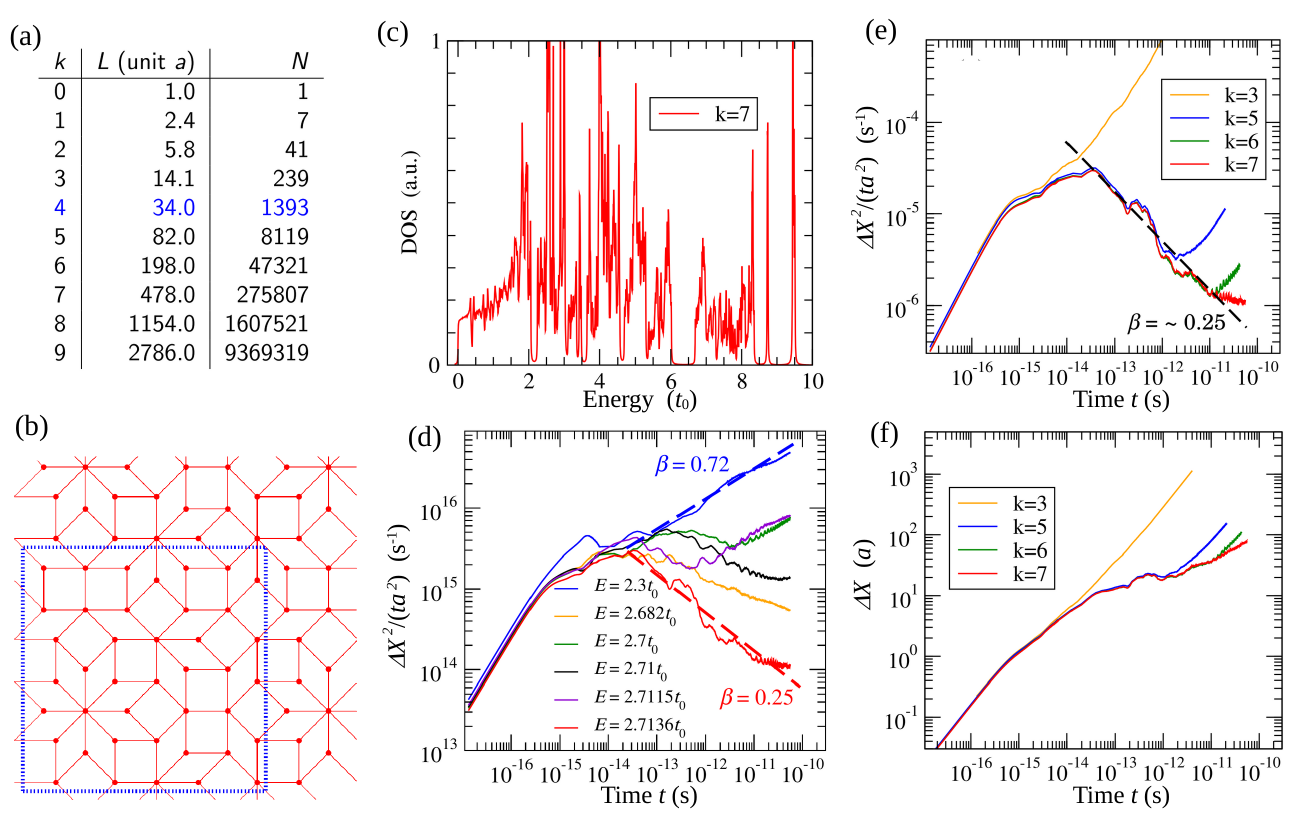}
\caption{
Density of states and quantum diffusion in octagonal approximants ``with potential'' (see text). 
(a) Series of periodic approximants proposed by M.\ Duneau et al.\ \cite{Duneau89} indexed by $k$, with increasing unit cell size $L$. $N$ is the number of nodes per cell and $a$ the length of the edges. 
(b) Sketch of the tiling $k=4$.
The blue dashed line is the cell.
To simulate schematically a possible effect of the presence of different chemical elements, the on-site energies $\epsilon_i$ are proportional to the coordination of atoms (see text case ``with potential'').
(c) Total DOS $n$ in approximant $k=7$, due to recursion method $n$ is convoluted by a Lorentzian of width 5\,meV.  
(d) Diffusion coefficient $2 {D}(t) = \Delta X(t)^2 / t$ in approximant $k=9$ for various energies $E$.
(e) $2 {D}(t)$ and 
(f) average spreading $\Delta X(t)$ at $E=2.7136t_0$ for different approximants. 
To have a realistic order of magnitude of time in seconds we used $t_0 = 1$\,eV. 
The  approximant tilings have been kindly provided by C.\ Oguey.
For the numerical method see \ref{Method_Real}.
Panels (c) and (d) are from\,\cite{Trambly17}.
}
\label{Fig_octogonal}
\end{center}
\end{figure}

\begin{figure}[t]
\begin{center}
\includegraphics[width=0.98\textwidth]{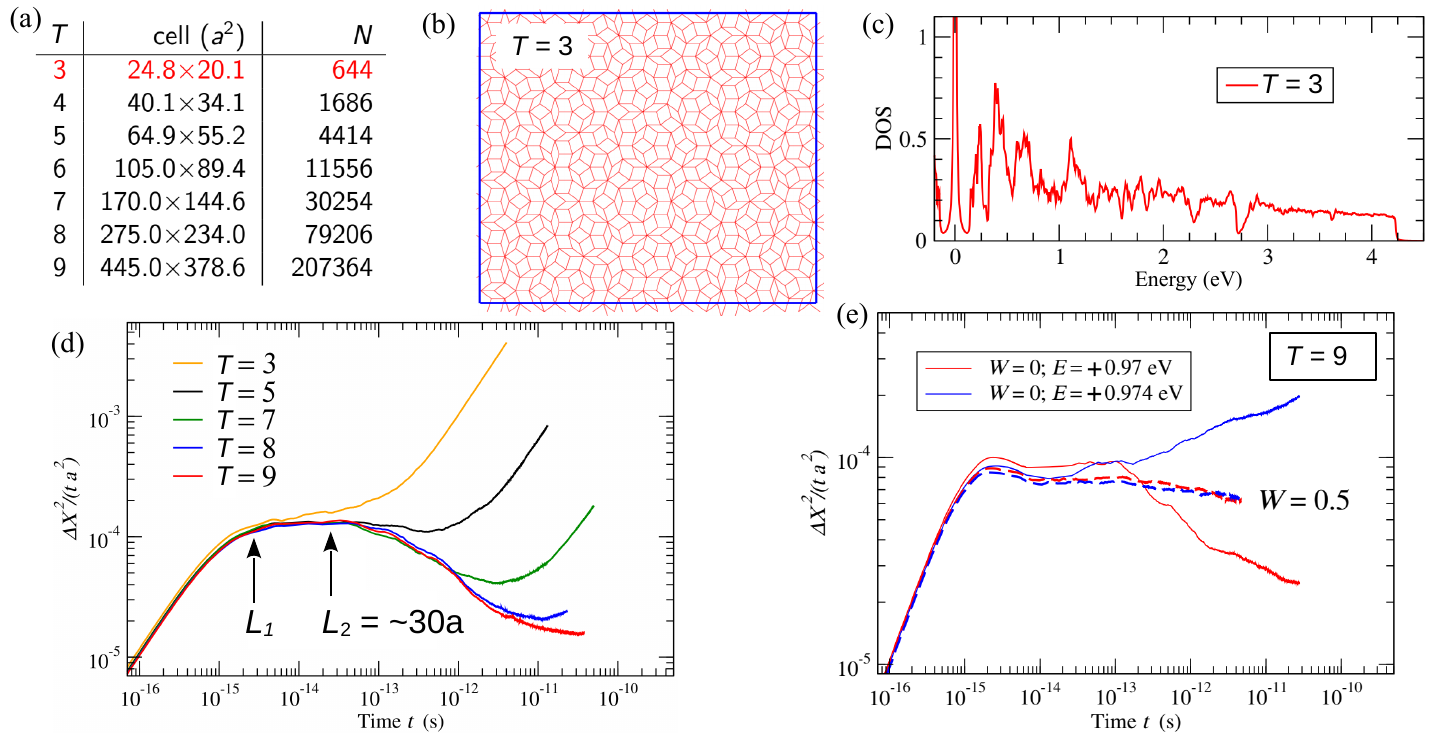}
\end{center}
\caption{\label{Fig_Penrose}
Density of states and quantum diffustion in approximants of 2D Penrose tiling with ``pure hopping'' without defects (except dashed line of panel (e)): 
(a) Series of periodic approximants proposed by M.\ Duneau and M.\ Audier \cite{Duneau94} indexed by an integer $T$, with increasing unit cell size $L$. $N$ is the number of nodes per cell and $a$ the size of the edges. 
(b) Sketch of a cell of the approximant $T=3$.
For pure hopping Hamiltonian ($\epsilon_i = 0$ and $t_0 =1$\,eV):
(c) Total DOS of the approximant $T=3$  \cite{Zijlstra00,Trambly14_QC}.  
(d) Diffusion coefficient $2{D}(t) = \Delta X^2 / t$ for a series of approximants at the same energy $E=0.047$\,eV.
See text for the definition of $L_1$ and $L_2$.
(e) Effect on $2 {D}(t)$ of static defects in large approximant ($T=9$):
${D}(t)$ for two energies corresponding to sub-diffusive ($\beta < 0.5$) and super-diffusive ($\beta > 0.5$) states, respectively; (solid line) without defects, and (dashed line)
with random distribution of Anderson defects $\epsilon_i \in [-W/2;W/2]$. 
To have a realistic order of magnitude of time in second we used the hopping term $t_0 = 1$\,eV.
The atomic positions of the approximants have been kindly provided by A.\ Sz\'all\'as and A.\ Jagannathan \cite{Jagannathan07_penroseAF,Szallas08}.
For the numerical method see \ref{Method_Real}.
Panel (c) is from\,\cite{Trambly14_QC}.
}
\end{figure}

We consider a simple non-interacting  tight-binding (TB) Hamiltonian,
\begin{equation}
\hat{H}_0 = \sum_{i}\, \epsilon_i\, c_i^\dag c_i + \sum_{\langle i,j \rangle} t_0\, c_i^{\dag}c_j,
\label{Eq_hamiltonian}
\end{equation} 
where
$i$ indexes the orbitals located on the vertices, 
$c_i^\dag$ and $c_i$ are the creation and annihilation operators,
and $t_0$ 
is the strength of the hopping between orbitals.
The sum on $\langle i,j \rangle$ is a double sum on the nearest-neighbors at tile edge distance $a$. 
To obtain realistic time values, we use $t_0 = 1$\,eV which is the order of magnitude of 
the hopping parameter in real intermetallic compounds.
For the on-site energy $\epsilon_i$ two cases are studied: 
\begin{itemize}
\item 
To simulate schematically a possible effect of the presence of different chemical elements (``with potential''),
the on-site energy $\epsilon_i$ is proportional to the coordination $\eta_i$ of the site $i$: 
$\epsilon_i = \eta_i t_0 $.
\item To simulate a ``pure hopping'' system: $\epsilon_i = 0 $ $\forall i$.
\end{itemize}
Some results for octogonal approximants ``with potential'' \cite{Trambly17} and Penrose approximants ``pure hopping'' are presented in Figs.\,\ref{Fig_octogonal} and \ref{Fig_Penrose}, respectively.
Note that qualitatively similar results are obtained for octogonal approximants ``pure hopping'' and Penrose approximants ``with potential''.

In the framework of the Kubo-Greenwood approach\,\cite{Kubo57,Greenwood58,Kubo91} 
for calculation of the conductivity,
we use the polynomial expansion method developed by D.\ Mayou, S.~N.\ Khanna, S.\ Roche, and 
F.\ Triozon \cite{Mayou88,Mayou95,Roche97,Roche99,Triozon02} 
(see \ref{Sec_QD_and_dcCond} and \ref{Method_Real})
to compute the mean square spreading $\Delta X(E,t)$ (Eq.\,(\ref{Eq_DX2_def})) of the wave packet at time $t$ and energy $E$, 
and the diffusion coefficient, 
\begin{equation}
{D}(E,t) = \frac{\Delta X^2(E,t)}{2 t},
\label{Eq_DiffusionCoef}
\end{equation}
which tends towards diffusivity at large $t$ (\ref{Sec_QD_and_dcCond}).
${D}(E,t)$ is shown in Figs.\,\ref{Fig_octogonal} and \ref{Fig_Penrose} for the octagonal and the Penrose tiling, respectively.
The ballistic regime due to the periodicity of the approximant is reached at very large $t$, 
when $\Delta X(t) > L_{cell}$ where $L_{cell}$ is the approximant cell size;
then $L(t) = V_{\B} t$, where $V_{\B}$ is the Boltzmann velocity, 
i.e.,\ the intraband velocity in periodic approximants
(Eq.\,(\ref{Eq_calcul_VB}) in the appendix). 
For the approximants with the largest cell and at realistic time range,  Figs.\,\ref{Fig_octogonal} and \,\ref{Fig_Penrose} show that
this Boltzmann term is negligible and, for all purposes of this discussion, 
the largest approximants are similar to the quasiperiodic system.
For approximants with a smaller unit cell, at very large times, a ballistic behavior is clearly achieved which is due to the periodicity of the approximants. 
Indeed, we have checked that at these times, the mean spreading $\Delta X$ is larger than the cell size of the approximants. 
For the largest approximants, this regime has not yet been reached.     

In both cases, depending on the time values, three different regimes are observed at each energy (see Figs.\,\ref{Fig_octogonal}(d,e,f) and \ref{Fig_Penrose}(d,e)):
\begin{itemize}
\item
At very small times, typically when $\Delta X(t) < a$ 
where $a$ is the length of the base tile edge, the mean spreading increases linearly with $t$,  
$\Delta X(t) = V_0 t$ (ballistic behavior), where $V_0 > V_{\B}$ \cite{Trambly06,Trambly14_CRAS}.
Indeed, the effective velocity $V_0$ includes intraband terms of the velocity operator and interband terms  of the velocity operator between band states with small energy difference $\Delta E \lesssim \hbar / t$.
\item 
For larger times, corresponding to $\Delta X(t) \gtrsim {\rm a~few~distances~}a $, the propagation seems to become diffusive as the diffusion coefficient is almost constant, ${D}(t) \approx {D}_{dif}$. 
Therefore $L_1$, defined by
$L_1 = {{D}_{dif}}/{V_0} \approx {\rm a~few~}a $,
is a kind of effective elastic scattering length, but it is not due to static scattering events since we consider perfect tilings. 
The corresponding effective elastic scattering time is $t_1 = L_1 / V_0$. 
Roughly speaking, it seems that when $\Delta X(t) \gtrsim L_1$,
i.e.,\ $t \gtrsim t_1$, the wave packet feels a random tiling.
\item  
Another distance $L_2$ (respectively another time $t_2$, $\Delta X(t_2) = L_2$) appears.
For $\Delta X(t) > L_2 \approx {\rm a~few~}10a$, a new regime appears and ${D}(t)$ follows a power law.
It is thus characteristic of the medium and long-range quasiperiodic order. 
Figures\,\ref{Fig_octogonal}(d) and \ref{Fig_Penrose}(e)  shows that the $\beta$ value can switch from a sub-diffusive regime ($\beta < 0.5$) to a super-diffusive regime ($\beta > 0.5$) over a small variation of energy.
The $t_2$ values, $t_2 \approx 10^{-13}$--${10^{-14}}$\,s, have the order of magnitude of the scattering time above which measurements show unusual transport properties in quasicrystals \cite{Berger94}.
\end{itemize}
Both distances $L_1$  (a few $a$), $L_2$ (a few 10$a$), and the exponent $\beta$ at time $t> t_2$, depend a lot on the energy value $E$. 
$L_1 < L_2$, but at some energy it even seems that $L_1 \approx L_2$. 
Further analysis is necessary to understand this energy dependence. 

Since the sub-diffusive states and the super-diffusive states are due to long-range quasiperiodic order their diffusive properties should be very sensitive to disorder. 
This is verified by the ${D}(E,t)$ calculation in the presence of a weak disorder (Anderson disorder) showing (Fig.\,\ref{Fig_Penrose}(e)) that a sub-diffusive state and a super-diffusive state become diffusive in the presence of disorder. 
These numerical calculations clearly show that on one hand, at certain energies, when the state is sub-diffusive, the quasiperiodic order localizes the charge carriers sufficiently for their behavior to be similar to that of a semiconductor,
and then the increase of disorder that destroys the effect of quasiperiodic order leads to an increase in diffusivity and thus in conductivity. 
This regime is therefore a semiconductor like regime.  
On the other hand, for other energies, when the state is super-diffusive, the regime is metallic like and the increase of disorder leads to a decrease in conductivity. 

To summarize, in the octagonal tiling and the Penrose tiling,
the charge carrier propagation is determined by the wave-packet spreading in the quasiperiodic lattice. 
From numerical calculations, two length scales seem to characterize this quasiperiodic spreading. 
$L_1$, typically $L_1 = {\rm a~few~} a$, above which the propagation is almost diffusive in spite of the absence of static defects. 
$L_2$, typically $L_2 = {\rm a~few~} 10a$, above which specific quasiperiodic symmetries lead to a 
power-law dependence of the root mean square spreading, $\Delta X(t) \propto t^{\beta}$. 
For some energies,
states are super-diffusive or diffusive, i.e.,\ $\beta \ge 0.5$, 
whereas for other energies, a sub-diffusive regime, i.e.,\ $\beta < 0.5$, sets in as expected for critical states characteristic of quasiperiodicity. 
This sub-diffusive regime is the generalization to quasicrystals of the non-Boltzmann propagation (section \ref{Sec_SVR}) found in
realistic approximants of i-AlMnSi and i-AlCuFe, the complex intermetallic alloys $\lambda$-AlMn (section \ref{Sec_SVR_approx}), and in small approximants of octagonal and Penrose tilings \cite{Trambly11_QC,Trambly14_QC}.
This sub-diffusive regime, which is not created by disorder in the atomic structure, but by the quasiperiodic order itself, results in a particular localization that corresponds to the SVR and that can be destroyed by disorder, resulting in semiconductor like behavior. 

In conclusion, the diffusion properties of charge carrier states in quasicrystals and approximants are very complex. The absence of periodicity in quasicrystals prevents the definition of flat bands. However, numerous studies and the results that we have presented here show the existence of confinement mechanisms specific to the quasipersodic structure, similar to flat-band states found in approximants. 
These states seem to be confined on multiple scales in real space, probably linked to the scale properties of the quasiperiodicity. 
It seems that these states can be found at all energies while being very close to non-localized states (whose regime is metallic).
New and more detailed studies still need to be carried out to understand the links between these states with very different transport regimes.

\subsection{Twisted bilayer graphene (TBG)}

\label{Sec_QD-TBG}

Experimentally, graphene can be formed in multilayers on SiC \cite{Ohta06,Brihuega08,Hass06,Hass08b,
Hass08,Sprinkle09,Hicks11} but also on metal surfaces such as Ni \cite{Luican11} and in exfoliated flakes \cite{Li10}  where hopping terms between successive layers play a crucial role. 
While on the Si face of SiC, multilayers have an AB Bernal stacking with electronic properties different from graphene\,\cite{Latil06,Ohta06,Brihuega08,Varchon08,
McCann13,Rozhkov16}, 
on the C-face multilayers are twisted multilayers of graphene  with various angles of rotation between two successive layers.
For large rotation angle $\theta$ between two layers, multilayers show graphene-like properties even when they involve a large number of graphene layers. Indeed, as shown by ARPES \cite{Hass08,Sprinkle09,Hicks11}, STM \cite{Miller09,Brihuega12}, transport \cite{Berger06}, and optical transitions \cite{Sadowski06}, their properties are characteristic of a linear graphene-like dispersion at K points of the Brillouin zone (Dirac cone). 
Therefore, in TBG interlayer hopping terms do not systematically destroy graphene-like properties, but they can lead to the emergence of new behavior induced by the moir\'{e} patterns\,\cite{Campanera07,Mele10,Gratias23} (Fig.\,\ref{Fig_moire}(a)) that is accentuated
when the  angle $\theta$  is very small\,\cite{LopesdosSantos07}.
Since the early 2010s, theoretical studies \cite{Trambly10,Bistritzer10, SuarezMorell10,Bistritzer11} and experimental evidences \cite{Li10,Brihuega12} have shown that moir\'e TBGs can confine conduction electrons in a controlled manner.

\subsubsection{Geometry of twisted bilayer graphene}
\label{Sec_TB-TBG}

\begin{figure}[t!]
\begin{center}

{(a)} ~~~~~~~~~~~~~~~~~~~~~~~~~~~~~~~~~~~~
~~~~~~~~~~~~~~~~~~~~~~~
{(b)}~~~~~~~~~~~~~~~~~~~~~~~~~~~~~~~~~~~~~~~~~~~~~~~~~{(c)}~~~~~~~~~~~~~~~~~~~~~~~~~~~~~~~~~~

\vskip -0.5 cm
\includegraphics[width=5.3cm]{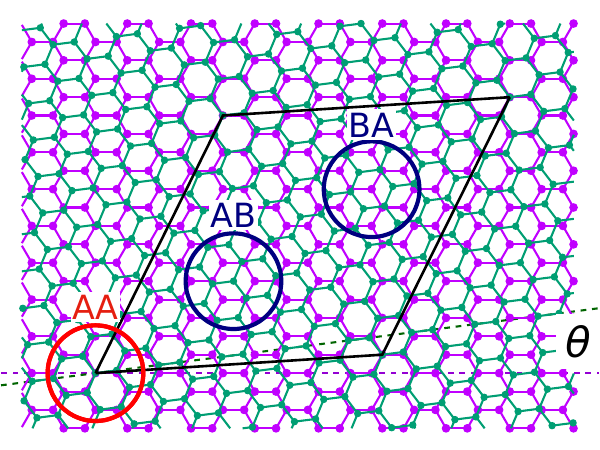}
~~\includegraphics[width=4.4cm]{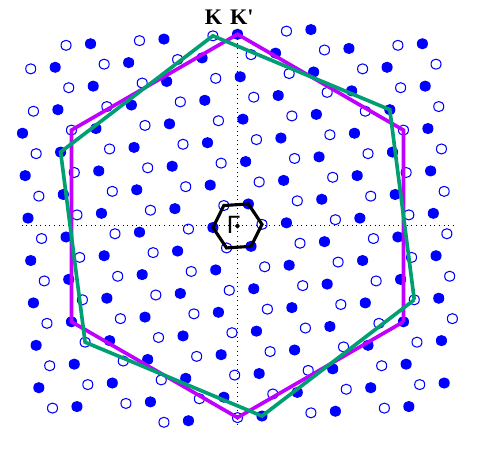}
~~\includegraphics[width=3cm]{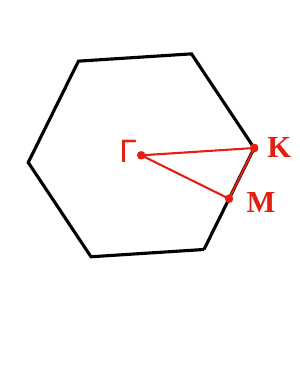}

~~~~~~~~~~~~~~~~~~~~~~~~~~~~~~~~~~~~~~~~~~~~~~~~~

\vskip -0.5 cm
\includegraphics[width=7cm]{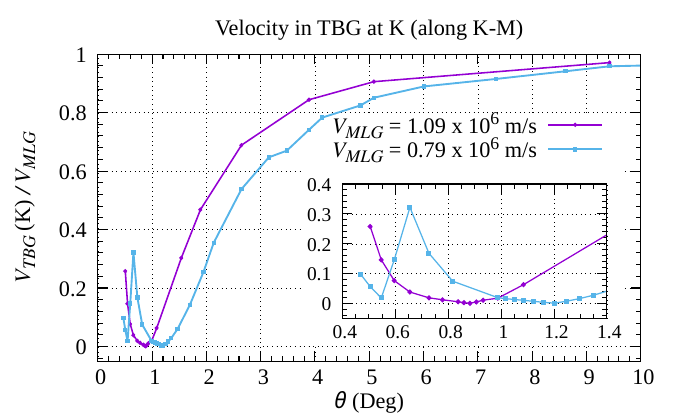}
\includegraphics[width=9.2cm]{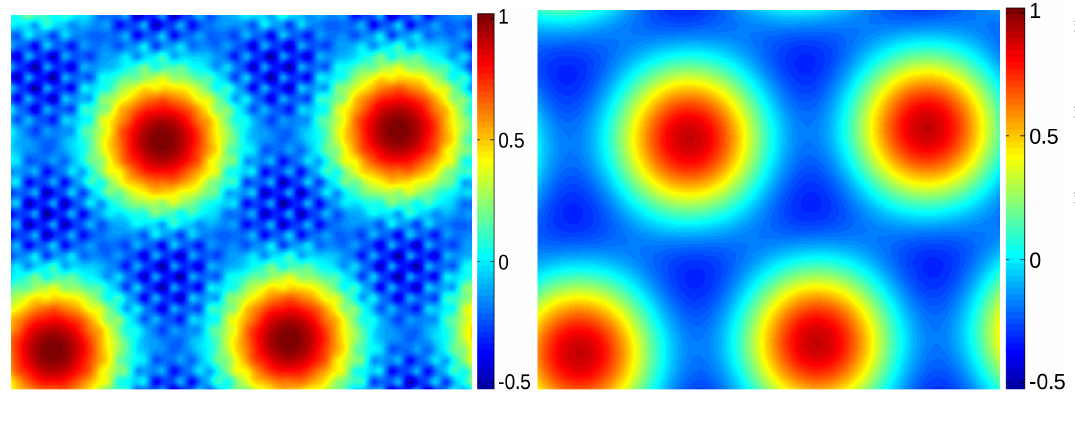}

\vskip -4.3cm
{(d)} ~~~~~~~~~~~~~~~~~~~~~~~~~~~~~~~~~~~~~~~~~~~~~~~~~~~~~~~~~~~~~~~~~~~~~~{(e)}~~~~~
~~~~~~~~~~~~~~~~~~~~~~~~~~~~~~~~~~~~~~~~~~~(f)~~~~~~~~~~~~~~~~~~~~~~~~~~~~~~~~~~~~~~~~~~~~~~
\vskip 3.9cm

{(g)} ~~~~~~~~~~~~~~~~~~~~~~~~~~~~~~~~~~
~~~~~~~~~~~~~~~~~~~~~~~~~~~~~~~~~~~~~~~~~~~~~~~~~~~~~~~~~~~~~~~~~~
~~~~~~~~~~~~~~~~~~~~~~~~~~

\vskip -0.7 cm
\includegraphics[width=0.7\linewidth]{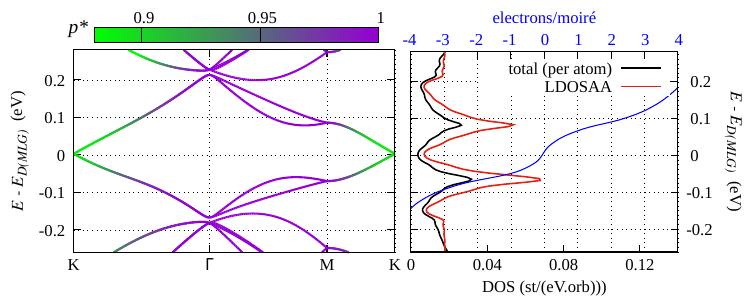}

\caption{\label{Fig_moire}
Commensurate TBG with rotation angle $\theta$ between the two layers.
(a)  Sketch of moir\'e pattern for (4,5) TBG, i.e.,\ $\theta= 7.34^\circ$. 
The black line shows the moir\'e cell containing $N=244$ C atoms.
(b) K and K' points of the (4,5) TBG reciprocal lattice.  
The black line shows the first Brillouin zone of the TBG, the green and magenta lines are the first Brillouin zone of the two graphene layers.
(c) Zoom on the TBG Brillouin zone.
(d) Intraband velocity at Dirac energy $E_D$ (computed from Eq.~(\ref{Eq_calcul_VB_derivee}) at the K point) versus the rotation angle $\theta$, for two TB models corresponding to 
$V_{MLG} = 1.09\cdot10^6$\,ms$^{-1}$ which correctly reproduces the position of the Van Hove singularity found experimentally in 
TBG as a function of the angle \cite{Brihuega12},  and 
$V_{MLG} = 0.79\cdot10^6$\,ms$^{-1}$, which reproduces DFT calculations \cite{Trambly10,Trambly12}, respectively. 
Some data from Refs. \cite{Trambly10,Trambly12,Mesple21}.
(e,f,g) Electronic structure of (12,13) TBG ($\theta = 2.66^\circ$):
SK-TB local DOS (LDOS) (arbitrary units) in real space for energy close to $E_D$, 
(e) SK-TB calculation (f) analytical model (Eq.\,(\ref{eqLDOS}))~\cite{Omid20};  
and 
(g-left) SK-TB Bands, the line color is the average layer participation ratio $p^*$, 
$p^* = 0.5/(P_1^2 + P_2^2)$, where $P_i$ it the weight of the eigenstate on layer $i$ \cite{Trambly16}. 
$p^*=0.5$ when eigenstates are located on one layer only, and $p^*=1$ when eigenstates are equally distributed on the two layers. 
(g-right) SK-TB total DOS, LDOS at the center of an AA stacking region, and (blue line) integrated total DOS with respect to the neutral situation (half filling).
(e,g) are calculated with the SK-TB model
corresponding to a monolayer intraband velocity 
$V_{MLG} = 0.79\cdot10^6$\,ms$^{-1}$ (Sec.\,\ref{Sec_ElecStrucCal}).
TBGs are rigid, i.e., without atomic relaxation. 
}
\end{center}
\end{figure}

A single layer of graphene consists of carbon atoms arranged in a honeycomb lattice such that the unit cell includes two  symmetrically equivalent
carbon atoms. 
Our starting point is AA bilayer graphene, where all the atoms in one layer are on top of an atom in the other layer; 
we choose the rotation origin O at an atomic site,
and one of the two layers is rotated by an angle $\theta$ with respect to the other layer.
A periodic 
commensurate bilayer structure parameterized by two integers $n$, $m$ can be constructed using the method of Refs.\,\cite{LopesdosSantos07,Campanera07,Mele10,Trambly10,Gratias23}.  
The rotation angle for  such a commensurate structure is then given by,
\begin{equation}
\cos\theta=\frac{n^2+m^2+4mn}{2(n^2+m^2+mn)} \, ,
\label{eq:angle}
\end{equation}
and the vectors of the TBG superlattice are 
$\mathbf{t}_1=n \mathbf{a}_1+m \mathbf{a}_2$ and 
$\mathbf{t}_2=-m \mathbf{a}_1+(m+n)\mathbf{a}_2$,
where $\mathbf{a}_1$ and $\mathbf{a}_2$ are the
lattice vectors of the non-rotated single graphene layer, 
$\mathbf{a}_{1,2} = a (\sqrt{3},\pm1)/2$, 
with $a=2.456$\,{\rm \AA}.
The number of atoms in the moir\'e cell is given by $N=4(n^2+m^2+mn)$.
The rotation angle $\theta$ is a good parameter to describe the system, but the number of atoms $N$ is not since cells of equivalent size
can be found for different angles. 
For $\theta$ values less than $\sim 15^\circ$,
TBG forms a moir\'e pattern with (pseudo)period,
\begin{equation}
P = \frac{a}{2 \sin(\theta/2)}.
\label{Eq_periodeMoire}
\end{equation}
When $|n-m|=1$, the periodic unit cell contains 1 moir\'e cell, 
and the smallest commensurable structure with rotation angle close to a given small $\theta$ is obtained for TBG
 $(n,m=n+1)$~\cite{Gratias23}.
Figure\,\ref{Fig_moire}(a) shows the resulting moir\'e pattern for $(n,m)=(4,5)$. 
AA stacking regions are at the corners of the moir\'e cell. 
Along its long diagonal we find Bernal AB and BA stacking regions, where half of the atoms in one layer are on top of an atom in the other layer.

It is also possible to add translation to the rotated layer in addition to rotation. 
For intermediate angles, this can slightly affect the electronic structure without qualitatively changing the results. 
For small angles, when the moir\'e  cell is very large, such a translation does not significantly modify the electronic structure, in line with what the crystallographic analysis suggests~\cite{Gratias23,Gratias20}. 

For $\theta$ angle values that do not correspond to a commensurate structure, e.g. $\theta = 20^\circ$, $30^\circ$ (...), the TBGs are quasiperiodic structures\,\cite{Gratias24}.

\subsubsection{Tight-binding electronic structure of twisted bilayer graphene}
\label{Sec_ElecStrucCal}

Given the large number of atoms per TBG cell, DFT studies of the electronic structure were not originally possible for small angles $\theta$. The first theoretical studies were therefore carried out using simplified models. Two main approaches have been used since the beginning of research on TBGs.
One approach is based on
tight-binding models that include $p_z$ orbitals of all carbon atoms\,\cite{LopesdosSantos07,Trambly10,SuarezMorell10,Bistritzer10,Bistritzer11,Trambly12,Santos12}.
Following this approach, we proposed in 2010 a Slater-Koster tight binding (SK-TB) Hamiltonian \cite{Trambly10,Trambly12} which has been reused in many works in the literature (see for examples Refs.\,\cite{Moon12,Moon13,Wang12,Gargiulo17,
Andelkovic18,Kerelsky2019,Klebl19,Goodwin19,Pathak22,Nguyen22,Sinha24,guerrero24_tBLG_disorder_Kubo,Guerrero25}).
Other TB models have been also proposed (see, e.g., Ref.\,\cite{Lin18}).
Another approach, originally  developed by R.\ Bistritzer and A.~H.\ MacDonald\,\cite{Bistritzer10,Bistritzer11}, consists in a continuum model for the low-energy bands by coupling the Dirac cones of the two graphene layers. This continuum model has been enormously successful and has led to major advances in our understanding of the properties of flat bands.
Some examples are Refs.\,\cite{Weckbecker16,Guinea18,Tarnopolsky19,Guinea19,tbgI,Mora19,andrei2020graphene,XIONG2023129048,Pantaleon21,Mannai21,Haddad23,Cances23,Escudero24,Yu2025}.
DFT studies for TBGs close to the magic angle have been carried out\,\cite{Uchida14,Yndurain19,Leconte22,Pathak22}, as well as other DFT-based TB models \cite{Jung14} and approaches using Wannier functions\,\cite{Fang16,Koshino18,Kang18,Vafek19}.
More recently, a topological heavy-fermion model, combining correlated localized flat-band ($f$)  electrons and itinerant conduction electrons, has also been proposed to understand the unconventional electronic behavior of magic-angle electron states\,\cite{Song22,Rai24}.
It is not possible in this review to cite and detail all these approaches, we will focus on the approach that we have used and developed based on a SK-TB Hamiltonian. 

In order to be used numerically for systematic study, an SK-TB model requires a commensurable structure (Sec.\,\ref{Sec_TB-TBG}) and angle-independent Slater-Koster parameters. 
The non-interacting SK-TB Hamiltonian is\,\cite{Trambly10,Trambly12}, 
\begin{equation}
\hat{H}_0  = \sum_{i} \epsilon_i \, c_{i}^\dag c_{i} + \sum_{\langle i,j \rangle} t_{ij} \,c_{i}^{\dag}c_{j} ,
\label{Eq_hamilt}
\end{equation}
where $c_{i}^\dag$ and $c_{i}$ are the creation and annihilation operators of an electron on orbital $p_z$ at site $i$,
and $\langle i,j\rangle$ is 
the double sum on indices $i$ and $j$ with $i\ne j$.
The coupling matrix element, $t_{ij}$, 
between two $p_z$ orbitals located at $\mathbf r_i$ and $\mathbf r_j$ is calculated by the Slater-Koster formula\,\cite{Slater54},
\begin{equation}
t_{ij} 
~=~   n^2 V_{pp\sigma}(r_{ij}) ~+~ (1 - n^2) V_{pp\pi}(r_{ij}),
\end{equation}
where $n$ is the direction cosine of 
$\mathbf{r}_{ij} = \mathbf{r}_j - \mathbf{r}_i$ along the $z$ axis 
and $r_{ij}$ the distance between the orbitals,
$n={z_{ij}}/{r_{ij}}$, with
$z_{ij}$ is the $z$-coordinate of $\mathbf r_{ij}$. 
$z_{ij}$ is either equal to zero or 
to a constant because the two graphene layers have been kept flat in our model.
The same distance dependence of the two SK parameters, $V_{pp\pi}$ and $V_{pp\sigma}$, is used:
\begin{eqnarray}
V_{pp\pi}(r_{ij})    = V_{pp\pi}^0 \,{\rm e}^{q_{\pi}    \left(1-\frac{r_{ij}}{a_0} \right)}  \,F_c(r_{ij}), \label{eq:tb0}
~~~~~
V_{pp\sigma}(r_{ij}) = V_{pp\sigma}^0 \,{\rm e}^{q_{\sigma} \left(1-\frac{r_{ij}}{a_1} \right)} \,F_c(r_{ij}) ,
\label{eq:tb1}
\end{eqnarray}
where $a_0$ is the nearest-neighbor distance within a layer, 
$a_0=a\sqrt{3}=1.418$\,{\rm \AA}, and $a_1$ is the
interlayer distance, $a_1=3.349$\,{\rm \AA}. 
First neighbors interaction in
a plane is taken equal to the commonly used value, $V_{pp\pi}^0 = - t_0 = -2.7$\,eV\,\cite{Castro09_RevModPhys}.
Second neighbors interaction $t_0'$ 
in a plane is set  to
$0.1 t_0$ \cite{Castro09_RevModPhys} 
which fixes the value of the ratio $q_{\pi}/{a_0}$ in equation (\ref{eq:tb0}).
The interlayer coupling between two $p_z$ orbitals in $\pi$ configuration 
is $V_{pp\sigma}^0$. $V_{pp\sigma}^0$ is fixed to obtain a good fit with DFT 
calculation around the Dirac energy in AA stacking 
and AB bernal stacking  which results in 
$V_{pp\sigma}^0=0.48$\,eV\,\cite{Trambly10,Trambly12}.
We chose the same coefficient of the exponential decay 
for $V_{pp\pi}$ and $V_{pp\sigma}$,
\begin{equation}
\frac{q_{\sigma}}{a_1} ~=~ \frac{q_\pi}{a_0} 
~=~ \frac{{\ln} \left(t_0/t_0' \right)}{a - a_0}
~=~ 2.218\,\mbox{\AA}^{-1},
\end{equation}
with $a = 2.456$\,{\rm \AA} the distance between second neighbors in a monolayer of graphene. 
In (\ref{eq:tb0}), 
a smooth cut-off function\,\cite{Mehl96} is introduced to avoid the unrealistic cut-off effects that can lead to the formation of false mini-gaps\,\cite{Omid20},
\begin{equation}
\ds F_c(r) ~=~  \left( 1 + {\rm e}^{ \frac{r - r_c}{l_c} }  \right)^{-1},
\label{Eq_cutoffFunction}
\end{equation}
with
$l_c= 0.5$\,a.u.\,
$ = 0.265$\,{\rm \AA},
and the cut-off distance is
$r_c = 2.5 a = 6.14$\,{\rm \AA}.
Without asymmetric doping, all $p_z$ orbitals have the same on-site energy $\epsilon_i$ 
(Eq.\,(\ref{Eq_hamilt})).
$\epsilon_i$ is set to $= -0.7834$\,eV so that
the energy $E_{D}$ of the Dirac point K is equal to zero in monolayer graphene ($E_{D(MLG)}=0$). 
$\epsilon_i$ is not zero because the intralayer coupling between atoms beyond first neighbors 
breaks the electron/hole symmetry and then shifts $E_{D}$.

As explained above, the SK parameters given below have been tuned to obtain good agreement with low-energy DFT band states for the monolayer graphene, Bernal AB bilayer, and the AA bilayer\,\cite{Trambly10,Trambly12}. 
However, it is well known that these bands yield an intraband Fermi velocity of monolayer graphene, $V_{MLG} = 0.79\cdot10^6$\,ms$^{-1}$, which does not correspond to the experimental values.
Consequently, in order to make comparisons with experimental results, in particular STM spectroscopy results, we have used a SK-TB model with the same parameters as above except $V_{pp\pi}^0 = -3.7$\,eV\,\cite{Brihuega12}, to get $V_{MLG} = 1.09\cdot10^6$\,ms$^{-1}$.
That SK-TB model has been used to successfully reproduce many experimental results for various angles\,\cite{Brihuega12,Cherkez15,Huder18,Huder18b,Mesple21,Mesple23}.
It works very well for TBGs where the two graphene layers remain flat, with or without in-plane relaxation or heterostrain (Sec.\,\ref{Sec_TBG_Heterostrain}).

The best proof of the validity of this SK-TB model for describing real TBGs, and in particular those at the first magic angle, is provided by the excellent agreement between experimental LDOS 
measurements (by STM) and TB calculations without heterostrain \cite{Brihuega12,Cherkez15}, and in the presence of various heterostrains \cite{Huder18,Mesple21}.
These comparisons also show that, contrary to what is often stated in the literature, it does not seem necessary to introduce out-of-plane relaxation into the calculations to simulate the experimental results (for twist angles greater than or equal to the first magic angle, see also Sec.\,\ref{Sec_TBG_Heterostrain}).
We also have shown an excellent agreement between this SK-TB model and STM experiments for 
the effect of a strong in-plane relaxation in the case of super-moiré of several million atoms formed by biaxial heterostrain \cite{Mesple23}.
It is therefore reasonable to assume that this model is valid for analyzing the effects of heterostrain and in-plane relaxation.
However, it is not clear that this SK-TB model is correct in the presence of out-of-plane relaxation, as to our knowledge it has not been validated by detailed comparisons with DFT calculations. Indeed it is possible that the exponential decay (Eq.\,(\ref{eq:tb1})) overestimates the effect of the out-of-plane relaxation.
This is why we do not discuss the effect of out-of-plane relaxation in detail in this review.
However, this does not seem to be a significant issue for the study of TBGs around the first magic angle, as comparisons with the experimental results cited above show that the SK-TB model without out-of-plane relaxation is valid.

\begin{figure}[t!]
\centering
\includegraphics[width=0.72\linewidth]{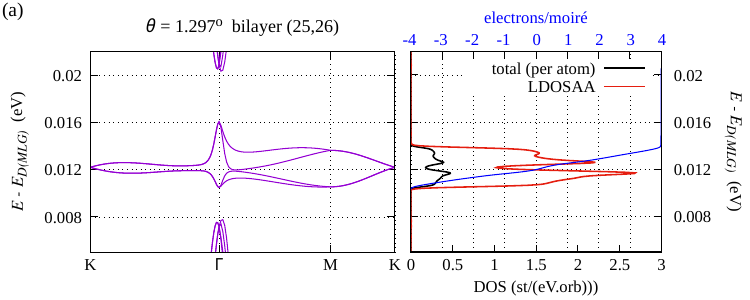}

\includegraphics[width=0.72\linewidth]{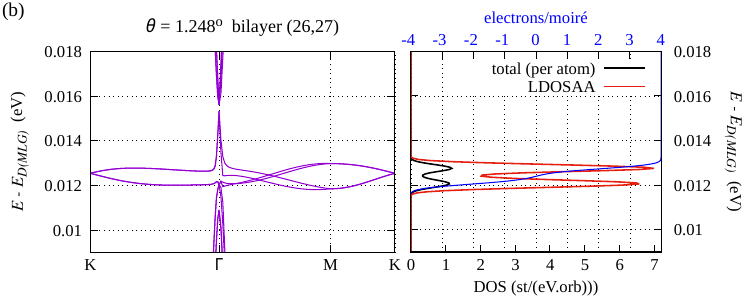}

\includegraphics[width=0.72\linewidth]{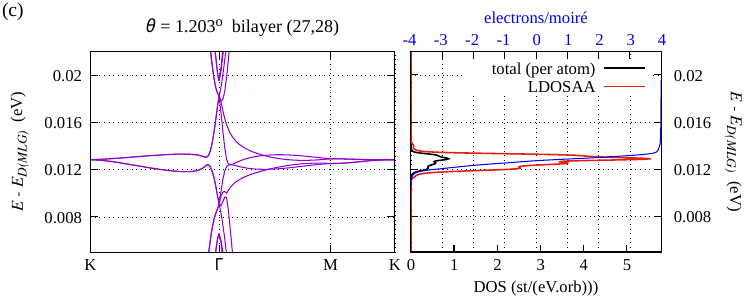}

\includegraphics[width=0.72\linewidth]{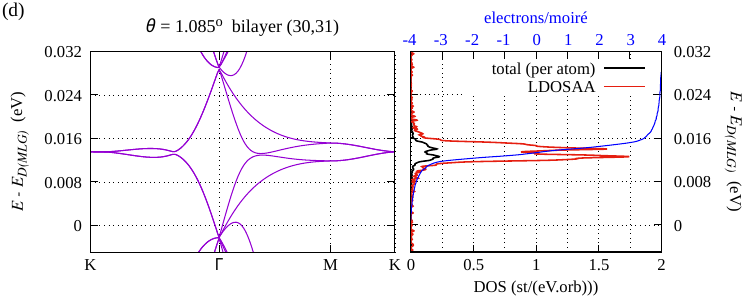}

\caption{\label{Fig_ElecStruc_TBG}
Electronic structure of the four flat bands of non-interacting electon states in magic-angle TBGs with rotation angle $\theta$ close to the first magic angle: TB low-energy bands, total DOS, and local DOS at the center of AA zone, and (blue line) integrated total DOS (electrons per moir\'e) with respect to the neutral situation (half filling). 
TBGs are rigid, i.e., without atomic relaxation. 
The SK-TB model is the one which results in an intraband monolayer velocity $V_{MLG} = 0.79\cdot10^6$\,ms$^{-1}$ (Sec.\,\ref{Sec_ElecStrucCal}).
DOSs are computed by Gaussian broadening of the spectrum calculated by diagonalization using a 48x48 k-grid in the reciprocal cell. 
The Gaussian broadening is 0.1\,meV. 
}
\end{figure}

For rigid TBGs, i.e.,  TBGs without atomic relaxation, theoretical studies have predicted \cite{LopesdosSantos07,Trambly10,SuarezMorell10,Bistritzer10,Bistritzer11,Trambly12,Santos12} the existence of three domains depending on the $\theta$ values: 
\begin{itemize}
\item[$(1)$]
For large rotation angles ($\theta > 20 ^\circ $) the layers are ``decoupled'' (in the sense that the electronic eigenstates are located in only one layer) and then behave as two isolated graphene layers. 
This result, which now seems to surprise no one, is not so easy to understand and it has aroused interest before 2010 when it was established experimentally and by DFT calculations (see, e.g.\,\cite{Hass08}).
TB calculations show that to obtain this monolayer behavior, it is essential to take into account a large number of interlayer couplings, i.e., a large cut-off distance $r_c$ in (\ref{Eq_cutoffFunction}).
Otherwise, a non-physical minigap appears in place of the Dirac cone\,\cite{Trambly16}.
In the monolayer, the Dirac cone comes from the triangular symmetry and the symmetry between the two sub-lattices of graphene (atoms A and B). One might think that the latter is broken in the TBGs, but it seems that long-range interlayer coupling restores it on average.

\item[$(2)$]
For intermediate angles, $ \sim 2^\circ < \theta< 20^\circ $, the dispersion, around the Fermi energy $E_F$, remains linear, but the velocity is reduced (see Fig.\,\ref{Fig_moire}(d,g)). 
Several analytical models~\cite{LopesdosSantos07,Bistritzer11,Santos12,Omid20} have been proposed to simulate this reduction, for example a perturbation of the Dirac cone for intermediate angles leads to a velocity  ratio between TBG and the monolayer,\,\cite{Omid_these,Omid20}
\begin{equation}
\label{Eq_Vrenormalization}
\frac{V_{TBG}}{V_{MLG}} = \frac{  1 }{1+\left(\frac{\theta_0}{\theta}\right)^2 } 
{\rm ~~~with~~}
\theta_0 = \frac{3}{\sqrt{2}\pi}  \frac{t_k}{t_0}
\approx 1.7^\circ,    
\end{equation}
where $t_k \approx 0.1$\,eV is the interlayer coupling term between the closest Dirac cone states of the two layers. 
Consequently, the energies $E_-$ and $E_+$ of the two van Hove singularities are shifted to the Dirac energy $E_D$ when $\theta$ decreases, as it has been shown experimentally \cite{Luican11,Brihuega12,Cherkez15}.
As $\theta$ decreases, the low-energy states are more and more localized in the AA stacking region of the moir\'e pattern, and thus the local DOS (LDOS) in the AA regions increases (Fig.\,\ref{Fig_moire}(e,f,g)). 
The analytical model based on a perturbation of the Dirac cone gives the variation of the  LDOS with respect to the position $\mathbf{r}$ in real space,\,\cite{Omid_these,Omid20}
\begin{equation}
\Delta n(E,\mathbf{r} ) \approx 
n_{MLG}(E)\,
\left( \frac{\theta_{1}}{\theta} \right)^2  \sum\limits_{j=1}^{6}  \, \cos(\mathbf{G}_j \cdot \mathbf{r}),
\label{eqLDOS}
\end{equation}
where $n_{MLG}$ is the DOS of monolayer graphene, $\mathbf{G}_j$ are  6 equivalent vectors of the reciprocal space of the moir\'{e} lattice and $\theta_{1} = \theta_0/\sqrt{3} \approx 1^\circ$.
This model (Fig.\,\ref{Fig_moire}(f)) reproduces very well the LDOS obtained by the SK-TB model (Fig.\,\ref{Fig_moire}(e)), with the exception of the contact between the two types of atoms with different LDOS in AB regions. 

\item[$(3)$] 
For the lowest $\theta$, with $\theta< \sim 2^\circ$ (Fig.\,\ref{Fig_ElecStruc_TBG}),  almost flat bands appear and result in electronic localization in AA stacking regions: states of similar energies, belonging to the Dirac cones of the two layers interact strongly.
In this regime, the velocity of states at the Dirac point goes to almost zero for specific angles, the so-called magic angles \cite{Trambly10,Bistritzer11,Trambly12}.
The following sections are devoted to the properties of these TBGs, which are characterized by low-energy flat bands.
\end{itemize}

In the following sections we consider rigid TBGs, except in section\,\ref{Sec_TBG_Heterostrain} where the effects of relaxation are discussed. 

\subsubsection{Low-energy flat bands in magic angle twisted bilayer graphene}
\label{Sec_flatBands_in_tBLG}

In the domain of small angles ($\theta < 2^\circ$), not all TBGs have low-energy flat bands.
For certain rotation angles, the so-called ``{\it magic angles}'' \cite{Bistritzer11}, the low-energy states, 
i.e.,\ the states with energy close to the Fermi energy for the neutral system, 
are strongly localized in the AA stacking regions of the moir\'e pattern (Fig.\,\ref{Fig_moire}(a))
and form flat bands\,\cite{Trambly10,SuarezMorell10,Bistritzer11}.\footnote{In this section we focus on electronic properties of perfect TBGs, i.e.,\ without defect, at the magic angle. The effect of resonant defects on TBG's conductivity with various rotation angles $\theta$ is presented in more detail in Sec.\,\ref{Sec_TBG_rand}.}
A specificity of this localization by the TBG moir\'e is that it contains very few electronic states: only four bands per moir\'e  (Fig.\,\ref{Fig_ElecStruc_TBG}), i.e.,\ four states over more than 10000 $p_z$ 
orbitals. 
Moreover, these states are located on large areas of the moir\'e (AA stacking regions), i.e.,\ on thousands of orbitals. 
The spatial extent 
of the corresponding wave packet is thus much larger than the interatomic distance. 
By means of a simple quantization argument \cite{Trambly12} in the AA zone of the moir\'e,  it is possible to find a simple condition for the confinement of Dirac electrons by the moir\'e pattern that allows us to predict the magic angle series found numerically from a continuum model\,\cite{Bistritzer11}: 
\begin{equation}
\theta_{magic} \approx \frac{1^\circ}{n} 
{\rm ~~~with~~}n = 1, 2, 3, \dots
\end{equation}
In 2018, it was experimentally proven that this electronic localization by the moir\'e geometry can induce strong electronic correlations and a superconducting state for some fillings of the four flat bands \cite{Cao18a,Cao18b,DINDORKAR2023}.
This new and exciting condensed matter physics joins in many aspects recent studies under the  bicrystallography \cite{Gratias20,Gratias23}, 
i.e.,\ systems
for which the atomic structure results from commensurable or non-commensurable modulation between two periodicities (or pseudo-periodicity), Eq.\,(\ref{Eq_periodeMoire}), of very different length (interatomic-scale periodicity and moir\'e-scale periodicity). 
In those complex systems, the consequences of crystallographic symmetries on the electronic properties are not yet well understood. 

Before detailing the electronic properties of flat bands in TBGs, it is interesting to note that these flat bands are not systematic when there is a moir\'e pattern. 
For example, we have not been able to obtain similar flat bands by calculations using the TB model for bilayer graphene with the similar moir\'e size
but resulting from biaxial strain of one of the two layers (hetero-biaxial strain) without atomic relaxation. 
On the other hand, a very small hetero-biaxial strain (with atomic relaxation) can produce a moir\'e of nearly four million atoms \cite{Mesple23}. 
It leads to strong atomic relaxation with the AA regions almost disappearing, the appearance of swirls that delimit large AB stacking zones, and new electronic confinements in and between these zones.   

Obviously, as shown by the calculation of the intraband velocity $V_{TBG}({\rm K})$ at K (Fig.\,\ref{Fig_moire}(d)), the magic angle value depends on the TB model:  
$\theta_{1^{st} magic} \approx 1.2^\circ$ and 
$\approx 0.9^\circ$ for  
$V_{MLG} = 0.79\cdot10^6$\,ms$^{-1}$
and $1.09\cdot10^6$\,ms$^{-1}$ for the SK-TB model, respectively (Sec.\,\ref{Sec_ElecStrucCal}).
It should also be noted that $V_{\rm K}$ is very low over a fairly wide range of angles of the order of $\Delta \theta \approx 0.2^\circ$. 
This is well confirmed by the fact that the bands are fairly flat, without ever being strictly flat, for a similar range of angles around the so-called magic angle (Fig.\,\ref{Fig_ElecStruc_TBG}). 
In this range of angles, other effects such as uniaxial heterostrain have a very strong effect on the bands \cite{Huder18,Bi19,Mesple21} (see Sec.\,\ref{Sec_TBG_Heterostrain}).

Figure\,\ref{Fig_ElecStruc_TBG} shows typical examples of the four flat bands for different angles around the first magic angle. 
Depending on the value of the angle, these bands can be isolated by minigaps or not strictly isolated, although in the latter case they are almost isolated by pseudogaps with a DOS very close to zero. 
So, from figure\,\ref{Fig_ElecStruc_TBG}, for $\theta=1.248^\circ$, $1.203^\circ$ and $1.085^\circ$
the four flat bands are not separated by minigaps in the rest of the spectrum, whereas this is the case for $\theta=1.297^\circ$.
It is sometimes argued in the literature that atomic relaxation is necessary to obtain these minigaps, but our results show that this is not the case, and that the shape of the bands vary strongly with angle, while remaining fairly flat overall. 
These bands are never perfectly flat, and it is not easy to define a ``good'' criterion for determining the most flat bands. 
Usually, as previously discussed, the intraband velocity of the Dirac cone is used, and thus the magic angle is obtained when this velocity goes to zero. 
However, other criteria such as bandwidth may also be relevant in some cases and may give a different value of $\theta$ for the magic angles.  

The local DOS in the AA zone confirms the main localization of the flat-band states in the AA zone. 
Detailed analysis of these flat-band states, in particular by the participation ratio calculation~\cite{Trambly16}, shows that they are all equally distributed on the two layers, whereas for intermediate angles, the distribution of each state is mainly in one of the two layers (see participation ratio in Fig.\,\ref{Fig_moire}(g)).

\begin{figure}[t!]
\begin{center}
\includegraphics[width=0.95\textwidth]{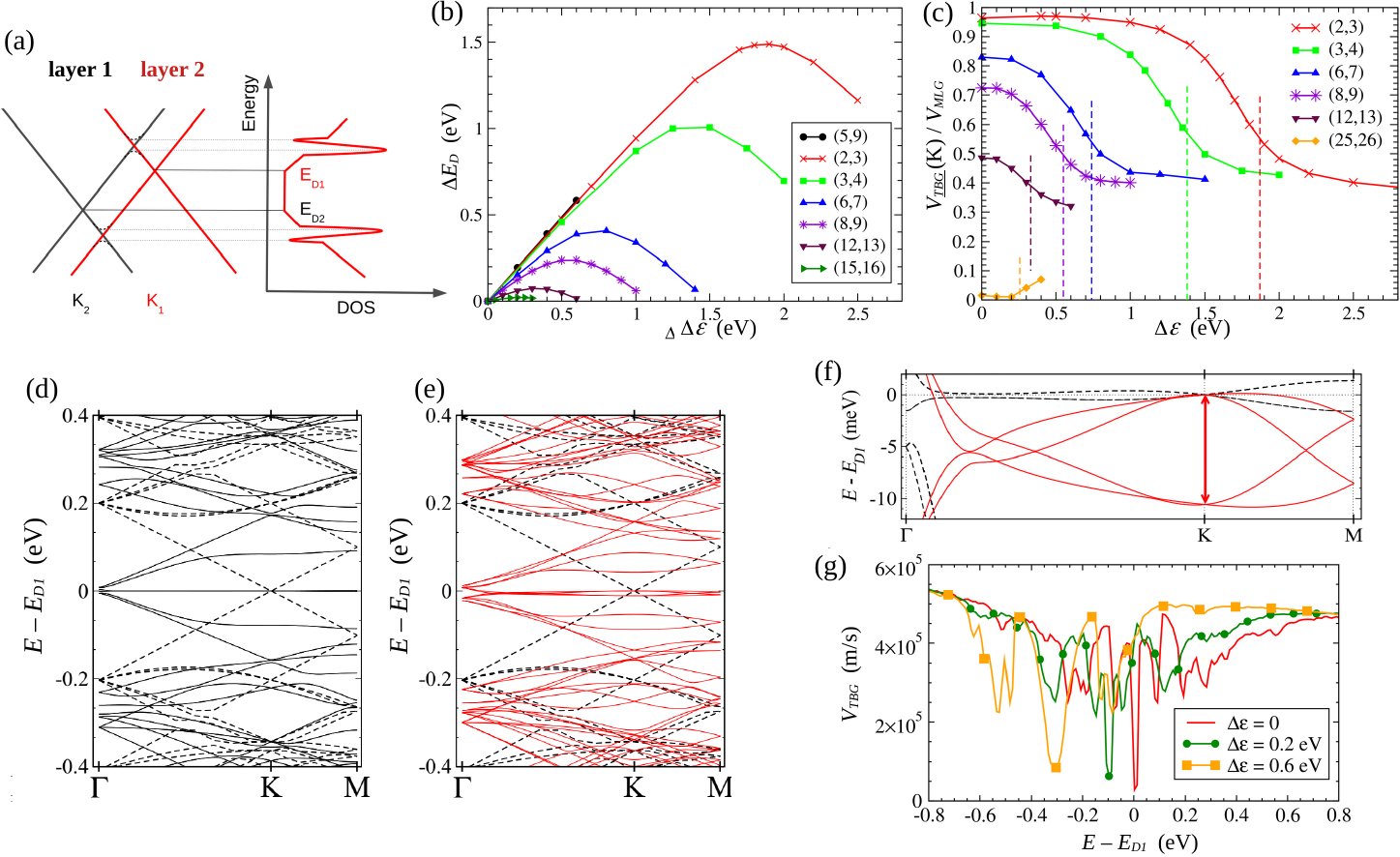}
\end{center}
\caption{\label{Fig_TBG_dope}
Electronic properties of asymmetrically doped TBGs. 
TB calculation with an on-site energy difference 
$\Delta \epsilon$ between the two layers. 
(a) Sketch of the Dirac cones and total DOS of a doped rotated bilayer.
(b) Energy difference $\Delta E_D$ between the energies of  two Dirac cones in TBG versus $\Delta \epsilon$. 
For $\Delta \epsilon$ small,
$\Delta E_D \approx \Delta \epsilon$ for large angles whereas $\Delta E_D < \Delta \epsilon$ for small angles.
(c) Band velocity at the K point versus $\Delta \epsilon$.
The vertical dashed lines show the value $\Delta \epsilon$ for which $\Delta E_D(\Delta \epsilon)$ is maximum (panel (b)).
(d), (e) Bands in (25,26) TBG for $\Delta \epsilon = 0$ and  $\Delta \epsilon = 0.2$\,eV, respectively.
Dashed lines are the monolayer bands.
(f) Zoom of panel (e). 
(g) Average band velocity $V_{TBG}$ versus energy $E$ for (25,26) TBG for different $\Delta \epsilon$ values ($V_{TBG}$ is the Boltzmann velocity computed from the velocity operator and Eq.\,(\ref{Eq_calcul_VB})).
TBG: 
$(5,9)$ $\theta = 18.73^\circ$,
$(2,3)$ $\theta = 13.17^\circ$,
$(3,4)$ $\theta = 9.43^\circ$,
$(6,7)$ $\theta = 5.08^\circ$,
$(8,9)$ $\theta = 3.89^\circ$,
$(12,13)$ $\theta = 2.13^\circ$,
$(25,26)$ $\theta = 1.297^\circ$.
SK-TB model is the one which results in an intraband monolayer velocity $V_{MLG} = 0.79\cdot10^6$\,ms$^{-1}$ (Sec.\,\ref{Sec_ElecStrucCal}).
TBGs are rigid, i.e., without atomic relaxation. 
From\,\cite{Trambly16}.
}
\end{figure}

\label{Sec_TBG_dope}
We now consider asymmetrically doped TBG, 
i.e.,\ TBG with an on-site energy difference 
$\Delta \epsilon=\epsilon({\rm layer\,1})-\epsilon({\rm layer\,2})$ between the two layers (Fig.\,\ref{Fig_TBG_dope}).
It may be due to asymmetrical doping of the two graphene monolayers (e.g., from substrate doping) or to the action of an external electric field perpendicular to the plane, which can be controlled by a gate. 
For large twist angles $\theta$, when the states of layer 1 and layer 2 are decoupled, the asymmetric doping simply shifts the two Dirac cones by $\Delta \epsilon$, so the difference in the two Dirac energies is $\Delta E_D \approx \Delta \epsilon$ for $\Delta \epsilon$ small.  
For smaller $\theta$, intermediate angles $\theta$ with $\theta \gtrsim 3^\circ$, the main effect of a small $\Delta \epsilon$ is to bring the energies of the Dirac cones of each layer closer together (so $\Delta E < \Delta \epsilon$) thus increasing the reduction in electron velocity \cite{Cherkez15,Trambly16} (Fig.\,\ref{Fig_TBG_dope}(b,c)).
For smaller angles close to the magic angle, 
$\Delta \epsilon$ shifts the band energies and modifies them, but without making flatness disappear and the sharp reduction in band velocity is still very much present (Fig.\,\ref{Fig_TBG_dope}(e,f,g)).

\subsubsection{Quantum diffusion}
\label{Sec_TBG_QD}

\begin{figure}[t]
\begin{center}

\includegraphics[width=6cm]{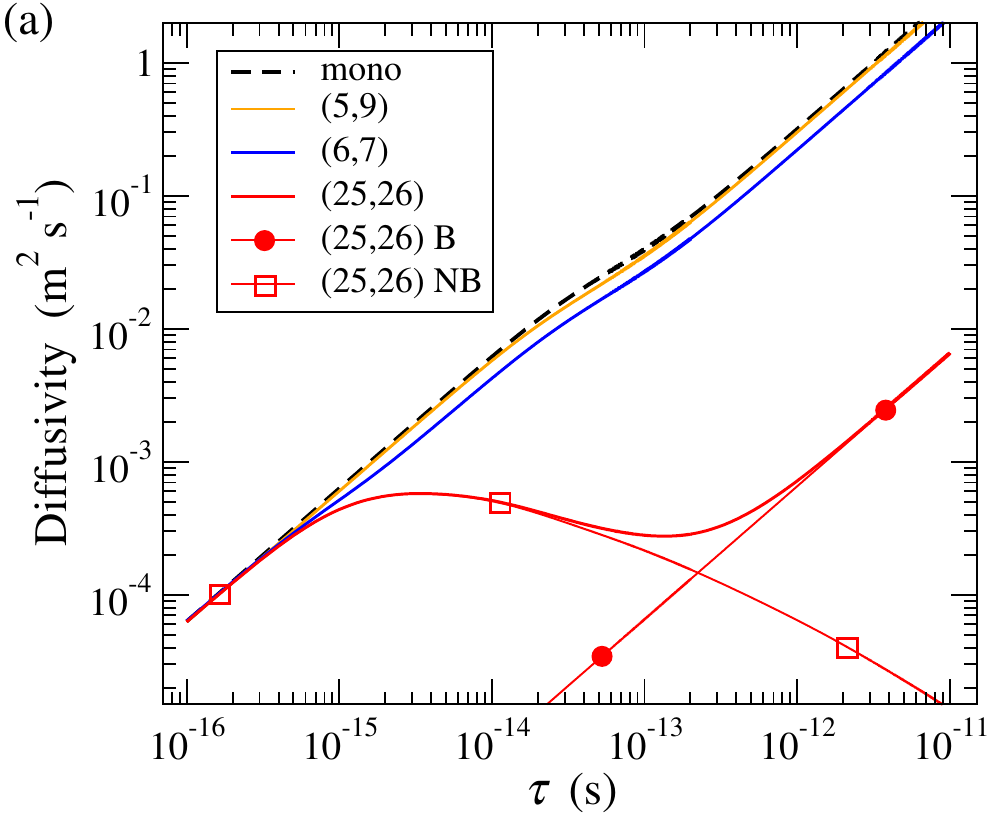}
~\includegraphics[width=6cm]{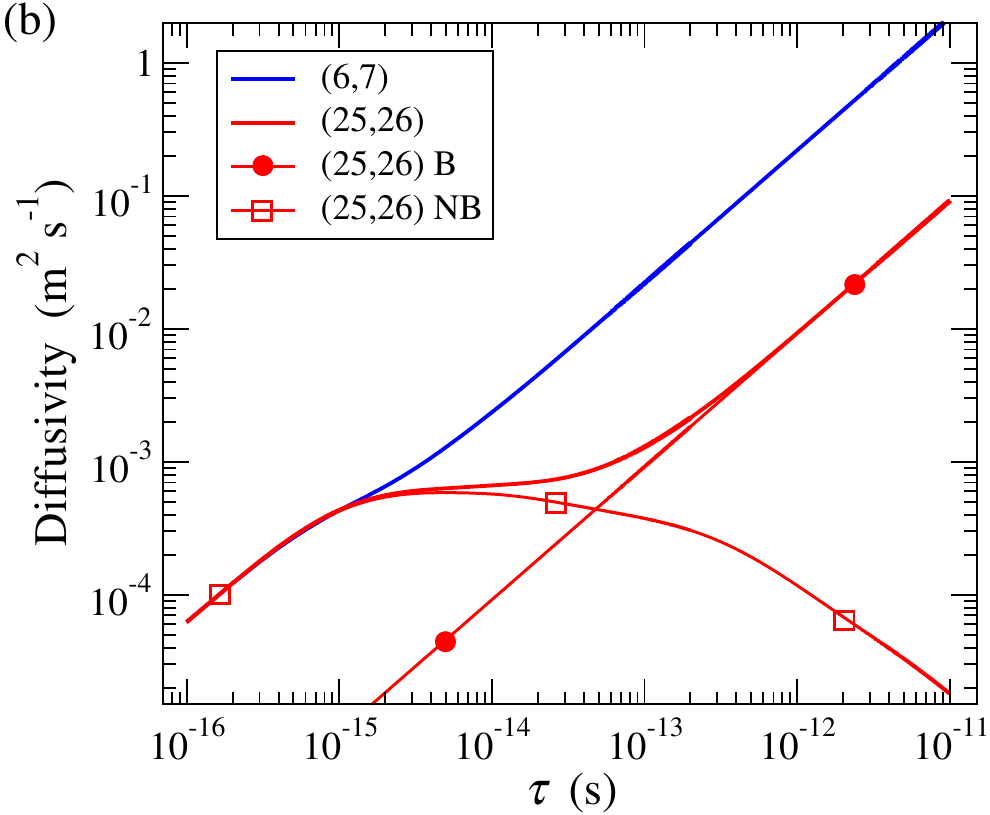}

\caption{
Diffusivity $\mathcal{D}$ versus scattering time $\tau$
in graphene, TBGs (5,9), (6,7), and (25,26), with rotation angle 
$\theta = 18.73^\circ$, $5.08^\circ$ and $1.297^\circ$, respectively.
For (25,26) TBG the Boltzmann (B) term and non-Boltzmann (NB) term are shown,
$\mathcal{D}(E_F,\tau) = \mathcal{D}_{\rm B}( E_F,\tau) + \mathcal{D}_{\rm NB}(E_F,\tau)$ 
(see Eq.\,(\ref{Eq_Dif_B_nonB})). 
These calculations are performed with an energy broadening of 10 meV around $E_F$. 
For (25,26) TBG  that corresponds mainly to the flat bands (Fig.\,\ref{Fig_ElecStruc_TBG}(a)), i.e., the minimum of $V_{\B}$ (Fig.\,\ref{Fig_TBG_dope}(g)).
(a) $\Delta \epsilon=0$, $E_{F}= 0.12$\,eV 
(see bands and DOS in Fig.\ref{Fig_ElecStruc_TBG}(a)), 
and
(b) $\Delta \epsilon=0.2$\,eV, $E_{F}= -0.1$\,eV. 
$\Delta \epsilon$ the energy difference between the on-site energy in the two layers (Sec.\,\ref{Sec_flatBands_in_tBLG}). 
SK-TB model is the one which results in an intraband monolayer velocity $V_{MLG} = 0.79\cdot10^6$\,ms$^{-1}$ (Sec.\,\ref{Sec_ElecStrucCal}).
TBGs are rigid, i.e., without atomic relaxation. 
For the numerical method see \ref{Sec_NumMeth_withoutDef}.
From\,\cite{Trambly16}.
}
\label{Fig_Transp_Dif}
\end{center}
\end{figure}

In this section, we describe the consequences of the particular localization of TBG flat band states on their quantum diffusion properties\,\cite{Trambly16,Andelkovic18,Sinha24,guerrero24_tBLG_disorder_Kubo,GobboKhun25}.
Here we are not considering the defect effect, which produces particular states, and  will be studied in the next section (Sec.\,\ref{Sec_TBG_rand}). 
We are only considering quantum diffusion in the defect-free medium, and we are treating the effect of disorder by a scattering time $\tau$. 
As explained in section \ref{Sec_SVR} (see also \ref{Method_Reciprocal}),
the diffusivity $\mathcal{D}$ at every energy $E$ and scattering time $\tau$ is the sum of a Boltzmann term, 
$\mathcal{D}_{\rm B}(E_{F},\tau) = V_{\B}^2 \tau$, 
and a non-Boltzmann term $\mathcal{D}_{\rm NB}$.  
For large $\tau$,  $\mathcal{D}_{\rm NB}$ decreases when $\tau$ increases, 
and $\mathcal{D}_{\rm NB}\rightarrow 0$ 
when $\tau \rightarrow +\infty$. 
Thus in crystals, 
$\mathcal{D} \approx \mathcal{D}_{\rm B}$ when $\tau \rightarrow +\infty$.

The diffusivity, calculated by diagonalization (\ref{Method_Reciprocal}), for monalayer graphene and several cases of TBGs are presented  in Fig.~\ref{Fig_Transp_Dif} 
for different $E_{F}$ values and for doped or undoped bilayers. 
For graphene and bilayers with large and intermediate rotation angles,  
$\mathcal{D} \approx \mathcal{D}_{\rm B}$
at every energy. 
The only effect of the non-Boltzmann term is a change in the slope of $\mathcal{D}(\tau)$ at scattering time 
$\tau \approx \hbar/E $
as explained in the appendix of Ref.~\cite{Trambly16}.
Eventually at small scattering time, $\tau \ll \hbar/E$, the interband transition between the two bands of each Dirac cone contributes significantly. In the case of graphene, with a first-neighbor coupling Hamiltonian, the non-Boltzmann term $\mathcal{D}_{\rm NB}$ is equal 
to the Boltzmann term $\mathcal{D}_{\rm B}$ and then $\mathcal{D} = 2\mathcal{D}_{\rm B}$ for $\tau \ll \hbar/E$ \cite{Trambly16}. 
This effect is related to the phenomenon of jittery motion, also called Zitterbewegung \cite{Castro09_RevModPhys}, which is important in the optical conductivity. 
In graphene and bilayers with large rotation angles, it occurs for very small scattering time values that are too small to be significant experimentally.
However, in rotated bilayers with very small rotation angle $\theta$ close to the first magic angle, 
for states at energy where velocity is very small
(i.e.,\ energies close to the Dirac energy),  
the Boltzmann terms in the transport coefficients (see equations (\ref{Eq_SVR_D}) and  in \ref{Method_Reciprocal}) decrease and the non-Boltzmann terms  
become significant.
As we have seen in the case of quasiperiodic or approximant structures, this has a strong effect on the charge carrier diffusion regime and therefore on the effect of disorder. 
 
For instance, Fig.~\ref{Fig_Transp_Dif} shows that for 
(25,26) bilayers ($\theta = 1.297^{\rm o}$), the diffusivity
$\mathcal{D}$ is strongly affected by the non-Boltzmann term for realistic $\tau$ values\,\cite{Wu07}. 
It results in a smaller diffusivity with respect to the case of graphene, that is not at all ballistic;
indeed $\mathcal{D}$ is not at all linear as a function of $\tau$ for realistic values of the scattering time, $\tau \approx 10^{-14}$ -- $10^{-13}$s.
Even more,  at charge neutrality (i.e.,\ $E_{F}=0.012$\,eV for (25,26) TBG see Fig.\,\ref{Fig_ElecStruc_TBG}(a)) and without asymmetric doping between the two layers (Fig.\,\ref{Fig_Transp_Dif}(a)),
diffusivity decreases from 
$\mathcal{D}=6\cdot10^{-3}$\,m$^2$s$^{-1}$ to 
$3\cdot10^{-3}$\,m$^2$s$^{-1}$,
as $\tau$ increases from 
$\tau \approx 4\cdot 10^{-15}$\,s and $10^{-13}$\,s.
This decrease in conductivity by a factor of 2 due to increased disorder, which may be static (defects) or dynamic (temperature)
clearly shows that disorder can lead to an increase in conductivity of the flat-band states, as expected in the sub-diffusive regime (see Sec.\,\ref{Sec_SVR_tiling}).
Since flat-band states are localized over large areas of space (several hundreds atoms in AA zones) and their band velocity is very low, it is clear that these states satisfy the SVR condition (Eq.~(\ref{Eq_SVR})).
It also shows indirectly that flat-band states are fragile and can be destroyed by disorder.  
Recent real space calculations of conductivity using Kubo's formula have shown such behavior for a magic-angle TBG in the presence of Anderson disorder \cite{guerrero24_tBLG_disorder_Kubo}. 
In asymmetric doped TBG (Fig.~\ref{Fig_Transp_Dif}(b)), similar effects occur at energies with small Boltzmann velocity (Fig.\,\ref{Fig_TBG_dope}(g)). 
This regime, called small velocity regime (SVR), where non-Boltzmann terms dominate transport properties, has already been observed in systems with complex atomic structure such as quasicrystals  and complex metallic alloys (Sec.~\ref{Sec_SVR_approx}).
Let us recall that SVR is reached when the mean free path $V \tau$ of charge carriers is smaller than the spatial extent $L_w$ of the corresponding wave packet for large time $t$, i.e.,\ $t  \gtrsim \tau$. 

Magic-angle TBGs with very small rotation angle have a huge unit cell and a huge cell of the moir\'e pattern, the flat-band states are confined in the AA zone and have then a very small velocity (Figs.\,\ref{Fig_moire}(d) and \ref{Fig_TBG_dope}(g)). 
Typically the size of the AA zone is $\sim 0.5P$, where $P$ is the moir\'e period (Eq.\,(\ref{Eq_periodeMoire})), and then the extent of confined states in the AA zone is $l \lesssim 0.5P$.
It is therefore reasonable to assume that the length of the wave packet, $L_w$, defined in section \ref{Sec_SVR} is of the order of magnitude of $l$ as confirmed by our calculations (Fig.\,\ref{Fig_DX2-NB}(e)).
As the Boltzmann velocity is very low, the condition (\ref{Eq_SVR}), $ V_{\B} \tau < L_w $, for the small velocity regime (SVR) should be satisfied for magic angles.
In doped and undoped magic-angle TBGs, for which $E_F$  does not correspond to flat-band states, and for TBGs with large twist angle, the Boltzmann velocity is larger and  the non-Boltzmann effect is negligible  (Fig.\,\ref{Fig_Transp_Dif}). 

\subsubsection{Heterostrain and atomic relaxation}
\label{Sec_TBG_Heterostrain}

The angle of rotation $\theta$ between the two graphene layers is the main parameter for the existence of flat moir\'e bands, however for angles close to the magic angle, other parameters, 
such as atomic relaxation and strain,
are also very important and can strongly modify the band structure. 
There are numerous studies on atomic relaxation and the effect of strain in TBGs, which we cannot present here as they would require a review article devoted to them. We will only mention a few general aspects that we consider important in the context of this review.

Many experimental and theoretical studies have discussed the importance of atomic relaxation 
\cite{Uchida14,Wijk15,Dai16,Nam17,Gargiulo17,Choi18,Angeli18,Lin18,Lin18b,Lin18c,Koshino18,Lucignano19,Yoo19,Guinea19,Liu21,Nguyen21,Yananose21,Kazmierczak21,Leconte22,Ezzi24,Ceferino24}.
Indeed, the presence  of different stacking zones AA, AB, SP, etc.\
in a moir\'e pattern, which do not have the same stability in relation to each other\,\cite{Gargiulo17}, strongly suggests that atomic reconstruction (or atomic relaxation) relative to rigidly TBG is likely, and that it will be all the stronger as the size of the moir\'e pattern (Eq.\,(\ref{Eq_periodeMoire})) increases.
It is useful to distinguish between atomic relaxation in the graphene layers (in-plane relaxation) and perpendicular to the graphene layers (out-of-plan relaxation or $z$-modulation).
The latter has been well described\,\cite{Gargiulo17,Koshino18} and mainly stems from the fact that the interlayer distance is not the same for stable AA and AB stacking structures.
It can modify the electronic structure and it is often argued that it is the cause of minigaps that isolate the flat bands.
However, it should be noted that such minigaps can exist without out-of-plane relaxation (see, e.g., Fig.\,\ref{Fig_ElecStruc_TBG}(a)).
Furthermore, to our knowledge, the effect of this relaxation has not been well studied, in a systematic way, by DFT calculations, which are difficult to perform for such large systems.
Finally, it should be noted that SK-TB models have not really been validated to account for this out-of-plane relaxation (Sec.\,\ref{Sec_ElecStrucCal}).
However, the SK-TB model correctly accounts for the effects of the in-plane relaxation.
In the case of graphene twisted bilayers, it has been shown that this effect is very weak for angles close to a first magic angle ($\theta \simeq 1^\circ$) and therefore has no significant effect on the electronic structure. However, 
for angles smaller than the first magic angle, several studies show the importance of in-plane reconstruction\,\cite{Gargiulo17,Guinea19,Nguyen21} with a reduction in AA zones in favor of AB zones, which are energetically more stable. 
For a very large moir\'e of several millions of atoms, this atomic reconstruction is stronger and can lead to new phenomena such as giant swirls associated with new types of electronic localization\,\cite{Mesple23}.

However, in the case of TBG on SiC close to the first magic angle, a detailed study 
by Huder et al.\,\cite{Huder18} and Mesple et al.\,\cite{Mesple21}
shows that the effect of ``heterostrain,'' i.e.,\ an in-plane uniaxial
deformation of one layer with respect to the other, is more important in many samples than the atomic relaxation. 
It is now well established that the heterostrain is present in many samples of TBG, although not well controlled experimentally, and it has strong consequences for the electronic properties of flat bands\,\cite{Huder18,Bi19,Mesple21,Kerelsky2019,Gao21,Pantaleon21,Mannai21,Parker21,XIONG2023129048,Escudero24,Yu2025}.
To study it in a quantitative way,
it is then possible to calculate the local electronic structure from our SK-TB model (Sec.\,\ref{Sec_ElecStrucCal}), without other adjustable parameters. 
This approach has allowed to study several samples that have have been studied from STM images in the literature (see Refs.\,\cite{Huder18,Mesple21,Huder17_these,FlorieThese} and Refs.\ therein). 
The main result is that
for TBGs with angles close to the first magic angle, a small amount of heterostrain is an ubiquitous factor in strongly modifying the electronic structure of the low-energy flat bands.  
For example, the uniaxial heterostrain can induce the appearance of a third peak in the low-energy DOS in addition to two van Hove singularities \cite{Huder18}. 
Moreover heterostrain modifies also the electron-electron interactions in many samples that have Fermi levels in the flat bands \cite{Mesple21}. 
It therefore appears that for the first magic-angle TBGs studied experimentally, 
the heterostrain is a determining parameter, more important than the atomic relaxations. 


\subsubsection{Magnetism}
\label{Sec_TBG_Mag}

\begin{figure}[t!]
\centering
\includegraphics[width=0.95\linewidth]{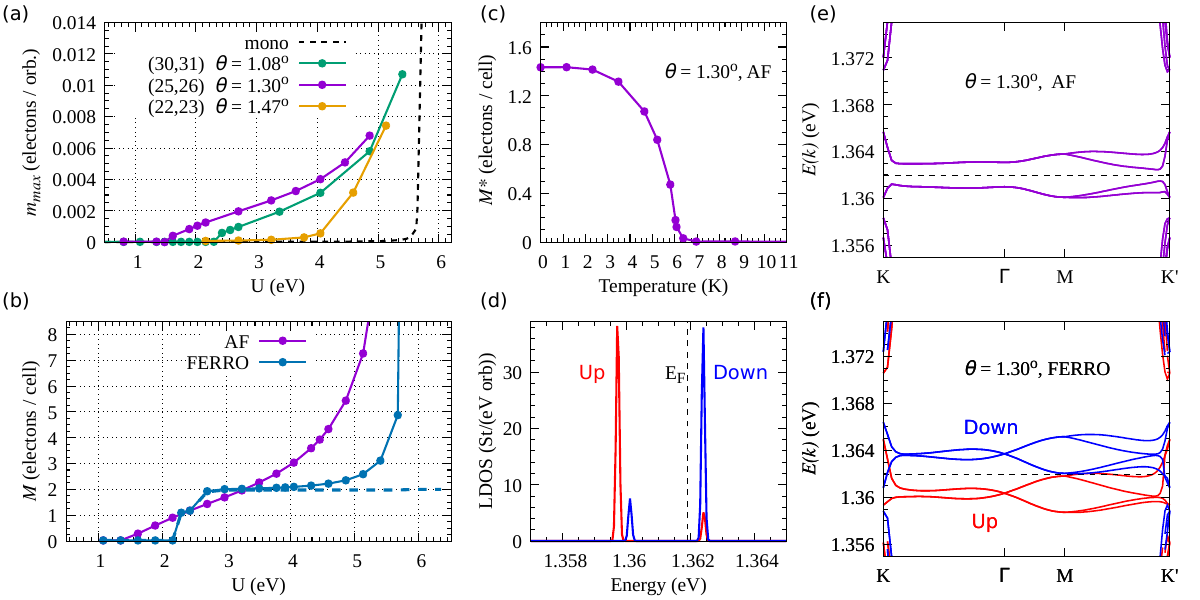}
\caption{MFT results for TBG: 
(a) Magnetization ($\mu_B$) at zero temperature ($T=0$) as a function of $U$ at rotation angles 
$\theta=1.085^\circ$ (bilayer (30,31), $N = 11164$), 
$\theta=1.297^\circ$ ((25,26), $N = 7804$), 
and $\theta=1.47^\circ$ ((22,23), $N = 6076$),
respectively. $N$ is the number of orbitals per moir\'e cell.
The monolayer case is calculated with similar conditions (similar supercell). 
For a monolayer, the critial $U$ above which the solution is magnetic is
$U_{Cm}\approx 2.09 t_0$, where
$t_0=2.7$\,eV is the first intralayer coupling.
The other panels are for $\theta=1.297^\circ$ (25,26) TBG:
(b) Magnetization per moiré cell: (solid line) $M_{tot}^*(U)$ for the AF and FERRO solutions, and (dashed line) $M_{tot}(U)$ for FERRO ($M_{tot} = 0$ for AF), see Eq.\,(\ref{Eq_defMmoire}). 
(c) $M^*_{tot}(U=t_0=2.7\,{\rm eV})$ versus temperature $T$ for the AF solution.
(d) Up and down  local DOSs on an orbital with up magnetization at the center of the AA stacking region for $U=t_0=2.7$\,eV for the AF solution.   
(e) and (f) Bands around $E_F$ of the AF and FERRO solutions for $U=t_0=2.7$\,eV. Up and down bands are degenerate for the AF solution. 
SK-TB model is the one which results in an intraband monolayer velocity $V_{MLG} = 0.79\cdot10^6$\,ms$^{-1}$ (Sec.\,\ref{Sec_ElecStrucCal}).
TBGs are rigid, i.e., without atomic relaxation. 
For details of the MFT calculations see 
Ref.\,\cite{Vahedi21_tBLG}.
}
\label{Fig_Mag}
\end{figure}

The large reduction in kinetic energy in flat bands increases the relative importance of interactions, making bilayer systems much more susceptible to correlation effects than a single layer \cite{Luis17,Cao18a,Sharpe19,Klebl19,Goodwin19,klebl2020,Cea20,Chen22,JimenoPozo23,Dale23,Wilhelm23,Rai24,Jiang25_gap_miniband,Kwan25} (see also \cite{Vahedi21_tBLG} and Refs.\ therein). 
One example is
the magnetization of edge states in graphene nanoribbons \cite{JPSJ.65.1920,JPSJ.67.2089,PhysRevB.59.8271,PhysRevLett.101.246803,Feldner10, Yazyev2010,Wakabayashi10,Tao11,Feldner11,FeldnerThese,Wakabayashi13}, TMD nanoribbons \cite{vahedi21_TMDedge}, or graphene nanoflakes \cite{PhysRevLett.99.177204,Bhowmick2008,Viana-Gomes2009,Feldner10,Yazyev2010,FeldnerThese,Roy2014,Valli2016,Thu20,Thu_these,Phung22}. 
However, in TBGs the flat bands at the origin of strong electronic correlations are not due to defects or edges, but to the particular symmetry of the moir\'e pattern. 
We have first investigated \cite{Vahedi21_tBLG} the question of magnetic instabilities at half filling of TBG using
our tight-binding description of the non-interacting bilayer systems (Eq.~(\ref{Eq_hamiltonian})) to which we add an on-site Hubbard interaction $U$ in order to model the Coulomb repulsion between electrons of opposite spin
\cite{Hubbard1963}, 
\begin{equation}
\hat{H} = \hat{H}_{0\uparrow} +  \hat{H}_{0\downarrow} + U \sum_{i\, \sigma} n_{i\sigma} 
n_{i \overline{\sigma}} - \mu_c \sum_{i\, \sigma} n_{i\sigma} ,
\end{equation}
where $\sigma =\, \uparrow$ ($\downarrow$) is the spin up (down), $\hat{H}_{0\sigma}$ is the non-interacting SK-TB Hamiltonian for each spin (Eq.\,(\ref{Eq_hamilt})),
and  $n_{i\sigma} =  c_{i\sigma}^{\dag}c_{i\sigma}$ 
is the number operator at site $i$ and $\mu_c$ the chemical potential. 
In the static mean-field (MFT) approximation, 
assuming that local magnetic moments are collinear to the $z$-direction (Hartree approximation), the Hamiltonian can be separated in two Hamiltonians for spin up and down, 
respectively (see Refs.~\cite{Feldner10,Yazyev2010,FeldnerThese,Wakabayashi13,Marcin19,Thu20} and Refs.\ therein),
\begin{equation}
\hat{H}_{\sigma} = \sum_{i} \epsilon_i c_{i\sigma}^\dag c_{i\sigma} + \sum_{\langle i,j \rangle} t_{ij} c_{i\sigma}^{\dag}c_{j\sigma} + U\langle n_{i \overline{\sigma}}\rangle\, n_{i\sigma} + {\rm Const}.
\label{Eq_Hamil_Hub}
\end{equation}
The site-dependent average densities $\langle n_{i\uparrow}\rangle$, $\langle n_{i\downarrow} \rangle$ are numerically determined self-consistently. This allows to calculate the magnetization of each site,
\begin{equation}
m_{zi} = \frac{1}{2} \left( \langle n_{i\uparrow} \rangle - \langle n_{i\downarrow} \rangle \right),
\end{equation}
the  maximum local magnetization,
\begin{equation}
m_{max} = {\rm Max} \left\{ |m_{zi}| \right\}_i,
\end{equation}
the total magnetization per moir\'e pattern, $M_{tot}$, 
and the 
sum of local moments per moir\'e pattern, $M^*_{tot}$,
\begin{equation}
M_{tot} = \sum_{i=1}^N  m_{zi} 
{\rm ~~~and~~}
M_{tot}^* = \sum_{i=1}^N  |m_{zi}| ,
\label{Eq_defMmoire}
\end{equation}
where $N$ is the number of orbitals per moir\'e cell.

\begin{figure}[t!]
\centering
\includegraphics[width=0.9\linewidth]{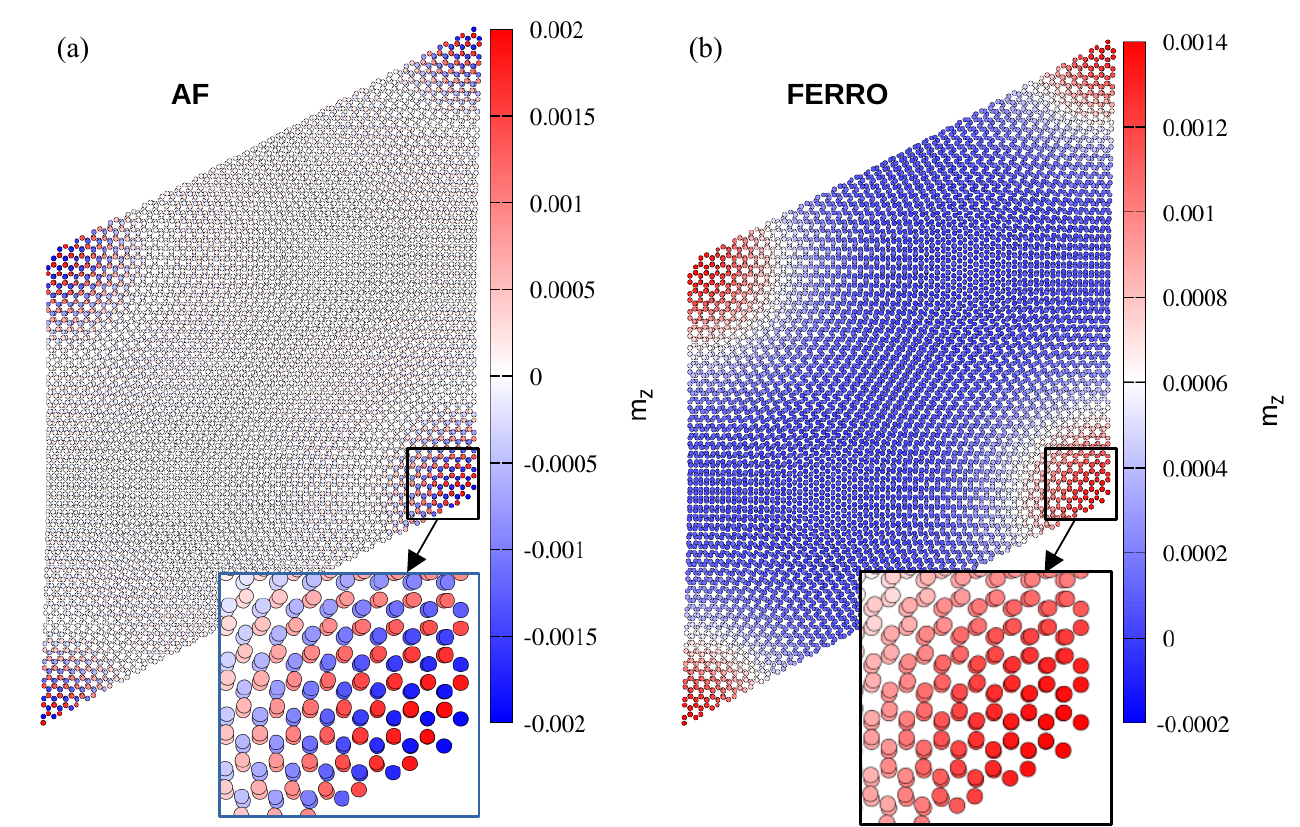}
\caption{MFT result for the spatial magnetization profile, $m_{zi}( \mathbf{r}_i)$, in a moir\'e cell (unit cell) of TBG at $\theta=1.297^\circ$ ((25,26) $N = 7804$) with $U=t_0=2.7$\,eV: 
(a) Antiferromagnetic (AF) solution 
($M_{tot} = 0$, with intralayer AF order and interlayer FERRO order), 
and (b) almost ferromagnetic FERRO  solution 
($M_{tot} \ne 0$, with a strong FERRO magnetization in the AA stacking region and a small opposite magnetization everywhere else).
The AA stacking regions are at the corners of the moir\'e cell.
}
\label{Fig_Mag2}
\end{figure}

To perform a more systematic study of the effect of parameters on magnetization, we had also used a rescaled TB model proposed by L.~A.\ Gonzalez-Arraga et al.\,\cite{Luis17} that simulates the magic-angle TBG flat band by a twisted bilayer containing only 869 atoms. It allows a more systematic study presented in Ref.\,\cite{Vahedi21_tBLG}. 
Figures\,\ref{Fig_Mag} and \ref{Fig_Mag2} show MFT results for actual TBGs, without the rescaled TB model. 
The main results of this work are summarized below.

Consider the critical Coulomb interaction $U_C$ above which  magnetic states emerge.
In monolayer graphene, mean-field theory yields  $U_{Cm}\approx 2.23t_0$ \cite{Sorella1992} and  $U_{Cm}\approx 2.09 t_0$
\cite{Vahedi21_tBLG} for 
only nearest-neighbor hopping and longer-range hoppings included, respectively.
For TBG, and
as excepted with flat bands, i.e., for twist angles close to the magic angle, $U_C$ is smaller than what would be required to render an isolated layer antiferromagnetic, i.e.,\ $U_C \ll U_{Cm}$ (Fig.\,\ref{Fig_Mag}(a)).
These static MFT results were qualitatively confirmed by dynamical mean-field (DMFT) calculations \cite{Vahedi21_tBLG} performed 
for the rescaled TB model.
Moreover, as shown in Fig.~\ref{Fig_Mag}(a),
$U_C(\theta = 1.30^\circ) <  U_C(\theta = 1.08^\circ)$,
while the velocity of the corresponding flat bands are both very low (Fig.~\ref{Fig_ElecStruc_TBG}). 
This shows the importance of the moir\'e cell size (and then the size of the AA region) in the value of $U_C$. There is therefore a complex competition between the low band dispersion and the moir\'e pattern size for the appearance of magnetism in TBG \cite{Vahedi21_tBLG}.

The real-space magnetic order for $U_C \le U \le U_{Cm}$ is a delicate issue.
Away from half filling, ferromagnetic states have been observed in experiments on doped TBG \cite{Sharpe19}, and predicted theoretically \cite{Wu19_FERRO,Seo19,pons2020flatband,Vidarte24}.
In our calculation, due to the large number of atoms of a moir\'e cell of magic-angle TBG ($N$ about 10000 orbitals), it is difficult to be sure that the convergence of the self-consistent process reaches the most stable magnetic state (lowest total energy). 
Indeed, for half filling (undoped system)
and $U_C<U<U_{Cm}$,
depending on initial conditions, 
the self-consistent calculation converges to an antiferromagnetic (AF) solution \cite{Vahedi21_tBLG}, or a ferromagnetic (FERRO) solution (Figs.\,\ref{Fig_Mag} and \,\ref{Fig_Mag2}). 
Note that, for $U=t_0$, the AF solution is in fact AF for intralayer order and FERRO for interlayer order (Fig.\,\ref{Fig_Mag2}(a)). 
The energy difference between the AF and FERRO solutions is very small. 
For example, considering the TBG with $\theta = 1.3^\circ$ (Fig.\,\ref{Fig_Mag}(b)), this difference is less than 0.1\,meV per orbital, which seems not really significant to permit a definite conclusion.
Let us just say that, according to this result, the FERRO solution seems more stable for small $U$, typically $U < 1.4 t_0$, and the AF solution seems to be more stable for larger $U$, $1.4t_0 < U \lesssim U_{Cm}$. 
In both solutions, the magnetization appears in the AA region of the moir\'e (Fig.\,\,\ref{Fig_Mag2}), as the flat-band states are located in the AA regions, and only flat bands are magnetic (Fig.\,\ref{Fig_Mag}(d,e,f)).
As expected, the AF solution is characterized by a small gap, while the FERRO solution is metallic with a splitting of up and down bands. 
It is also interesting to note that, for $U < U_{Cm}$, the total magnetization of the FERRO solution is $M_{tot} = 2$\,electrons per moir\'e cell, which corresponds exactly to the magnetization of the 4 flat bands at half-filling (see dashed line Fig.\,\ref{Fig_Mag}(b)).

For $U  \gtrsim U_{Cm}$, all carbon atoms become magnetic and the transition to full magnetization necessarily involves AB and BA stacking regions (Fig.\,\ref{Fig_moire}(a)) that are geometrically frustrated.
Therefore, a complex magnetic state is expected for strong $U$, $U \gtrsim U_{Cm}$, which has not yet been studied.

The above calculations assume that the periodicity of the magnetic order is equivalent to the periodicity of a unit cell containing a single moir\'e cell, thus allowing only the studies of magnetic order within a moir\'e cell. 
Moreover, an analysis of the noninteracting susceptibility suggests\,\cite{Vahedi21_tBLG} that in real space the magnetic order inside each moir\'e cell could be modulated from one moir\'e cell to another with a wave vector $\bf q$ (${\bf q} \ne 0$). 
Further investigations are needed to clarify this point.

Several theoretical studies have been performed for variable fillings (above half filling)\,\cite{Wu19_FERRO,Seo19,pons2020flatband,Vidarte24}.
A delicate point is the fact that beyond half filling it is possible for local non-collinear magnetic moments to stabilize the magnetic energy, as is the case for the honeycomb lattice for quarter doping per orbital, i.e.,\ when $E_F$ is at a Van Hove singularity \cite{Wang12_MC_graphene1-4,Jiang14,Maxime25}.
To calculate the magnetism in these cases, the Hubbard model must be treated within the framework of the Hartree-Fock approximation.

\subsection{Twisted bilayer MoS$_2$}
\label{Sec_TB_MoS2}

\begin{figure}[t!]
\begin{center}
\includegraphics[width=15cm]{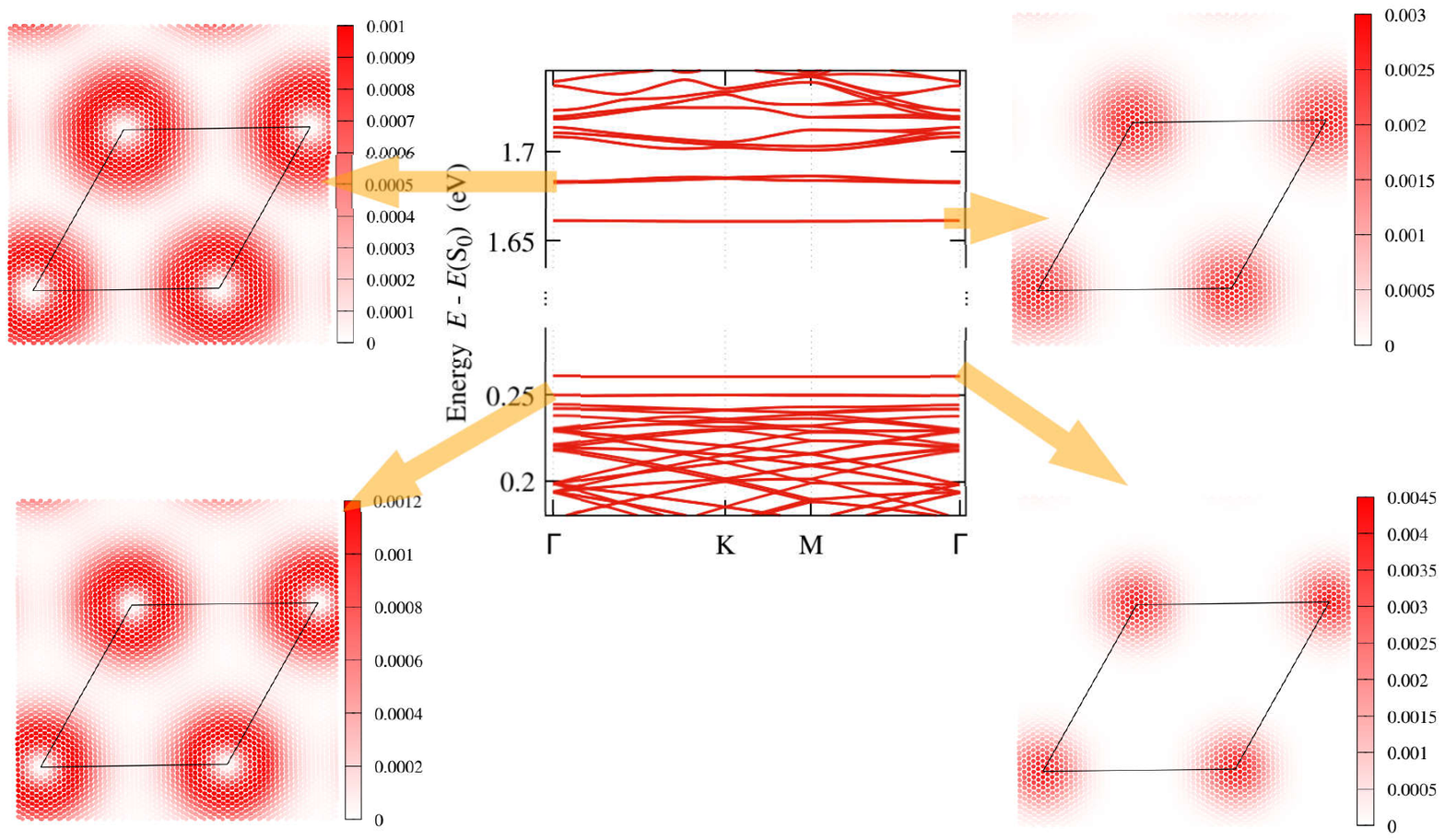}
\caption{ \label{Fig_MoS2}
(Center) TB band dispersion around the gap of rigid twisted bilayer MoS$_2$ with rotation angle $\theta = 1.61^\circ$. 
(Surrounding panels) Average weight of the eigenstates at K of the flat bands around the gap in real space.  
The color scale shows the weight of the
eigenstate on each $d_{z^2}$ orbital of the Mo atoms. The sum of these weights is more than 98\% and 95\% of each state for the valence and
conduction band, respectively.
Black lines show the unit cell containing 2522 Mo atoms. AA stacking regions are at the corners of this cell. 
TBGs are rigid, i.e., without atomic relaxation. 
From \cite{Venky20,Venky_these}.}
\end{center}
\end{figure}

The TBGs that we have presented so far correspond to twisted  massless 2D materials. 
It is also interesting to consider the appearance of flat bands in twisted massive 2D materials.
That is obtained in many systems based on a honeycomb lattice, when equivalence between the two sublattices is removed, thus resulting in a gap at $E_F$. 
Among the many examples now existing of such a twisted bilayer semiconductor,  
the broad family of  
transition metal dichalcogenides (TMDs) 
\cite{Wang15_reviewTMD,Liu15_reviewTMD,
Duong17_review}, which offers a 
wide variety of possible rotationally stacked bilayer systems, has also 
prompted numerous experimental 
\cite{vanderZande14,Liu14,Huang14,Huang16,Yeh16,Zhang17,Trainer17,Wilson17, 
Pan18,Zhang20_tTMD,Graham2021} and theoretical 
\cite{Roldan14b,Fang15,Cao15,Wang15,
Constantinescu15,Tan16,Lu17,Wilson17,Naik18,Conte19,
Maity19,Tang20,Wu20,Lu20,Pan20,
Zhan20,Zhang20,Naik20,Xian21,Angelie21,Graham2021,PhysRevB.110.235410,Soltero24} studies to understand 
confined moir\'e pattern states in semiconductor materials. Many 
of these studies analyze the interlayer distances, the possible atomic 
relaxation, the transition from a direct band gap in the monolayer system 
to an indirect band gap in bilayer systems, and more generally the effect 
of interlayer coupling in those twisted bilayer 2D systems at various rotation 
angles $\theta$. For small values of $\theta$, the emergence of flat bands 
has been established \cite{Naik18} from first-principles density 
functional theory calculations in twisted bilayer MoS$_2$ (tb-MoS$_2$), 
and observed in a 3$^\circ$ twisted bilayer WSe$_2$ sample by using 
scanning tunneling spectroscopy \cite{Zhang20_tTMD}. 
It has been also
shown numerically \cite{Lu20} that lithium intercalation in tb-MoS$_2$ 
increases interlayer coupling and thus promotes flat bands around the gap. 
There is also experimental and theoretical evidence that moir\'e patterns may give rise to 
confined states due to the mismatch of the lattice parameters in 
MoS$_2$-WSe$_2$ heterobilayers \cite{Pan18,Soltero24}.
Correlated electronic phases have been observed experimentally in twisted bilayer WSe$_2$ over twist angles ranging from 4 to 5.1$^\circ$\,\cite{Pan18}, and recently even superconductivity\,\cite{Guo25}.

\begin{figure}[t!]
\begin{center}
\includegraphics[width=11cm]{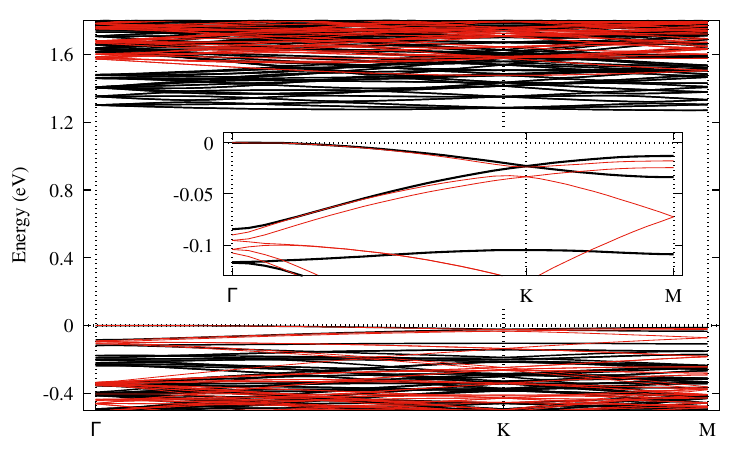}
\caption{ \label{Fig_MoS2_relax}
DFT band dispersion of (red) rigid (i.e., non-relaxed), and (black) relaxed twisted bilayer MoS$_2$ with rotation angle $\theta = 5.09^\circ$. 
The zoom show the highest valence bands. 
Calculations are performed with the ABINIT software \cite{Gonze16} with LDA pseudopotential. 
For calculation details and data see Ref.\,\cite{venky23}.
Similar results have been published previously see Refs.\,\cite{Naik18,Vitale21}.
}
\end{center}
\end{figure}

Most theoretical investigations of the electronic structure of bilayer 
MoS$_2$ are density-functional theory (DFT) studies 
\cite{vanderZande14,Liu14,Huang14,Zhang17,Roldan14b,Fang15,Cao15,Wang15,Constantinescu15, 
Huang16,Tan16,Lu17,Trainer17,Naik18,Debbichi14,Tao14,He14,sun2020effects,Naik20,Vitale21,venky23} 
with eventually a Wannier wave function analysis \cite{Fang15}. Those 
approaches provide interesting results, but they do not allow a systematic 
analysis of the electronic structure as a function of the rotation angle 
$\theta$, in particular for small angles, {i.e.}, very large moir\'e
cells, for which DFT calculations are not feasible. Several tight-binding 
(TB) models, based on Slater-Koster (SK) parameters \cite{Slater54}, have 
been proposed for monolayer MoS$_2$ 
\cite{Cappelluti13,Rostami13,Zahid13,Ridolfi15,SilvaGuillen16} and 
multi-layer MoS$_2$ \cite{Cappelluti13,Roldan14b,Fang15,Zahid13}.
Following these efforts, we propose \cite{Venky_these,Venky20} a SK set of parameters 
that match correctly the DFT bands around the gap of tb-MoS$_2$ with 
rotation angles $\theta > 5^\circ$. 
This Hamiltonian takes into account the five $d$ orbitals of the Mo atoms and the three $p$ orbitals of the S atoms. 
This SK-TB model, with the same 
parameters, is then used for smaller angles, in order to describe the 
confined moir\'e pattern states. We thus show that, for $\theta \lesssim 6^\circ$, 
the valence band with the highest energy is separated from the other 
valence states by a minigap of a few meV. 
In addition, the width of this 
band decreases as $\theta$ decreases so that the average velocity of these 
electronic states reaches 0 for $\theta \lesssim 2^\circ$ such that flat 
bands emerge at these angles (Fig. \ref{Fig_MoS2}). 
This is reminiscent of the vanishing of the 
velocity at certain ``magic'' rotation angles in bilayer graphene 
(Sec. \ref{Sec_flatBands_in_tBLG}) except that in the case of bilayer MoS$_2$ it arises for an interval of 
angles. Other minigaps and flat bands are also found in the conduction 
band. The confined states that are closest to the gap are localized in the 
AA stacking regions of the moir\'e pattern, as in twisted bilayer  graphene.
Note that there are two types of stacking for tb-MoS$_2$ and both yield flat bands for different, but fairly comparable values of the angles \cite{Venky20}.

More recent results \cite{Naik18,Naik20,Vitale21,venky23,Soltero24} have shown that atomic relaxation is important in tb-TMD, even for not too small twist angles.  
For angles $\theta$ close to 5$^\circ$, the atomic relaxations are mainly out-of-plane relaxations and they are well described by a simple model of atomic positions in a moir\'e cell \cite{venky23}. 
As a result, the flat bands are still present, but they are modified compared to the non-relaxed system, and their degeneracy is also modified (Fig.\,\ref{Fig_MoS2_relax}). 
We are therefore working on a Slater-Koster TB model that will correctly simulate these relaxation effects.
It should be noted that, to our knowledge, there is as yet no SK-TB model that correctly simulates the effect of this out-of-plane relaxation on the electronic structure. 
For smaller angles \cite{Naik18,Naik20,Vitale21,Walet21}, typically  2$^\circ$ or less, in-plane relaxation is very important and has been shown numerically to strongly modify the size of the AA and AB stacking regions.
The less energy-efficient AA stacking zone is greatly reduced in favor of the AB and BA zones.

\section{Flat bands due to static defects}
\label{Sec_FlatBands_Defects}

Electronic transport in graphene (see for example  Refs.\,\cite{Berger04,Novoselov2005,Zhang05,
Berger06,Castro09_RevModPhys,Yang18}) 
and and other graphene-related 2D materials is sensitive to
static defects that are, for example, frozen ripples, screened charged impurities, or local defects like vacancies or adsorbates. 
Adsorbates, 
which can be adatoms or admolecules (H, OH, CH$_3$, ...)
attached to the surface of graphene, are of great interest in the context of functionalization
to control the electronic properties\,\cite{Peres06,Pereira06,Pereira08a,Lherbier08,Robinson08,Wehling10,
Yuan10b,Bostwick09,Leconte10,
Skrypnyk11,Ferreira11,Leconte11,Lherbier11,
Lherbier12,
Trambly13,Roche13,Gargiulo14,Xu16,
Jorgensen16,VanTuan16,Yang18,Son20}. 
In particular, a functionalization by an adatom (or admolecule) covalently bound to an atom of the 2D material can be a resonant scatterer for conduction states which strongly affect electronic structure and transport properties.

This appearance of a resonant state by functionalization, so-called midgap states, is particularly important in the case of bipartite lattices, i.e.,\ systems in which atoms of  a sub-network $\alpha$ are bound to atoms of a sublattice $\beta$, and reciprocally.   
Considering a first-neighbor tight-binding (TB) Hamiltonian,
when the numbers of $\alpha$ orbitals and $\beta$ orbitals are different, isolated states appear at the energy $E=0$ (center of the band) \cite{Brouwer02,Peres06,Pereira06,Pereira08a}. 
Such a mechanism is exact globally for the whole structure. 
It also allows us to understand the appearance of resonant scatterers through a random functionalization, even if the number of $\alpha$ and $\beta$ orbitals is  nearly the same. 
Indeed let us consider a random distribution of defects (adatoms or admolecule), with a not too strong concentration $c$ of defects. 
Locally, typically around a defect, an unbalance occurs between the two sublattices and a confined state appears at $E=0$ around the defect. 
When defects are in the same sublattice, these states are coupled to each other and form a kind of midgap band, leading to a peak in the DOS of width that depends on the concentration of defects. 
This mechanism applies to all the 2D materials presented in this review that have a bipartite lattice.
Note that this yields also a simple picture to understand the appearance of edge states in a bipartite lattice, since an edge can produce a local unbalance between the two sublattices.

Theoretical studies of transport in the presence of local defects have dealt mainly either with the Bloch-Boltzmann formalism (semi-classical methods) (see for instance \cite{Guinea08,Castro09_RevModPhys,DasSarma11,
McCann13}). 
In these theories, a major length scale that
characterizes the electron scattering is the elastic mean free path $L_e$. These approaches indeed explain some experimental observations such as the quasi-linear variation
of conductivity with a concentration of charge carriers. Yet, these theories have important limitations and can hardly describe in detail the localization phenomena that have been reported in some experiments. 
Indeed, in the presence of a short-range potential, such as that produced by local defects, the electronic states are localized on a length scale $\xi$. 
A sample will be insulating unless some source of scattering, like electron-phonon interaction, leads to a loss of the phase coherence on a length scale 
$L_i < \xi$. Note that electron-electron interactions may also play a dephasing role but this occurs usually at the lowest temperatures and we do not discuss it specifically here \cite{Lee85,Polini20}. 
Therefore, in addition
to the elastic mean free path $L_e$, the inelastic mean free path
$L_i$ and the localization length $\xi$ also play a fundamental role for the conductivity of graphene with adsorbates.
Moreover, the study of transport by semi-classical methods is valid when the Fermi energy $E_F$ is far enough from the Dirac energy. 
However, for $E_F$ close to the Dirac energy, the effects of resonant scatterers can strongly modify the electronic structure (such as gap opening \cite{Missaoui18,Jouda21}...). 
These effects are important when the resonant scatterer concentration (defect concentration) is large with respect to the charge carrier concentration; 
indeed, each resonant scatterer creates one midgap state. 
Since these quantum effects, beyond the semi-classical behavior, are extremely dependent on the type of functionalization, systematic theoretical studies are needed to understand experimental results and stimulate new experimental studies. 

\subsection{Quantum transport calculation and calculated quantities}
\label{Sec_QT}

We used the 
real-space Kubo-Greenwood method, developed by 
D.\ Mayou, S.~N.\ Khanna, S.\ Roche, and F.\ Triozon \cite{Mayou88,Mayou95,Roche97,Roche99,Triozon02}, which has already been used to study quantum transport in disordered graphene, chemically doped graphene and bilayers (see for instance \cite{Lherbier08,Zhu10,Lherbier11,Leconte11,Leconte11b,Lherbier12,Roche12,Roche13,
Trambly13,Missaoui17,Missaoui18,Omid20,Jouda21,Chen23,Miranda23,Kaladzhyan23,guerrero24_tBLG_disorder_Kubo,Guerrero25}),
functionalized carbon nanotubes 
\cite{Latil2004,IshiiInelastic2010,Jemai19,Fan20}, 
and many other systems \cite{Fan14,Fan20} such as
quasicrystals \cite{Triozon02,Trambly17}, 2D molecular semiconductors \cite{Fratini17,Missaoui19}, and perovskites \cite{Lacroix20}.  
This numerical method connects the dc-conductivity $\sigma$,
$\sigma(E,t) = e^2 n(E) \mathcal{D}(E,t)$,
to the DOS $n$ and the diffusivity $\mathcal{D}$,
$\mathcal{D}(E,t)=\lim_{t\rightarrow \infty}{\Delta X^2(E,t)}/{2t}$ 
(see  \ref{Sec_QD_and_dcCond}),
where the average square spreading $\Delta X^2$
is calculated at every energy $E$ and time $t$ by  using a polynomial expansion method
(\ref{Method_Real}).  
This numerical approach has the advantage of using efficiently the  method  in real space. 
It takes into account all quantum effects due to a random distribution of static scatterers in a very large supercell containing more than $10^7$ orbitals.
Most of our calculations are performed in a supercell of 1000$\times$1000 or 1500$\times$1500 cells of the initial cell of the studied crystal (graphene (2 atoms), Bernal bilayer (4 atoms), TMD (3 atoms), phosphorene (4 atoms / layer)), with periodic boundary conditions. 
Such a calculation may be performed by the recursion method (Lanczos algorithm) where the Hamiltonian in real space is written as a tridiagonal matrix
\cite{Lanczos50,Cullum1984,Pettifor,Dagotto1994} of dimension $N_r$. 
Typically, we use $N_r = 1000$--$3000$ and we checked that 
results do not change significantly when $N_r$ increases.
One can then directly evaluate the DOS from the Lanczos coefficients \cite{Haydock1972,Pettifor,Gagliano1987,Dagotto1994,Schenk08}.
This method is known as recursion method and
has been used in some of our papers \cite{Trambly11,Trambly13,Trambly14,Missaoui17,
Missaoui18,Omid20}, but it
leads to a convolution of the DOS by a Lorentzian function which a small width $\epsilon$.  
Thus $\epsilon$ is a kind of energy resolution of the calculation. Usually we use $\epsilon = 1$ or $5$\,meV.
However, for systems with a gap, to avoid the tail expansion of the Lorentzian function in the gap, it is more efficient to diagonalize the tridiagonal
Lanczos matrix
of dimension $N_r\times N_r$ 
and to evaluate the DOS by Gaussian broadening of the spectrum\,\cite{Lacroix20,Lacroix_these}.
Most of the time, we use a Gaussian standard deviation of $\epsilon =5$\,meV. 
Note that for energies $E$ that are not close to the gap, the two methods give almost the same results.

The tight-binding (TB) Hamiltonian $\hat{H}$, written in a supercell, takes into account the effects of elastic collisions (static defects). 
Therefore, in the framework of a TB model, all quantum effects --including all multiple scattering effects-- are  taken into account to calculate the average square spreading $\Delta X^2(E,t)$ and the diffusion coefficient without inelastic scattering, i.e.,\ at zero temperature.
 
At finite temperature $T$, inelastic scattering caused by electron-phonon interactions are implemented by using the approximation of Relaxation Time Approximation (RTA). 
The details of the implementation of the RTA are given in \cite{Trambly13} (\ref{Sec_RTA}). 
The conductivity $\sigma$ and the inelastic mean free path $L_i$ in the $x$-direction are thus 
calculated at every energy $E$ and every inelastic scattering time $\tau_i$. 
$\tau_i$ is the time beyond which the velocity autocorrelation function (Appendix Eq. (\ref{RTA_C})) goes exponentially to zero, and
$L_{i}(E_F,\tau_{i})$ is the typical distance of propagation during the time interval $\tau_{i}$ for electrons at energy $E$.  
Thus $L_i$ is also the distance beyond which a wave packet loses  its phase coherence due to inelastic scattering processes, whereas elastic scattering events do not destroy the phase coherence. 
We know that $L_i$ decreases when the temperature $T$ increases, 
however the exact dependence of $L_i$ versus $T$ is unknown.
This is why we consider different cases according to different possible values of $L_i$. Indeed,
three different transport regimes may exist depending on the value of $L_i$ with respect to the elastic mean free path $L_e$, which is the average distance between two elastic scattering events. 
$L_e$ can be calculated approximately using the phenomenological formula (\ref{Eq_le}) of the Appendix.
The different transport regimes are: 
\begin{itemize}
\item
{\bf Localization regime}. 
When $L_i \gg L_e$,  multiple scattering effects (such as weak or strong localization) strongly affect the transport and the conductivity is ``macroscopic'' in the sense that it is established over large sample sizes. 
This happens at sufficiently low temperature $T$, 
and then $\sigma$ decreases when $L_i$ increases (i.e.,\ $T$ decreases).
At each energy $E$, the localization length $\xi$ can be extracted by extrapolation of the curve $\sigma$ versus $L_i$, 
when $\sigma(L_i = \xi) = 0$.
\item 
{\bf Diffusive regime}.
For smaller values of $L_i$, since $L_i \gtrsim L_e$, i.e.,\ at higher temperatures (typically room temperature), $\sigma(L_i)$ reaches a conductivity ``plateau'' close to the maximum $ \sigma$ value (see \ref{Sec_RTA}), 
\begin{equation}
\sigma(E_F,L_i) \approx
\sigma_M(E_F) = {\rm Max}\{ \sigma(E_F,\tau_i) \}_{\tau_i}
\end{equation}
This regime is called ``diffusive'' regime, where $\sigma(L_i)$ is almost independent of $L_i$ over a large $L_i$ range depending on the energy $E_F$. 
Usually the conductivity plateau corresponds to values of $L_i$ from a few nm to a few 10\,nm, which may correspond to high temperatures and room temperatures, respectively.  
In this case, the conductivity of a sample 
depends only on quantum scattering over small distances which are typically of the order of magnitude of the distances between static defects ($L_e$); this is the reason why  $\sigma_M$ is called ``microscopic'' conductivity. 
At each energy, the microscopic diffusivity $\mathcal{D}_M$ and microscopic conductivity $\sigma_M$ are defined as the maximum value of $\mathcal{D}(\tau_i)$ and $\sigma(\tau_i)$, respectively.
The corresponding diffusivity is also called the microscopic diffusivity, $\mathcal{D}_M$, and it  is possible to deduce from it the elastic mean free path $L_e$
 (Eq.\,(\ref{Eq_le})).
\item
{\bf Ballistic regime}. The situation $L_i < L_e$ is an extreme case that most of the time should not be reached 
in real materials at realistic temperature (room temperature). This corresponds to the case of very pure materials with very few static defects. The conductivity is independent of static defects, and thus $\sigma(L_i)$ increases when $L_i$ increases. 
\end{itemize}

\subsection{Mono and bilayer graphene with random distribution of defects}
\label{Sec_biL_rand}

We start from the TB Hamiltonian,
\begin{equation}
\hat{H} =  \sum_{\langle i,j \rangle} t_{ij}\, c_i^{\dag}c_j  ,
\label{Eq_hamiltonian2}
\end{equation} 
where $t_{ij}$ are the hopping terms between $p_z$ orbitals $i$ and $j$, 
and $\langle i,j\rangle$ is the double sum on index $i$ and $j$ with $i\ne j$.
We can include only the first-neighbor hopping terms (first-neighbor TB: intralayer hopping  $t_{ij}=-t_0=-2.7$\,eV and for Bernal bilayer graphene (BLG) interlayer hopping $t_{ij}=t_1=0.34$\,eV), or hopping terms beyond the first neighbors which is often more realistic using the same SK-TB parameters as in Sec.\,\ref{Sec_ElecStrucCal}. 
Both models correctly recover the monolayer graphene (MLG)  and Bernal bilayer graphene (BLG) energy band dispersions in the vicinity of the Dirac energy level $E_D = 0$ \cite{Trambly12,Missaoui18}. 
For MLG and BLG,
the spectrum is symmetric with respect to $E_D$ (electron-hole symmetry) for first-neighbor TB (see Sec.\,\ref{Sec_biAB_selecive_bipartite}), whereas it is slightly asymmetrical due to coupling beyond the first neighbors with the SK-TB (Sec.\,\ref{Sec_ElecStrucCal}).
It is important to note that another TB model exists for the BLG, in particular a well-known model deduced from the TB model for bulk graphite (see Ref.\,\cite{McCann13} and Refs therein). For the properties that we are studying here, the latter gives results comparable to the SK-TB  model (Sec.\,\ref{Sec_ElecStrucCal}).

\subsubsection{Monolayer graphene (MLG)}
\label{Sec_mono_rand}

Electronic properties in graphene with static defects such as 
vacancies, substitutions, or adsorbates has been intensively studied  (see for instance Refs.\,\cite{Peres06,Lherbier08,Lherbier08b,Peres09,Leconte10,Lherbier12,Roche12,Roche13,
Trambly13,Missaoui17,Missaoui18,Dutreix19,Jouda21}). 
We cannot go into all these studies in detail here, but we will just mention a few aspects that we consider essential. 
Roughly speaking, it is possible to  propose a unified description of transport in one graphene sheet with adsorbates that fully takes into account localization effects and loss of electronic coherence due to inelastic processes \cite{Trambly13}.
For this purpose, we distinguish between defects that are resonant scatterers or defects that are non-resonant scatterers. Examples of resonant scatterers are adsorbates that create a covalent bond with a carbon atom of a graphene layer, such as simple atoms or adsorbate molecules: H, OH, CH$_3$...
To simulate this covalent bond, we assume that the $p_z$ orbital of carbon, just below the adsorbate, is removed \cite{Pereira08a,Robinson08,Wehling10}. 
As explained in the introduction of this part\,\ref{Sec_FlatBands_Defects},
in a monolayer graphene (MLG), which is a zero-gap material with a bipartite lattice, such functionalization 
creates so-called ``midgap states'' at the Dirac energy $E_D=0$. 
If we imagine a periodic structure with a low concentration of resonant defects, so that the defect states do not interact with each other, the midgap states would  be associated with strict flat bands. 
These midgap states at energy $E_D$ are states that are only uncoupled from the rest of the band if we consider a Hamiltonian that includes only first-neighbor couplings. 
If couplings beyond the first neighbors are taken into account, the midgap states are coupled among themselves, and we then obtain a ``midgap peak'' whose width depends on the defect concentration, at an energy shifted with respect to $E_D$\,\cite{Trambly14,Missaoui17}. 
This midgap peak is associated to flat bands in a periodic system.
Examples of non-resonant adsorbates are adsorbates that create two covalent bonds with two neighboring C atoms of the MLG. 
This is the case for the adsorption of an oxygen atom (epoxy group). 
In generic terms, a non-resonant disorder can also be simulated by an Anderson disorder (random variation of the on-site energy of all orbitals). Such defects do not create resonant peaks, so they do not alter strongly  the DOS, but they do affect the transport properties of charge carriers \cite{Trambly13}.
As non-resonant defects are not associated with flat bands, we will not go into too much detail in this review. 

In MLG with a random distribution of resonant scatterers, for energy sufficiently far from the Dirac energy and at sufficiently small concentrations, the semi-classical Bloch-Boltzmann  theory can be a good approximation \cite{Trambly11,Trambly13,Trambly14,Missaoui17}. 
For a Fermi energy $E_F$  near the Dirac energy, different regimes of transport are obtained \cite{Trambly13}.  
Some universal aspects of the conductivity\,\cite{Peres06} are present with or without the hopping beyond nearest neighbors.
For small inelastic scattering length such as $L_i \approx L_e$ the conductivity is almost equal to the universal minimum (plateau) of the microscopic conductivity (similar to the semi-classical conductivity), $\sigma_M = {4 e^2}/(\pi h) = \frac{2}{\pi} G_0$, with G$_0 = 2e^2/h$, except for an energy exactly at the midgap-state energy (see Sec.\,\ref{Sec_CondMidGapSt}). 
For larger $L_i$, $L_i \gg L_e$, the conductivity follows a linear 
behavior with $\log L_i$. 
This allows us to estimate the relationship between the localization length and the elastic mean free path, and we have estimated  it, $\xi \approx 13 L_e$ \cite{Trambly13}.

\subsubsection{Bernal bilayer graphene (BLG)}
\label{Sec_biAB_rand}

\begin{figure}[t]
\centering
\includegraphics[width=0.98\textwidth]{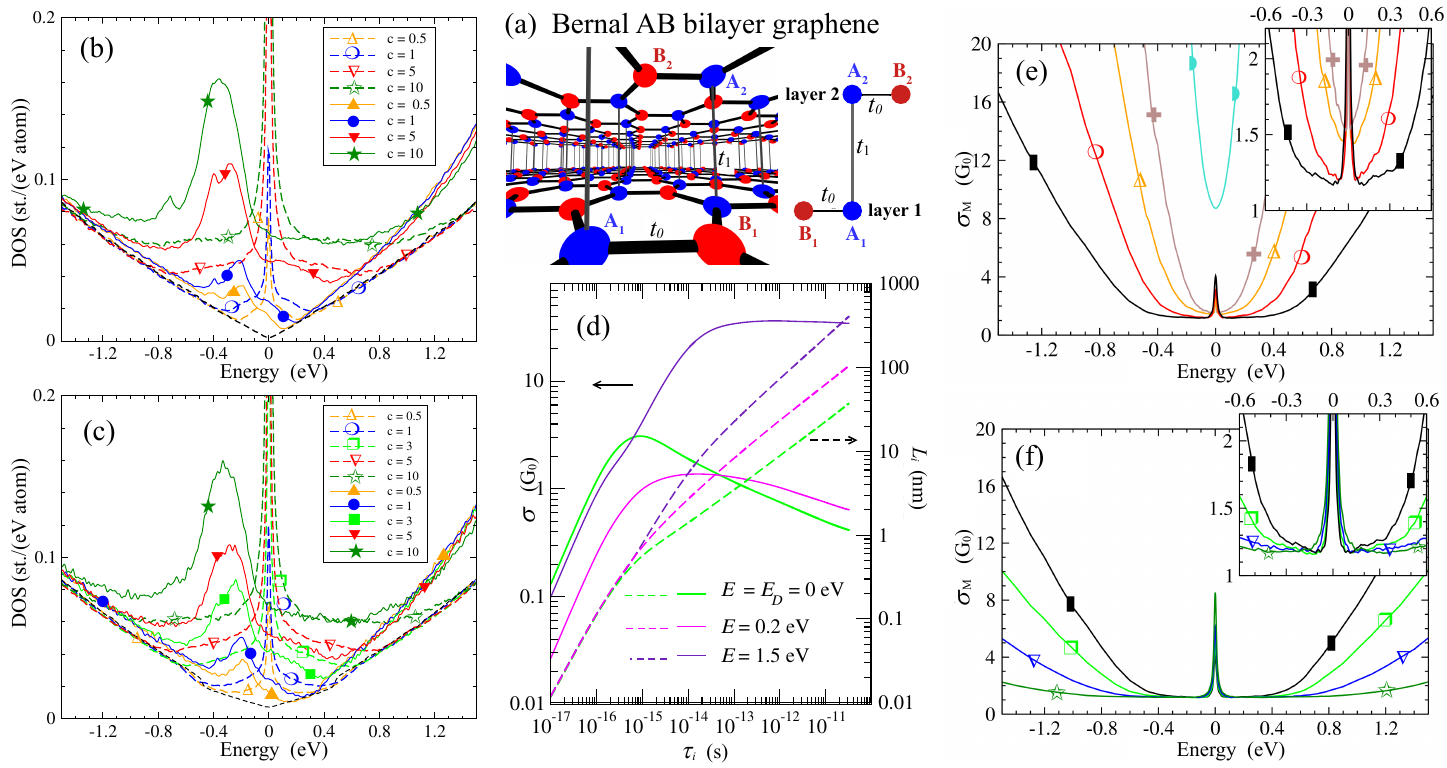}
\caption{\label{Fig_biAB_randVa}
Electronic properties of Bernal AB bilayer graphene (BLG).
(a) Sketch of the crystal structure of  BLG. Atoms A$_1$ and B$_1$ are on the lower layer (layer 1); A$_2$, B$_2$ on the upper layer (layer 2).
First-neighbor TB:
$V^0_{pp\pi} = - t_0 = -2.7$\,eV and $V^0_{pp\sigma} = t_1= 0.34$\,eV are the intralayer and interlayer first-neighbor hopping terms, respectively.
BLG is a bipartite lattice between sublattices $\alpha =\{{\rm A}_1,{\rm B}_2\}$ and $\beta =\{{\rm A}_2,{\rm B}_1\}$.
Total DOS of 
(b) Monolayer (MLG) and 
(c) BLG with random distribution of vacancies with concentrations $c$ (\%) of vacant atoms:
(dashed line) First-neighbor TB (see panel (a)),
and (solid line) SK-TB with hopping beyond first neighbor (see Sec.\,\ref{Sec_ElecStrucCal}).
The black dashed curve is pristine BLG ($c=0$), calculated  with  first-neighbor TB.
(d) Conductivity $\sigma$ (solid line) and inelastic scattering length $L_i$ (dashed line) in BLG versus inelastic scattering time $\tau_i$ for concentration $c = 1\%$ of resonant adsorbates (vacancies), and for 3 energy values, calculated with first-neighbor TB.
(e,f) Microscopic conductivity $\sigma_M$ 
calculated with  first-neighbor TB: 
(e) $c<c_l\sim 1.5$\% and $\sigma_{M}$ with a constant minimum value (“plateau”) which decreases as $c$ increases, 
(f) $c>c_l\sim 1.5$\% and $\sigma_{M}$ reaches a minimum value independent of the value of $c$  (see text).
Concentration of vacancies: semi circle $c=0.05\%$, cross $c=0.25\%$,  triangle up $c=0.5\%$, circle $c=1\%$, rectangle $c=2\%$, square $c=3\%$, triangle down $c=5\%$, star $c=10\%$.
Insert shows enlarged curves around $E_D=0$.
$G_0 = {2 e^2}/{h}$. 
For the numerical method see \ref{Method_Real}.
From\,\cite{Missaoui17,Missaoui_these}.
}
\end{figure}

The unit cell of 
BLG
contains four carbon atoms, A$_1$, B$_1$ in layer 1 and A$_2$, B$_2$ in layer 2 (Fig.\,\ref{Fig_biAB_randVa}(a)). 
Atoms A have three B first neighbors in the same layer and one A neighbor in the other layer, while atoms B have only three A first neighbors in the same layer.
Since BLG is a bipartite lattice (see Sec.\,\ref{Sec_bilayerGraphBipartite} below), an isolated functionalization by resonant scatterers creates a midgap state like in monolayer graphene (MLG).
We first consider now the case of a random distribution of resonant defects in the two layers in BLG\,\cite{Missaoui17}.
Figure \ref{Fig_biAB_randVa} shows the total density of states (total DOS) $n(E)$ for different concentrations $c = 0.5\%$ to $10\%$ of defects in MLG and BLG. 

Without defects, $c=0$, MLG  and BLG DOSs differ \cite{Trambly13,Missaoui17} for energies $E$ such as 
$-t_1 < E < t_1$ \cite{McCann13}, where $t_1$ is the interlayer hopping term (Fig.\,\ref{Fig_biAB_randVa}(b,c)). For small $c$ concentrations, $c< c_l \approx 1.5\%$, this distinction is still observed. 
Nevertheless, MLG and BLG have remarkably similar DOS for concentrations of defects $c$ larger than $c_l$. 
With the TB model (first-neighbor hopping only), midgap states occur at energy $E_{MLG}=0$. 
With the TB model including all neighbors (SK-TB), the midgap state is no longer at $E=0$, but it becomes a broad peak at negative energy $E_{MLG}$\,\cite{Missaoui17}. 
The value  of $E_{MLG}$ varies from $E_{MLG}\approx-0.2\,{\rm eV}$ to $E_{MLG} \approx-0.3\,{\rm eV}$ when the defect concentration increases from $c=0.5\%$ to $10\%$, as in monolayer graphene \cite{Castro09_RevModPhys,Trambly14}. 
For large values of energies  $E$ the DOS is weakly affected by the presence of disorder and near the Dirac energy there is an intermediate regime where the pseudogap is filled (``plateau''). 
We distinguish  three cases according to energy values, which correspond to different transport 
regimes (Fig. \ref{Fig_biAB_randVa}) \cite{Missaoui17}:
\begin{itemize}
\item[$(i)$] For sufficiently large energies,  MLG and BLG DOSs are similar and they are not strongly modified by the presence of resonant defects. 
Transport is well described by
the Boltzmann theory. For large values of $\tau_i$, the conductivity $\sigma$ is almost constant as expected in a diffusive regime. 
The conductivity is inversely proportional to the concentration $c$ of defects. 
\item[$(ii)$] For energies $E$ in the DOS ``plateau'' due to vacancies but not in the midgap states (Fig.~\ref{Fig_biAB_randVa}(a)),
the microscopic conductivity ($\sim$ room temperature conductivity) $\sigma_M$ (Fig.~\ref{Fig_biAB_randVa}(e,f))  shows a ``plateau'' independent of $E$.
For small $c$, $c<c_l \approx 1.5$\%, the conductivity plateau value depends on $c$ (Fig.~\ref{Fig_biAB_randVa}(e)). 
This effect results from interlayer hopping, 
whereas for large $c$ ($c>c_l$) the conductivity plateau value is independent of $c$ like in MLG, and $\sigma_M = 2 \sigma_{M,{MLG}}$, with $\sigma_{M,{MLG}} = 0.6$\,G$_0$ (Fig.~\ref{Fig_biAB_randVa}(f)). 
As explained in \cite{Missaoui17}, this transition from BLG to MLG behavior when the defect concentration $c$ increases, can be understood by comparing the elastic scattering length $L_e$ and the average traveling distance, $l_1 \approx 2$\,nm, in a layer between two interlayer hoppings.  

We also calculate the conductivity versus the inelastic length $L_i$. In the localization regime (very low temperature), as expected by the Anderson localization theory for 2D materials \cite{Lee85}, we get:
\begin{equation}
\label{Eq_localization}
\sigma(E,L_i) = \sigma_0(E) - \alpha G_0\, \ln \left(\frac{L_i}{L_e(E)} \right),
\end{equation} 
with a coefficient $\alpha$ almost independent of the energy $E$. The value of $\alpha$  is similar to that of MLG \cite{Trambly13}. 

\item[($iii$)] 
For energies in the midgap states, the conductivity is very low, and various anomalous behaviors are obtained as in MLG\,\cite{Trambly13} (see also Sec.\,\ref{Sec_CondMidGapSt}). Indeed, conduction by the midgap states is very specific to the TB model.
\end{itemize}

\subsubsection{Twisted bilayer graphene (TBG)}
\label{Sec_TBG_rand}

The geometry and the very rich physics of magic-angle TBG has been presented in the section \ref{Sec_QD-TBG},
in particular 
their quantum diffusion properties for defect-free systems (Sec. \ref{Sec_TBG_QD}).
Here, we analyze the diffusion properties in TBG with a random distribution of resonant defects such as adsorbed atoms or molecules, simulated by vacant atoms, like in MLG and BLG (see introduction of Sec. \ref{Sec_biAB_rand}).
We focus on the effect of the angle of rotation $\theta$ between the two graphene layers and we compare the case in which defects are distributed in one layer only (layer 2) to the case in which defects are distributed in the two layers (layers 1 and 2) \cite{Toai_these,Omid20}.

\begin{figure}[t]
\begin{center}
\includegraphics[width=15.5cm]{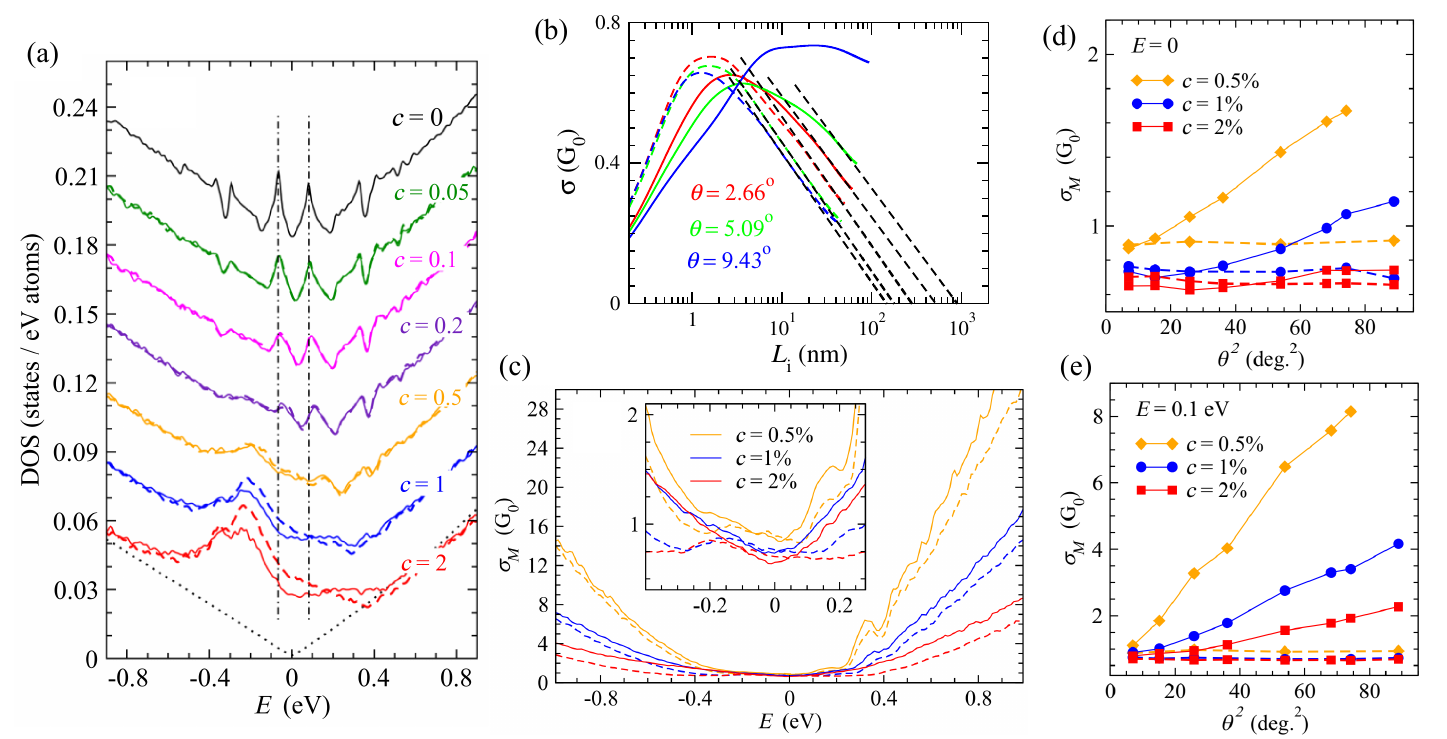}
\caption{\label{Fig_TBG_defects}
Electronic properties of TBG with defects (vacancies):
(a) Total DOS in (12,13) TBG ($\theta = 2.66^\circ$
, $N=1873$) for different concentration $c$ of vacancies. 
(b) Conductivity $\sigma$ versus inelastic mean free path $L_i$ at energy $E=0$ for 3 TBGs with $c=2$\,\%. 
(c) Microscopic conductivity $\sigma_M$ in TBG ($\theta = 2.66^\circ$). Insert: zoom around the Dirac energy $E_D =0$.
(d),(e)  $\sigma_M(E)$ versus $\theta^2$ for $E=0$ and $E=0.1$\,eV, respectively.
In all panels: (full lines) with
vacancies in layer 2 and (dashed lines) with vacancies in both layers.
The SK-TB model is the one which results in an intraband monolayer velocity $V_{MLG} = 0.79\cdot10^6$\,ms$^{-1}$ (Sec.\,\ref{Sec_ElecStrucCal}).
TBGs are rigid, i.e., without atomic relaxation. 
For the numerical method see \ref{Method_Real}.
$G_0 = {2 e^2}/{h}$.
From\,\cite{Omid20}.
}
\end{center}
\end{figure}

The effect of static defects (vacancies) on total density of states (DOS) and microscopic conductivity $\sigma_M$ (i.e.,\ room-temperature conductivity, Sec.\ \ref{Sec_QT}) are shown in Fig.\ \ref{Fig_TBG_defects}.
As expected, when disorder increases, i.e.,\ the concentration $c$ of vacancies increases, the two van Hove singularities are gradually destroyed 
(Fig.\,\ref{Fig_TBG_defects}(a)).

For vacancy distributions in the two layers $\sigma_M(E)$ is almost independent of the value of $\theta$ 
(see dashed lines in Fig.\,\ref{Fig_TBG_defects}(d,e)).
When the vacancy concentration $c$ is large, the behavior is similar to that of MLG and $\sigma_M \approx 2 \sigma_{M,{MLG}}$, 
where $\sigma_{M,{MLG}}$ is the conductivity for MLG \cite{Trambly13}, 
$\sigma_{M,{MLG}} \approx 0.6$\,G$_0$.
$\sigma_{M,{MLG}}$ reaches the well-known universal minimum of the conductivity so-called conductivity 
``plateau'' --independent of defect concentration-- at energies around $E_{D}$ \cite{Castro09_RevModPhys}. 
For smaller concentration $c$,  $\sigma_{M}$ increases when $c$ increases. 
These two regimes are similar to the one found in AB Bernal bilayer graphene \cite{Missaoui17} (see Sec. \ref{Sec_biAB_rand}).  
Roughly speaking, 
for large values of $c$ ,  
the elastic mean free path $L_e$ in MLG is smaller than the average traveling distance $l_1 \approx 1.5$\,nm within a layer, between two interlayer hoppings of the charge carriers, and thus carriers behave as they would in MLG. Whereas for small $c$ values, $L_e > l_1$ and thus interlayer hoppings are involved in the diffusive regime and BLG conductivity properties are different from MLG ones.   

For vacancies distribution in layer 2 only,
and large rotation angle 
(see solid lines in Fig.\,\ref{Fig_TBG_defects}(d,e)), 
the conductivity is larger than in the previous case. 
Indeed for large $\theta$, typically $\theta > 10^\circ$, eigenstates are located mainly in one layer (``decoupled'' layers)\,\cite{Trambly10,Trambly16} 
and thus conductivity of the bilayer is the sum of the conductivities of two almost independent layers, 
\begin{equation}
\sigma_M = \sigma_{M,1} + \sigma_{M,2},
\label{eq_sigma_M}
\end{equation}
corresponding to the conductivity of layers 1 and 2, respectively.  
The conductivity of a layer with defects is close to the MLG conductivity $\sigma_{M,2} \approx  \sigma_{M,{MLG}}$ and the conductivity of the layer without defects $\sigma_{M,1}$
is affected by the presence of defects in layer 2. 
With increasing $\theta$, the eigenstates are more and more located on one layer, thus layers are more and more decoupled, and 
$\sigma_{M,1}$ increases as layer 1 becomes more and more like a pristine MLG. 
Consequently the conductivity of the TBG increases when $\theta$ increases. 
In these cases numerical results 
(Fig.\,\ref{Fig_TBG_defects}(d,e)) 
show that $\sigma_{M}$ increases as $\theta^2$.  
This behavior has been clarified by a continuous analytic  model \cite{Omid_these,Omid20}, showing that for intermediate angles,
\begin{equation}
\sigma_{M}(E) \approx \sigma_{M,MLG} \left( 1 + \frac{n_1(E)}{n_2(E)} \frac{\theta^2}{\theta_0^2}  \right),
\label{eq_sigma_bi_Ftheta}
\end{equation}
where $\theta_0 \approx 1.7^\circ$ (Eq.\,(\ref{Eq_Vrenormalization})), and $n_1$, $n_2$ are the DOS in layer 1 and 2 that are different since the midgap state peak is not seen in a defect-free layer \cite{Omid20}.

For small angles, eigenstates are located almost equally on both layers for all energies around the Dirac energy \cite{Trambly16} (Sec. \ref{Sec_TBG_QD}); therefore they are affected in a similar way by the two kinds of vacancy distributions: ($i$)  in the two layers and ($ii$)  in layer 2. 
Conductivity is thus very similar in the two cases.

In the localization regime (i.e.,\ the low-temperature limit, $L_i \gg L_e$), the multiple scattering effects may reduce the conductivity with respect to the microscopic conductivity $\sigma_M$. The inelastic length $L_i$ thus satisfies  $L_i \gg L_e$ and $L_i \gg l_1$. 
To evaluate this effect we compute the conductivity $\sigma$ versus $L_i$ at every energy $E$ \cite{Omid20} for the two vacancy distribution cases: ($i$) and ($ii$). 
As expected in disordered 2D systems \cite{Lee85}, for large $L_i$, $\sigma(L_i)$ follows a linear variation with the logarithm of $L_i$ (Eq.\,(\ref{Eq_localization})), as in the case of MLG\,\cite{Trambly13,Trambly11} and BLG\,(Sec. \ref{Sec_biAB_rand}).
Similarly to $\sigma_M$, for intermediate-angle TBG, the localization length $\xi$ is almost independent of $\theta$ when the defects are located in both layers, but $\xi$ increases sharply  when defects are located in a single layer. 
Despite some studies, e.g.,\,\cite{Andelkovic18}, further studies are needed to understand the effect of resonant defects on the magic-angle TBG properties, in particular to find out how it can destroy the moir\'e flat bands.

\subsection{Selective distribution of defects in bipartite systems}
\label{Sec_SelectiveDefects}

As already mentioned, the remarkable electronic properties of MLG and BLG make them good candidates for many applications, 
but these applications are severely limited by the absence of a gap.
Hence, the band-gap opening and the control of bilayer graphene become essential for the applications in various electronic devices.
One way to create a gap in MLG is the selective functionalization. 
It has been done, for example, with hydrogen adsorption on a moir\'e pattern of graphene-Ir(111) \cite{Jorgensen16}.
BLG offers more possibilities for obtaining a gap.
One of its advantages  is the control of its gap by applying an external gate voltage \cite{Castro07,McCann06,McCann13}, which opens the way to multiple applications for nanodevices \cite{Zhang09,Overweg18,Kurzmann19}. 
On the other hand, the BLG  electrical conductivity can be tuned with using the influence of substrate \cite{Zhou07}, vacancies,  adatoms or admolecules adsorbed on the surface of BLG \cite{Leenaerts09,Mapasha12,VanTuan16,Missaoui17,
Missaoui18,Katoch18,VanTuan16,Pinto20,Son20,Joucken21,Jouda21}. 
For instance, single- and double-sided fluorination has been shown to affect conductivity strongly, exhibiting insulating and conducting behavior, respectively \cite{Son20}.
This suggests that the specific functionalization of only one of the two sublattices should be also an  important parameter for controlling a real gap or a mobility gap\,\cite{VanTuan16,Missaoui18,Joucken21,Jouda21}.

In this section, we study the effect of a selective functionalization in a bipartite lattice.  
``Selective functionnalization'' means that defects (vacancies or adsorbates) are randomly distributed in a specific sublattice of the system. 
MLG and BLG are bipartite lattices, i.e., their lattice is composed of two sublattices, $\alpha$ and $\beta$, such that any first neighbor of an $\alpha$ atom is a $\beta$ atom and conversely.
In Sec. \ref{Sec_biAB_selecive_bipartite}, we focus on the density of states (DOS) when defects are located only on sublattice $\alpha$  or sublattice $\beta$ . 
In this case an effective Hamiltonian can be written to simply understand the effect of the defects.
In Sec. \ref{Sec_QL_selective_BLG},  we present a detailed numerical study of BLG transport properties when one or two types of adsorption site are present simultaneously. 
In Sec. \ref{Sec_CondMidGapSt}, we focus on the very particular conductivity of midgap states at finite temperature. 

\subsubsection{Effective Hamiltonian for defects in a single sublattice of a bipartite lattice}
\label{Sec_biAB_selecive_bipartite}

Considering in this section a bipartite lattice and the first-neighbor TB Hamiltonian $\hat{H}$ (Eq.\,(\ref{Eq_hamiltonian2})),  $\hat{H}$ couples only $\alpha$ states ($\beta$ states)  with $\beta$ states ($\alpha$ states).
Quite generally an eigenstate $\varphi$ with energy $E$ of $\hat{H}$ (with or without vacancies),  can be written as 
$\varphi =  \varphi_\alpha  + \varphi_\beta $, 
with states  $\varphi_\alpha $ ($ \varphi_\beta  $) belonging to the  sublattice  $\alpha$ ($\beta$).  
It is easy to  show that $\varphi_\alpha$ and $\varphi_\beta$ are eigenstates of the effective Hamiltonian,
\begin{equation}
\hat{\tilde{H}} = \hat{H}^2, 
\end{equation}
with eigenvalue,
\begin{equation}
\tilde{E} = E^2.
\end{equation}
For non-zero energy the DOS $\tilde{n}(\tilde{E})$ of  $\hat{\tilde{H}}$ is related to the DOS $ n_\alpha(E)$ and $n_\beta(E)$ on sub-parts  $\alpha$ and $\beta$ by, 
\begin{equation}
n_\alpha(E) =n_\beta(E) =  2\, |E|\, \tilde{n}(\tilde{E}). 
\label{eq_n_ntilde}
\end{equation}
These relations show that the spectrum of a bipartite lattice with first-neighbor TB coupling is necessarily symmetrical with respect to the on-site energy of the orbitals (here $E_D=0$).
This symmetry is lifted if we introduce TB couplings beyond the nearest neighbors, but the results we present in the following still give good qualitative behavior, as we have verified in the case of the BLG\,\cite{Missaoui18}.

We now analyze the average local density of states $\tilde{n}_i(\tilde{E})$ (LDOS) of $\hat{\tilde{H}}$ in the presence of vacancies located in a single sublattice, where $i$ is the index of the sublattice. 
Vacancies in a single sublattice ($\alpha$ or $\beta$) create midgap states at $E=0$ that are not coupled 
\cite{Brouwer02,Pereira08a}. 
Those midgap states at $E=0$ are not drawn in DOS shown in this section.

\begin{figure}[t]
\centering

~\raisebox{6mm}{\includegraphics[width=3.7cm]{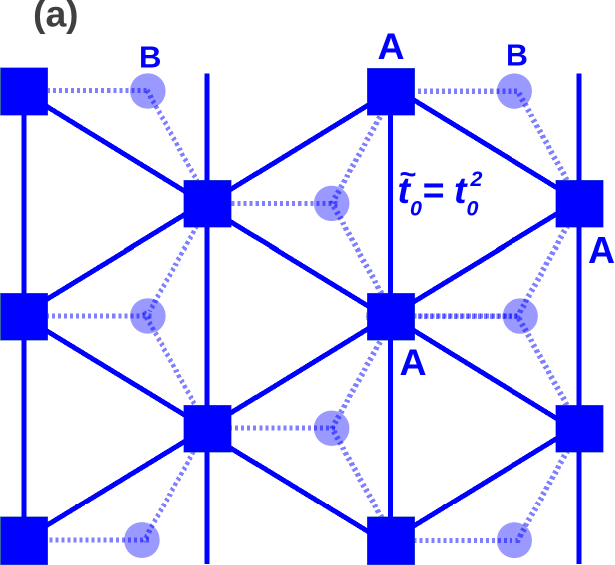}}
~~~~\includegraphics[height=4cm]{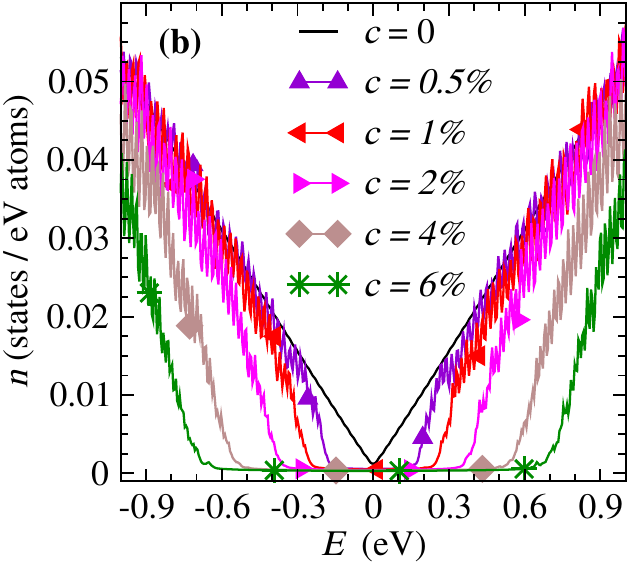} 

\includegraphics[height=4cm]{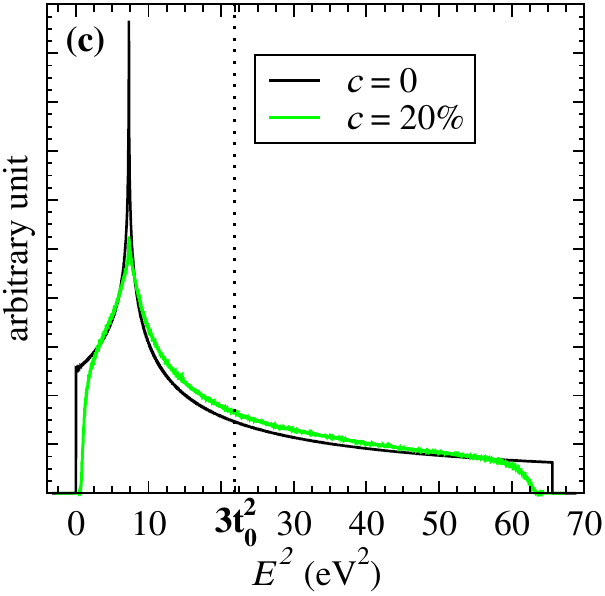}
~~~~\includegraphics[height=4cm]{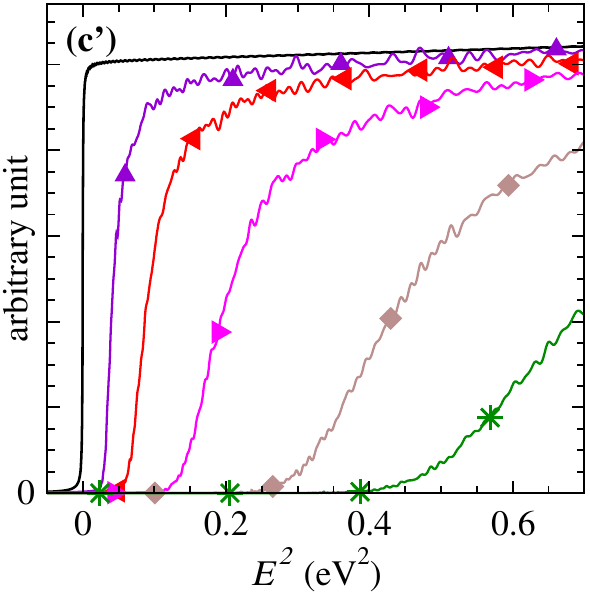} 
~~~~\includegraphics[height=4cm]{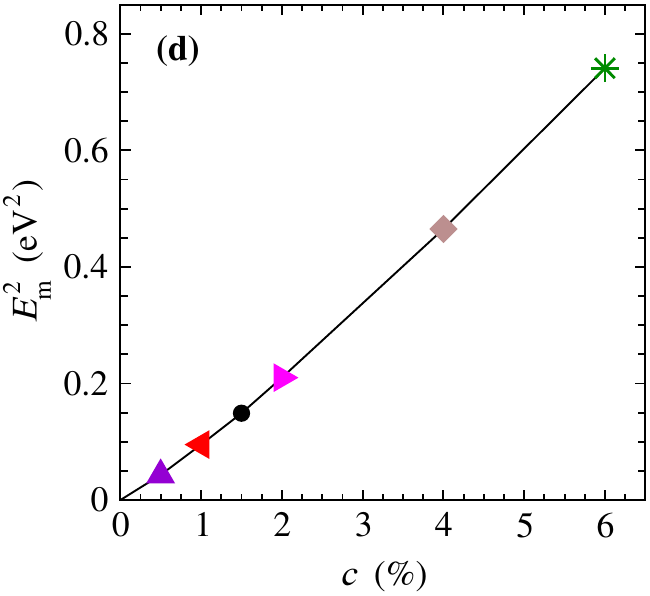}

\caption{ \label{Fig_DOS_bipartite_graphene}
Monolayer graphene (MLG) with vacancies in the A sublattice:  
(a) A sublattice in MLG (without vacancy) 
with the sketch of $\hat{\tilde{H}}=\hat{H}^2$ hopping parameters, where
$\hat{H}$ it the first-neighbor TB Hamiltonian (hopping term $t_0=2.7$\,eV).
The A sublattice is a triangular lattice.
(b) Total density of states DOS $n$ of $\hat{H}$ 
(the black line is without vacancies $c=0$).
The midgap states at $E=0$ (located on the B sublattice) are not shown. 
Note that the average local DOS on sublattices A and B are the same 
for $E\ne0$  (Eq.\,(\ref{eq_n_ntilde})), therefore per atom:
$n = n_A = n_B({\rm without~peaks~of~midgap~states~at~}E=0)$. 
(c) Average local DOS on triangular sublattices A, $\tilde{n}_A(\tilde{E}=E^2)$, 
of the Hamiltonian $\hat{\tilde{H}} = \hat{H}^2$;
(c') is a zoom of (c) at low energies for the same concentration of vacancies as in panel (b).
(d) Minimum value of the spectrum of $\hat{H}^2$, $\tilde{E}_{m} = E^2_{m}$, versus $c$ (the line is a guide for the eyes). 
$c$ is the concentration of vacancies with respect to the total number of atoms in MLG. 
For the numerical method see \ref{Method_Real}.
From\,\cite{Missaoui18,Missaoui_these}.
}
\end{figure}

\vskip 0.5cm
\noindent
{\it Effective Hamiltonian for monolayer graphene (MLG).}

For MLG \cite{Missaoui18}, $\alpha$ states are $p_z$  orbitals of A atoms and  $\beta$ states are $p_z$ orbitals of B atoms.
The effective Hamiltonian ${\hat{\tilde{H}}}$ is a Hamiltonian of a triangular lattice of the $\alpha = {\rm A}$ sublattice, 
\begin{equation}
\hat{\tilde{H}}_{{\rm A}}  = \sum_i \tilde{\epsilon}_{\alpha} \, c^\dag_{{\rm A} i} c_{{\rm A} i}^{ }  
+ \sum_{\langle i,j \rangle} \tilde{t}_{0}\,  c_{{\rm A} i}^\dag c_{{\rm A} j}^{ } ,
\label{Eq_Htilde_alpha}
\end{equation}
with,
\begin{equation}
\tilde{\epsilon}_{{\rm A}} = 3 t_0^2 {\rm ~~and~} \tilde{t}_{0} = t_0^2,
\end{equation}
where $t_0 = 2.7$\,eV is the first-neighbor hopping term.
The hopping terms $\tilde{t}_{0}$ are sketched in Fig.\,\ref{Fig_DOS_bipartite_graphene}(a).
The middle of $\hat{H}$ band ($E=0$) corresponds to lowest energy of the $\hat{\tilde{H}}$ band ($\tilde{E}=0$). 
The $n(E)$ calculated numerically from $\hat{H}$ is shown in Fig.\,\ref{Fig_DOS_bipartite_graphene}(b), and
$\tilde{n}(\tilde{E})$ calculated from $\hat{\tilde{H}}$ is shown in Fig.\,\ref{Fig_DOS_bipartite_graphene}(c).
These results verify equation (\ref{eq_n_ntilde}).

The effect of vacancies in sublattice ${\rm A}$ on the DOS away from zero energy  can be understood \cite{Missaoui18} by considering $\hat{\tilde{H}}_{{\rm A}}$ (Eq.\,(\ref{Eq_Htilde_alpha})),
but with functionalized ${\rm A}$ sites which are simply deleted. Without vacancies the coordination $\eta$ of each atom of the ${\rm A}$ sublattice is 6. With a small concentration $c$ of vacancies in the ${\rm A}$ sublattice, the average coordination is, 
\begin{equation}
\eta \approx 6 \left( 1-c \right). 
\end{equation}
The center of the ${\rm A}$ band is fixed by on-site energies $\tilde{\epsilon}_{{\rm A}}$ and it is not affected by vacancies; but the width of the band will decrease  when $\eta$ decreases (i.e.,\ when $c$ increases) (see Figs.\,\ref{Fig_DOS_bipartite_graphene}(c,c')). 
As expected from this simple tight-binding argument, the minimum values,
$\tilde{E}_{m}= E^2_{m}$, of the spectrum of $\hat{\tilde{H}}$, found numerically 
(Fig.\,\ref{Fig_DOS_bipartite_graphene}(d)),  is almost proportional to the average coordination number $\eta$ (average number of A-A  nearest neighbors of the A sublattice of the bipartite lattice). 
Consequently the average ${\rm A}$ DOS, $\tilde{n}_{{\rm A}}$, has a gap induced by vacancies for $0 \le \tilde{E} \le \tilde{E}_m$. 
This means that the DOS in the $\alpha ={\rm A}$ and $\beta = {\rm B}$ sublattices of MLG also presents a gap for $-E_{m} \le E \le E_{m}$, with ${E}_m = \sqrt{\tilde{E}_m}$ (Fig.\,\ref{Fig_DOS_bipartite_graphene}(b)).

\vskip 0.5cm
\noindent
{\it Effective Hamiltonian for Bernal bilayer graphene (BLG).}
\label{Sec_bilayerGraphBipartite}

\begin{figure}
\centering
 
~\includegraphics[width=4cm]{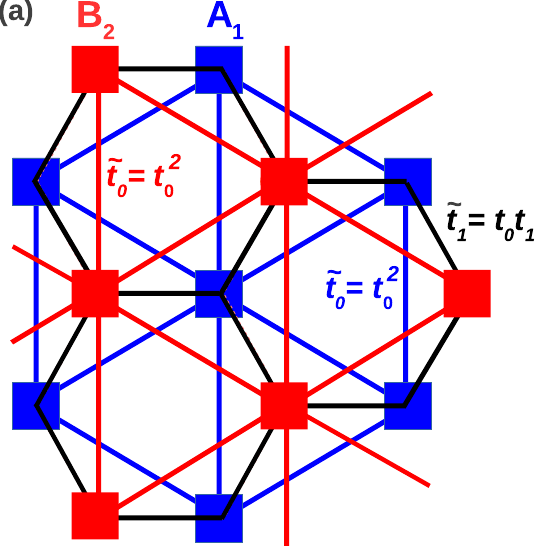}
~~~\includegraphics[width=4.1cm]{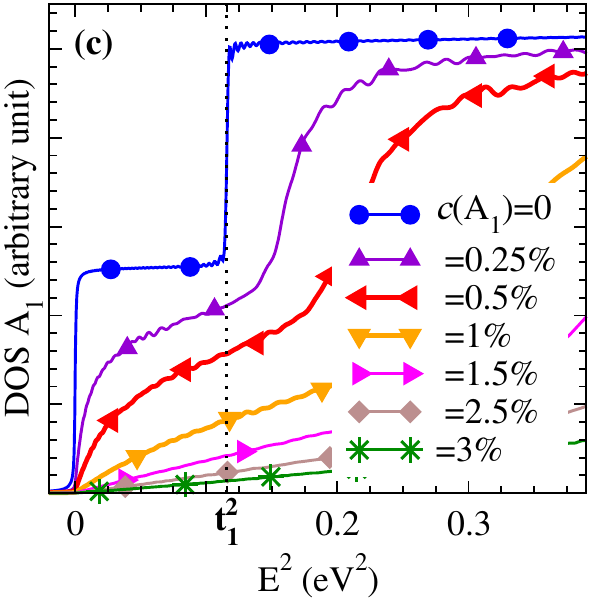}
~ \includegraphics[width=4.1cm]{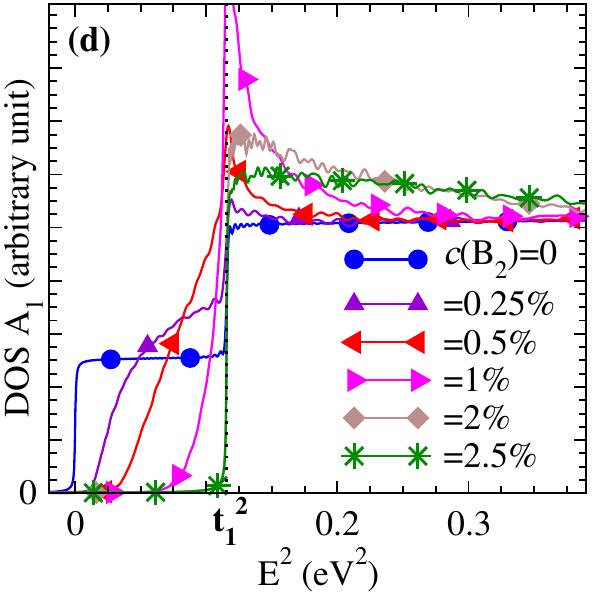} 

\includegraphics[width=4.2cm]{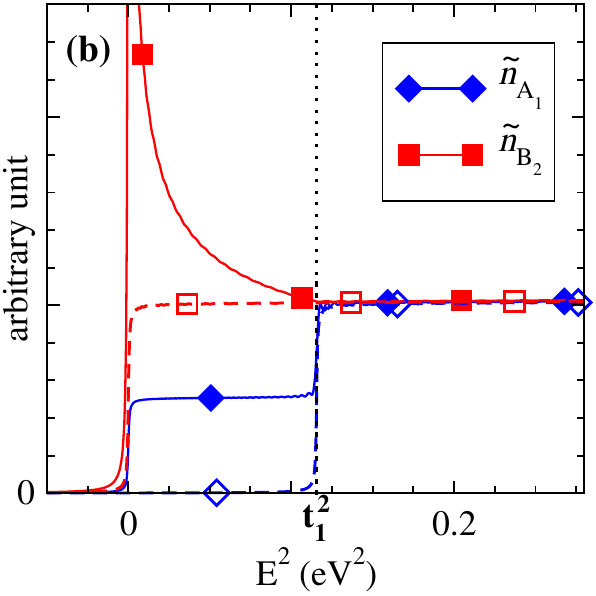}
~~~\includegraphics[width=4.1cm]{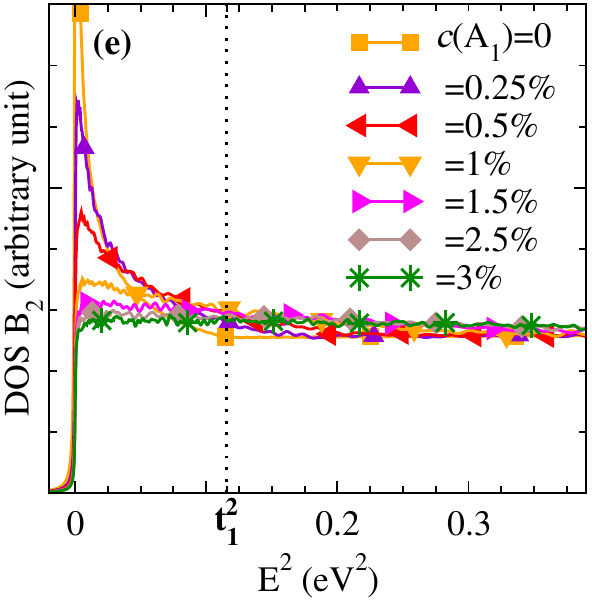}
~ \includegraphics[width=4.1cm]{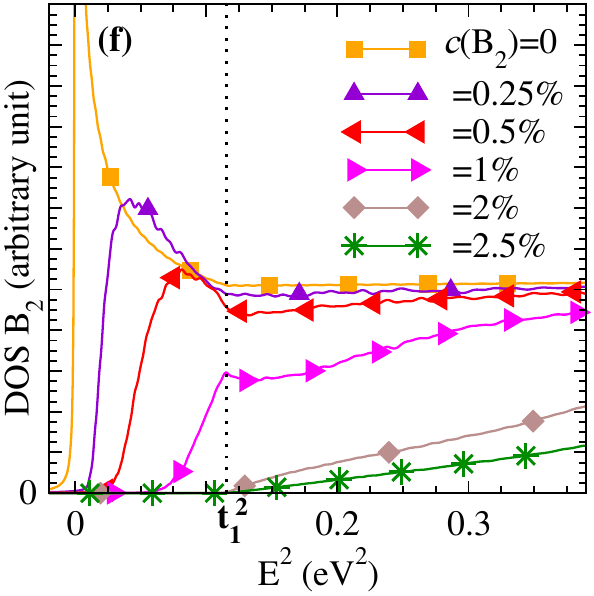}

\caption{ \label{Fig_DOS_bipartite_bilayer}
BLG with vacancies in the $\alpha$ sublattice:   
(a) $\alpha = \{ {\rm A}_1 ; {\rm B}_2\}$ sublattice in BLG (without vacancy) with $\hat{H}^2$ hopping parameters.
(b) Average LDOS of $\hat{\tilde{H}} = \hat{H}^2$ of pristine BLG on sublattice A$_1$, $\tilde{n}_{{\rm A}_1}(\tilde{E})$, 
and on sublattice B$_2$, $\tilde{n}_{{\rm B}_2}(\tilde{E})$. (dashed line) show LDOS 
without A$_1$-B$_2$ hopping, $\tilde{t}_{{\rm A}_1{\rm B}_2} = 0$.
Average LDOS on (c) atoms A$_1$, $\tilde{n}_{{\rm A}_1}(\tilde{E})$, 
and (e) atoms B$_2$, $\tilde{n}_{{\rm B}_2}(\tilde{E})$, with vacancies on the A$_1$ sublattice. 
Average LDOS on (d) atoms A$_1$, $\tilde{n}_{{\rm A}_1}(\tilde{E})$, 
and (f) atoms B$_2$, $\tilde{n}_{{\rm B}_2}(\tilde{E})$, with vacancies on the B$_2$ sublattice. 
$c$ is the concentration of vacancies with respect to the total number of atoms in BLG. 
The corresponding DOS of $\hat{H}$, $n(E)$, and the microscopic conductivity are shown Fig.\,\ref{Fig_biAB_B1B2}(a,b).  
For the numerical method see \ref{Method_Real}.
From\,\cite{Missaoui_these}.
}
\end{figure}

Using the above results for MLG, we also analyze  the electronic structure of the BLG. 
The unit cell of  BLG contains four carbon atoms, A$_1$, B$_1$ in layer 1 and A$_2$, B$_2$ in layer 2 (Fig.\,\ref{Fig_biAB_randVa}(a)). 
Atoms A have three B first neighbors in the same layer and one A neighbor in the other layer, while atoms B have only three A first neighbors in the same layer.
Therefore, BLG is a bipartite lattice with $\alpha =\{{\rm A}_1,{\rm B}_2\}$ and $\beta =\{{\rm A}_2,{\rm B}_1\}$.

Let us first consider the case without defect (i.e.,\ without vacancy).
The effective Hamiltonian $\hat{\tilde{H}}$ in  the subspace $\alpha$, represented in Fig.~\ref{Fig_DOS_bipartite_bilayer}(a), is,
\begin{equation}
\hat{\tilde{H}}_\alpha =  \hat{\tilde{H}}_{{\rm A}_1} + \hat{\tilde{H}}_{{\rm B}_2} + \hat{\tilde{H}}_{{\rm A}_1{\rm B}_2} 
\end{equation}
with
\begin{eqnarray}
\hat{\tilde{H}}_{{\rm A}_1}  &=& \sum_i \tilde{\epsilon}_{{\rm A}_1} \, c^\dag_{{\rm A}_{1i}} c_{{\rm A}_{1i}}^{ } 
+ \sum_{\langle i,j \rangle} \tilde{t}_{0}\,  c_{{\rm A}_{1i}}^\dag c_{{\rm A}_{1j}}^{ } + h.c. , 
\\
\hat{\tilde{H}}_{{\rm B}_2}  &=& \sum_i \tilde{\epsilon}_{{\rm B}_2} \,c^\dag_{{\rm B}_{2i}} c_{{\rm B}_{2i}}^{ } 
+ \sum_{\langle i,j \rangle} \tilde{t}_{0}\,  c_{{\rm B}_{2i}}^\dag c_{{\rm B}_{2j}}^{ } + h.c. , 
\\
\hat{\tilde{H}}_{{\rm A}_1{\rm B}_2}  &=& \sum_{\langle i,j \rangle} \tilde{t}_{1}\,  c_{{\rm A}_{1i}}^\dag c_{{\rm B}_{2j}}^{ } + h.c. ,
\end{eqnarray}
where $\langle i,j \rangle$ represents the corresponding nearest-neighbor pairs of occupied sites. 
$\hat{\tilde{H}}_{{\rm A}_1}$ and $\hat{\tilde{H}}_{{\rm B}_2}$ are the Hamiltonians of A$_1$ and B$_2$ states respectively; they are coupled by $\hat{\tilde{H}}_{{\rm A}_1{\rm B}_2}$.
The coupling parameters of these Hamiltonians are deduced from coupling parameters in the BLG first-neighbor TB 
Hamiltonian (intralayer first-neighbor coupling $t_0 = 2.7$\,eV and the interlayer first-neighbor coupling $t_1 = 0.34$\,eV, see Sec.\,\ref{Sec_biL_rand}).
For the triangular sublattice A$_{1}$, the on-site energies  are  
\begin{equation}
\tilde{\epsilon}_{{\rm A}_{1}} = 3 t_0^2 + t_1^2 = 21.9856\,{\rm eV^2},
\label{eq_t01tilde}
\end{equation}
for the triangular sublattice ${\rm B}_{2}$, the on-site energies are
\begin{equation}
\tilde{\epsilon}_{{\rm B}_{2}} = 3 t_0^2 = 21.87\,{\rm eV^2}, 
\end{equation}
for both sublattices A$_1$ and B$_2$, the coupling terms are (Fig.~\ref{Fig_DOS_bipartite_bilayer}(a)),
\begin{equation}
\tilde{t}_0 =
\tilde{t}_{{\rm A}_1{\rm A}_1} = \tilde{t}_{{\rm B}_2{\rm B}_2} = t_0^2 = 7.29\,{\rm eV^2},
\end{equation}
and the two sublattices are coupled through nearest-neighbor hopping  between A$_1$ and B$_2$ with hopping parameter,
\begin{equation}
\tilde{t}_1 = \tilde{t}_{{\rm A}_1{\rm B}_2} = t_0 t_1 = 0.981\,{\rm eV^2}.
\end{equation}

The local density of states without vacancies $\tilde{n}_{{\rm A}_1}(\tilde{E})$ and $\tilde{n}_{{\rm B}_2}(\tilde{E})$ on an atom A$_1$ and on an atom B$_2$, respectively,
are shown in Fig.~\ref{Fig_DOS_bipartite_bilayer}(b).
If the A$_1$-B$_2$ coupling is turned off, $\tilde{t}_{1} = 0$, while maintaining $t_1 \ne 0$ in Eq.\,(\ref{eq_t01tilde}),
(see dashed lines in Fig.~\ref{Fig_DOS_bipartite_bilayer}(b)), 
$\tilde{n}_{{\rm A}_1}(\tilde{E})$ and $\tilde{n}_{{\rm B}_2}(\tilde{E})$ correspond to two independent bands; 
at low energies they are constant for $ \tilde{E} \ge \tilde{\epsilon}_{{\rm A}_{1}} - 3 \tilde{t}_{{\rm A}_1{\rm A}_1} = t_1^2 = 0.1156$\,eV$^2$ and 
$ \tilde{E} \ge \tilde{\epsilon}_{{\rm B}_{2}} - 3 \tilde{t}_{B_{\rm 2}{\rm B}_2} = 0$, respectively. In that case, from Eq.\,(\ref{eq_n_ntilde}), 
$n_{MLG}(E) = 2 {|E|} \tilde{n}_{{\rm B}_2}(E^2)$ is the DOS of monolayer graphene (MLG). 

We now consider the case with selective resonant defects (vacancies).
Similarly to the MLG case, vacancies in the A$_1$ sublattice (or the B$_2$ sublattice) create a pseudogap at the lowest energies of A$_1$ DOS (B$_2$ DOS) (Fig.\,\ref{Fig_DOS_bipartite_bilayer}(c,f)). 
Note that this pseudogap is a real gap for $t_1=0$ (MLG case). 
In BLG, 
thanks to this pseudogap the low-energy DOS of the 
clean sublattice ($\beta$) recovers its value in the absence of the effective hopping $\tilde{t}_1=\tilde{t}_{{\rm A}_1{\rm B}_2}$. Therefore the eigenstates of the
clean sublattice are simply obtained by suppressing the defected sublattice. According to the basic relations in a bipartite Hamiltonian explained above, the component of the eigenstate in sub-part  $\beta = \{ {\rm A}_2,{\rm B}_1\}$  is also obtained by suppressing the contribution of the defected sublattice. Therefore our main conclusion is that the electronic structure of BLG in the low-energy region where the pseudogap occurs in the defected sublattice is simply given by deleting the defected sublattice. 
We can consider two extreme cases \cite{Missaoui18}: 
\begin{itemize}
\item
When vacancies are on the A$_1$ sublattice, we therefore get an effective decoupling between the layer 2 (A$_2$, B$_2$) and the defected layer 1. We recover therefore  the DOS of pure graphene at low energies in the lower layer 2, and of course a high conductivity (Fig.\,\ref{Fig_DOS_bipartite_bilayer}(c,e)). 
\item
When vacancies are on the sublattice B$_2$, we get an effective system which is the layer 1 coupled only to the sublattice A$_2$ of the layer 2. This system presents a gap up to an energy $t_1 = 0.34$\,eV. 
Above this energy the system is metallic with a high mean free path due to the decoupling of defected sublattice (Fig.\,\ref{Fig_DOS_bipartite_bilayer}(d,f)). 
\end{itemize}
These results are also consistent with the those obtained from numerical calculation performed with a Hamiltonian including hopping terms beyond nearest neighbors \cite{Missaoui18}.

\subsubsection{Quantum localization effects monitored by selective functionalization in Bernal bilayer graphene (BLG)}
\label{Sec_QL_selective_BLG}

\begin{figure}[t]
\centering
\includegraphics[width=0.9\textwidth]{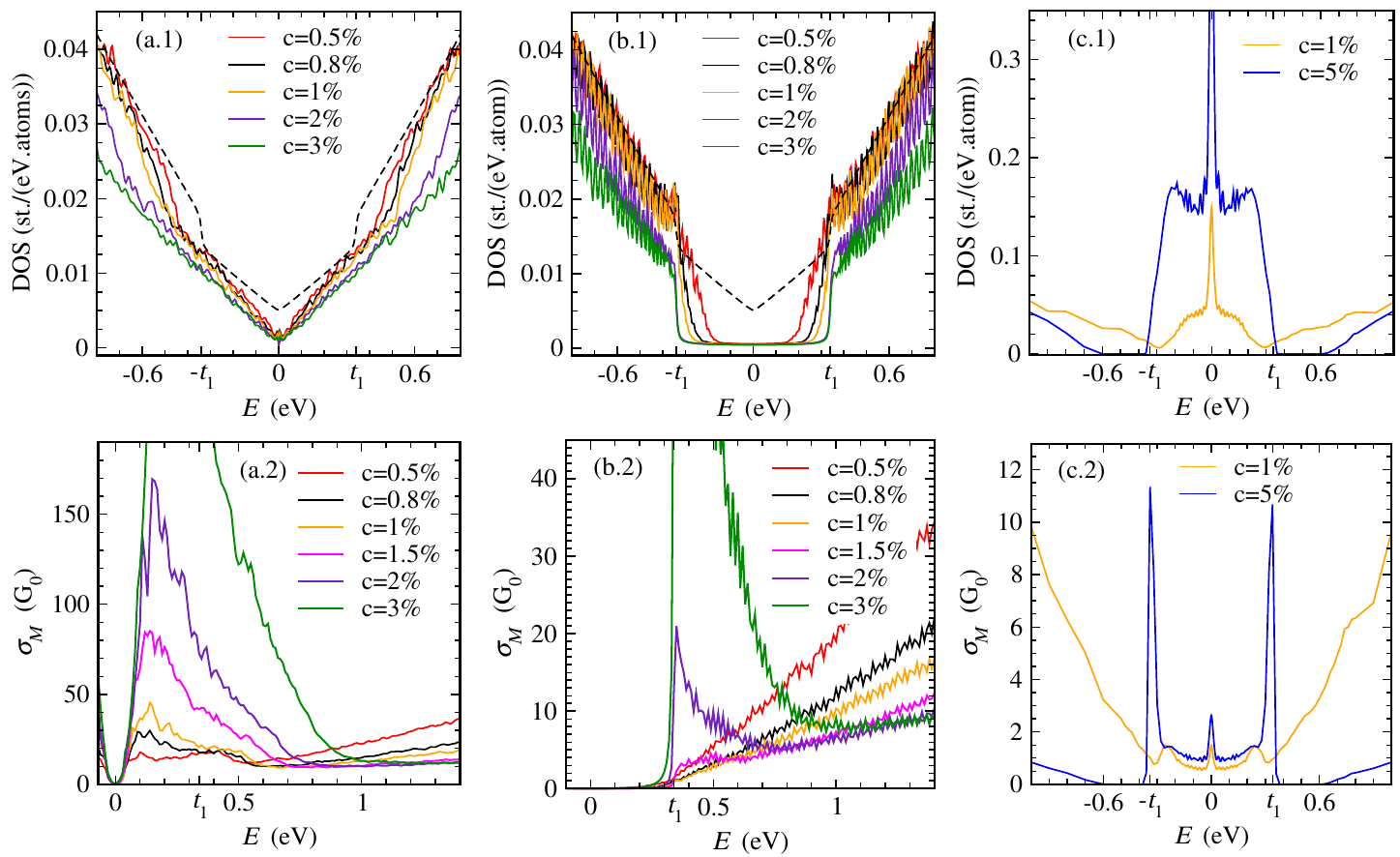}
\caption{ \label{Fig_biAB_B1B2}
Examples of total DOS and microscopic conductivity $\sigma_M$ in Bernal AB bilayer with selective distribution of defects.
(a.1) (a.2) Vacancies randomly distributed on the A$_1$ sublattice.
(b.1) (b.2) Vacancies randomly distributed on the B$_2$ sublattice.
(c.1) (c.2) Vacancies randomly distributed randomly on B$_1$ and B$_2$ sublattices.
In the case $c=5\%$, an isolated midgap-states band is found. At the border of this bands, i.e.,\ $E \approx t_1$, states have a spectacular large conductivity.
Calculation with the first-neighbor TB model (see caption of Fig.\,\ref{Fig_biAB_randVa}).
$c$ is the concentration of defects (vacancies) with respect of the total number of atoms. 
In panels (a) and (b) the midgap states at $E=0$ (see text) are not included.
$ {\rm G}_0 = 2 e^2 / h$.
For the numerical method see \ref{Method_Real}.
From\,\cite{Missaoui18,Jouda21}.
}
\end{figure}

In BLG the local environment of A and B atoms is different,
indeed the B atoms  have only three first neighbors (A atoms) in the same layer, while the A atoms have in addition a neighbor A in the other layer (Fig.\,\ref{Fig_biAB_randVa}(a)). 
Thus 
the probability that an atom or molecule will stick to an atom A or an atom B should be different,
and it is reasonable to think that the functionalization of B atoms is favored.  
This simple argument has been confirmed by DFT calculations \cite{Moaied14} showing that the H adsorption energy difference between A site and B site is about $\Delta E = 85$\,meV in favor of B site, when the number of adsorbates is very low. 
For a larger number of adsorbates, one can therefore expect competition between two contradictory effects: on the one hand preferential adsorption on the B-sites of the bilayer, 
and on the other hand adsorption on different sublattices of the same layer as expected in MLG \cite{Boukhvalov08,Moaied14}.
Indeed in MLG, there exists an interaction between defect states that favors configurations with adsorbates on different sublattices. 
Such  asymmetric adsorption properties between sublattice A and sublattice B have been recently suggested by experimental measurements \cite{Katoch18,Joucken21}, where the distribution of hydrogen adsorbates on the sublattices is adequately controlled. 

Overall, the BLG lattice is a bipartite lattice of the two sublattices  $\alpha$ \{A$_1$,B$_2$\} and $\beta$ \{A$_2$,B$_1$\}, from which one expects very specific electronic properties produced by selective functionalization\,\cite{Missaoui18,Jouda21,Kaladzhyan23}. 
In the previous section \ref{Sec_biAB_selecive_bipartite} and Ref.\,\cite{Missaoui18}, we discussed the limiting cases where adsorbates are randomly distributed only on A$_1$ sublattice (B$_2$ sublattice) of layers 1 (layer 2)  while  the other layer remains pristine. 
As shown in Fig.\ref{Fig_biAB_B1B2}(a.1) and \ref{Fig_biAB_B1B2}(b.1) it leads 
to the absence or the presence of
a gap, respectively. 

On one hand, such a selective functionalization leads to the creation of a gap when sublattice B$_2$ is functionalized. 
This gap is a
fraction of one eV, but of at least 0.5 eV for a concentration $c$ of adsorbates larger than 1\% of the total number of atoms. 
On the other hand, functionalization of sublattice A$_1$ decreases the effective coupling between layers, and thus the conductivity increases when $c$ increases, since the pristine layer is less perturbed by the disordered layer when $c$ increases. 
These two types of selective functionalization exhibit very different and unusual behaviors. 
This opens the way to the control of electronic properties through selective functionalization, which is experimentally feasible \cite{Katoch18}. 
However, these extreme cases (A$_1$ or B$_2$ functionalization only) 
seem too simple to correspond to the experimental situation. 
Indeed, the complexity of the bipartite BLG lattice requires further theoretical studies of other selective adsorbate distributions.  
This is why we have also studied a combined functionalization of two sublattices\,\cite{Jouda21}. 
In particular, we have considered cases where midgap states are coupled to each other (vacancies on A$_1$, B$_1$ sublattices, or vacancies on B$_1$, B$_2$ sublattices) and thus form a midgap band, 
leading to new diffusivity properties that are not a simple combination of the extreme situations studied in Ref.\,\cite{Missaoui18}, in which midgap states are not coupled. 

A truly exciting case is the quantum diffusion by the midgap bands formed by vacancies randomly distributed in the B$_1$ and B$_2$ sublattices \cite{Jouda21}. 
B$_1$- and B$_2$-midgap states are distributed over the entire structure with different weights on the atoms A$_1$, A$_2$, B$_1$, and B$_2$ \cite{Missaoui17}.
They form a band since B$_1$-vacancy midgap states and B$_2$-vacancy midgap states are coupled by the Hamiltonian (Fig.\,\ref{Fig_biAB_B1B2}(c.1)).    
Several regimes are present  depending on both energy $E$ and vacancy concentration $c$: 
\begin{itemize}
\item
For small concentrations $c$, e.g., $c = 1\%$, 
there is no gap in the DOS (Fig.\,\ref{Fig_biAB_B1B2}(c.1)) and states around $E_D$ form a narrow midgap-states band. 
The corresponding microscopic conductivity $\sigma_M$ presents a plateau  at a value independent on $c$, 
$\sigma_{M}\approx 2\sigma_{M,MLG}$ (Fig.\,\ref{Fig_biAB_B1B2}(c.2)). 
\item
For high concentrations $c$, e.g., $c = 5\%$,  the density of states (Fig.\,\ref{Fig_biAB_B1B2}(c.1)) around $E_D$ increases significantly, and as a direct consequence, the plateau of conductivity increases $\sigma_{M}> 2\sigma_{M,MLG}$ (Fig.\,\ref{Fig_biAB_B1B2}(c.2)).
As explained above (Sec.\,\ref{Sec_biAB_selecive_bipartite}), in each layer the gap due to B vacancies (B-vacancy) increases when $c$ increases, therefore
for large $c$ the midgap-states bandwidth becomes smaller than the gap, and the midgap-states band becomes isolated from other states by small gaps at $|E| \gtrsim t_1 $, 
where $t_1 = 0.34$\,eV is the interlayer hopping term between first-neighbor A atoms. 
The width of this isolated band is $\Delta w \approx 2t_1$, i.e.,\ $E \in [-t_1,t_1]$.
For large concentrations  $c$, the edge states ($E\approx\pm t_1$) have a very exotic conductivity $\sigma_M$ which strongly increases when $c$ increases, whereas the DOS does not change too much. 
Roughly speaking this spectacular behavior can be explained by considering the coupling between the B$_1$-vacancy monolayer midgap states and the B$_2$-vacancy monolayer midgap states.
In a monolayer, B-vacancy midgap states are located on the A sublattice around a B vacancy. 
B-vacancy midgap states of each layer are not coupled to each other.  
However, since each A orbital is coupled to an A orbital of the other layer, a B$_1$-vacancy midgap state is coupled to a B$_2$-vacancy midgap state, with a hopping term $\gamma_{B_1-B_2}$. 
For the smallest $d_{B_1-B_2}$ distance between the B$_1$-vacancy and the B$_2$-vacancy (typically first neighbor $B_1$-$B_2$) one can estimate that $\gamma_{B_1-B_2} \approx t_1$, and $\gamma_{B_1-B_2}$  decreases when $d_{B_1-B_2}$ increases. 
When $c$ increases, the average  $d_{B_1-B_2}$ 
distance decreases and thus the average  $\gamma_{B_1-B_2}$ value increases. As a result, by a kind of percolation mechanism between monolayer B-midgap states of the two layers, the conductivity through the BLG  increases strongly when $c$ increases.  
\end{itemize}

To conclude, by a systematic studied of selective functionalization \cite{Missaoui17,Jouda21}, we prove theoretically that controlled functionalization can be an excellent way to tune BLG conductivity. 
This is in agreement with recent experimental results \cite{Katoch18,Son20,Joucken21} showing that it is possible to control the functionalization with an adsorbate concentration of the order of 1\% of the total number of atoms and to fabricate single- and double-side adsorbed bilayer graphene. 
We find a wide variety of original behaviors and have classified them according to the functionalized sublattices, the adsorbate concentration $c$, and the energy. 
For example, we give the conditions for opening a mobility gap of several 100\,meV. 
Experimentally,  and according to Ref.~\cite{Katoch18}, Hydrogen adsorption on B atoms in one layer is energetically favored. For this reason, the study of the specific cases of B$_1$-B$_2$-adsorbates is very interesting. An isolated midgap-states band is predicted. 
Spectacularly, for $c\gtrsim 1 \%$, its edge states have a high electrical conductivity due to the large diffusivity of charge carriers, which deserves further investigation.
As the functionalization of atoms can be performed experimentally, one can imagine that those of the B$_1$-B$_2$-adsorbates can also be carried out, which makes it possible to control the conductivity. 

\subsubsection{Universal conductivity of the midgap states}
\label{Sec_CondMidGapSt}

\begin{figure}[t]
\begin{center}
\includegraphics[width=6.6cm]{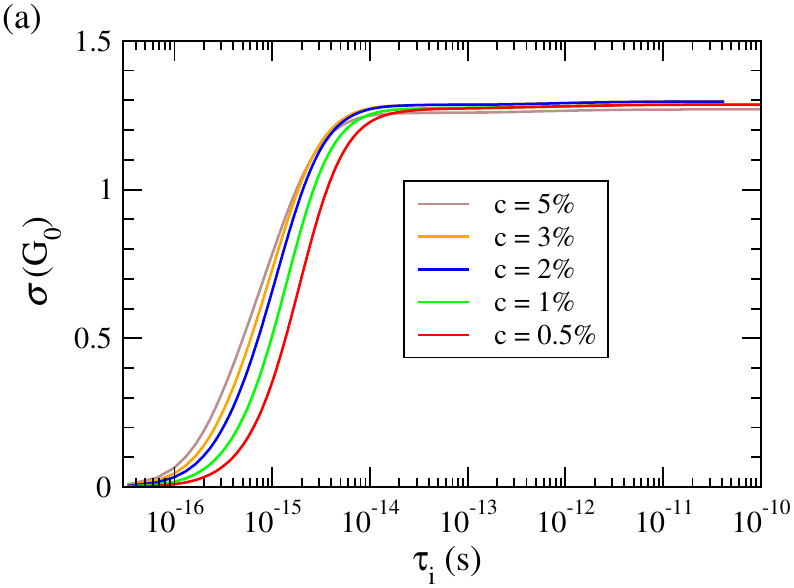} ~~~~
\includegraphics[width=6.3cm]{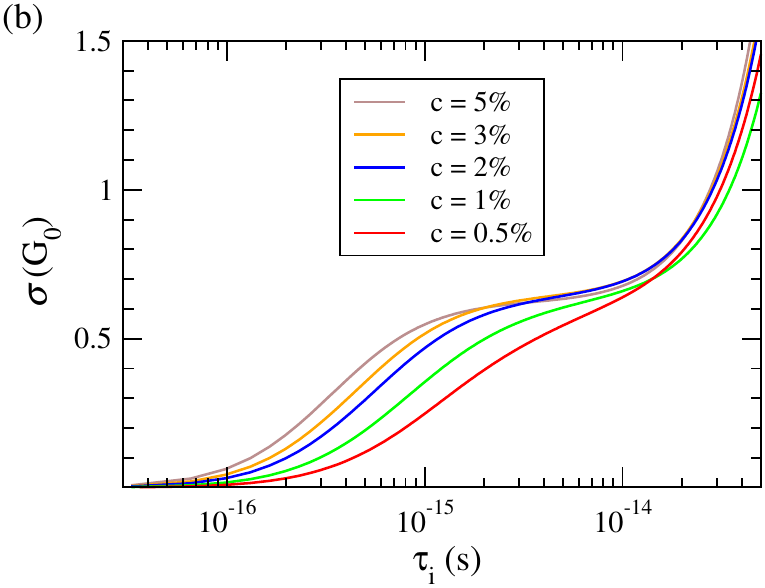}
\end{center}
\caption{\label{Fig_sigma_FTaui_E0} Midgap states conducity, $\sigma(E=E_D=0,\tau_i)$,  in Bernal AB bilayer as a function of inelastic scattering time $\tau_i$ and for different defect concentrations $c$, calculated by the formula (\ref{Eq_sigma_FTaui_E0}) 
of the main text. (a) Vacancies randomly distributed on the atoms B$_{1}$, (b) Vacancies randomly distributed on the atoms A$_{1}$. In both cases midgap states are uncoupled states at $E_D=0$.
In (a) B$_{1}$ vacancies case these states are isolated by gaps, 
whereas in (b) A$_1$ Vacancies case they are in the continuum metallic band of the pristine layer (layer 2) \cite{Missaoui18}.
$G_0 = 2e^2/h$.
For the numerical method see \ref{Method_Real} and main text.
From\,\cite{Jouda21}.
}
\end{figure}

It is also interesting to focus on the conduction by flat-band midgap state themselves, i.e., here midgap states at energy $E_D=0$ when they are not coupled to each other by the Hamiltonian. 
This is done, for example, when vacancies are randomly distributed in one of the two sublattices of MLG\,\cite{Missaoui18} or when vacancies are randomly distributed in the A$_1$ and A$_2$ atoms (or A$_1$ and B$_2$ atoms) in BLG\,\cite{Jouda21}. 

In these midgap states, the average velocity is zero but conduction is possible due to the inherent quantum fluctuations of the
velocity which are due to the interband contributions of the velocity correlation function\,\cite{Bouzerar20,Bouzerar21}.
Indeed, in the presence of inelastic scattering these fluctuations are modified and do not cancel completely at large times which allows  electronic diffusion. 
This yields a non-Boltzmann conductivity, similar to the one found in approximants of quasicrystals (Sec.\,\ref{Sec_QC}) and magic-angle twisted bilayer graphene (Sec.\,\ref{Sec_TBG_QD}), and graphene with particular defects inducing flat bands \cite{Bouzerar20,Bouzerar21}.
The microscopic conductivity $\sigma_M$ of those midgap states 
is negligible for rather small inelastic mean free time $\tau_i$. However, at large $\tau_i$ (large $L_i$), the Kubo-Greenwood conductivity of midgap states is\,\cite{Bouzerar21}, 
\begin{equation}
\sigma(E,\tau_i) \approx e^2 \,n_i(E,\tau_i)\, \mathcal{D}(E,\tau_i),
\end{equation}
where $n_i$ and $\mathcal{D}$ are the DOS and the diffusivity (Eq.\,(\ref{EquationDiffusivityTaui})) in the presence of inelastic scattering. 
Since midgap states are non-dispersive states at $E=0$, $n_i$ is the broadening of the Delta function, $c \delta(E)$, by a Lorentzian with a width at half maximum $\eta$, $\eta = \hbar/\tau_i$. Thus at the Dirac energy $E_D=0$,
\begin{equation}
\sigma(E=0,\tau_i) \approx \frac{16}{S} G_0\, c\, \tau_i\, \mathcal{D}(E=0,\tau_i),
\label{Eq_sigma_FTaui_E0}
\end{equation}
where $S$ is the surface of the unit cell.
As an example, figure\,\ref{Fig_sigma_FTaui_E0} shows two extreme cases of conductivity through the midgap state at $E=E_D$ in BLG, where
resonant defects (vacancies) are randomly distributed in B$_1$ sites or A$_1$ sites, respectively. 
For the B$_1$ vacancies case (Fig.\,\ref{Fig_sigma_FTaui_E0}(a)), the midgap states at $E_D=0$ are isolated by gaps;  therefore, for large $\tau_i$, $\sigma(E=0,\tau_i)$ reaches a universal constant  value, independent of the defect concentration $c$, which equals two times the graphene one\,\cite{Bouzerar21}, $\sigma(E=0) \approx 1.3\,G_0$, as it was found also for  vacancies in A$_1$, A$_2$ sites, and  for vacancies in A$_1$, B$_2$ sites\,\cite{Jouda21}. 
For the A$_1$ vacancies case (Fig.\,\ref{Fig_sigma_FTaui_E0}(b)), the situation is completely different because the midgap states are located in layer 1 only \cite{Missaoui18}, whereas layer 2 remains pristine.
Therefore, at intermediate $\tau_i$ values, and for sufficiently large defect concentration $c$, the conductivity of the bilayer is driven by the midgap-states plateau value of layer 1, $\sigma(E=0,\tau_i) \approx 0.65 \,G_0$. At large $\tau_i$ values or small concentrations $c$, the conductivity is dominated by the ballistic conductivity through layer 2 and thus $\sigma(E=0,\tau_i) \propto \tau_i^{2}$. 
Numerically this contribution of layer 2 may be  due to the accumulation of small numerical errors. Yet further investigations are needed as we cannot exclude that this behavior is intrinsic to the system as is observed \cite{Bouzerar21} for the dice model, 
where the peak of localized states is not in a true gap but just at the edge of the continuum. 

\subsection{Metallic carbon nanotubes}
\label{Sec_Transp_nanotubes}
 
\begin{figure}[t]
\includegraphics[width=16.3cm]{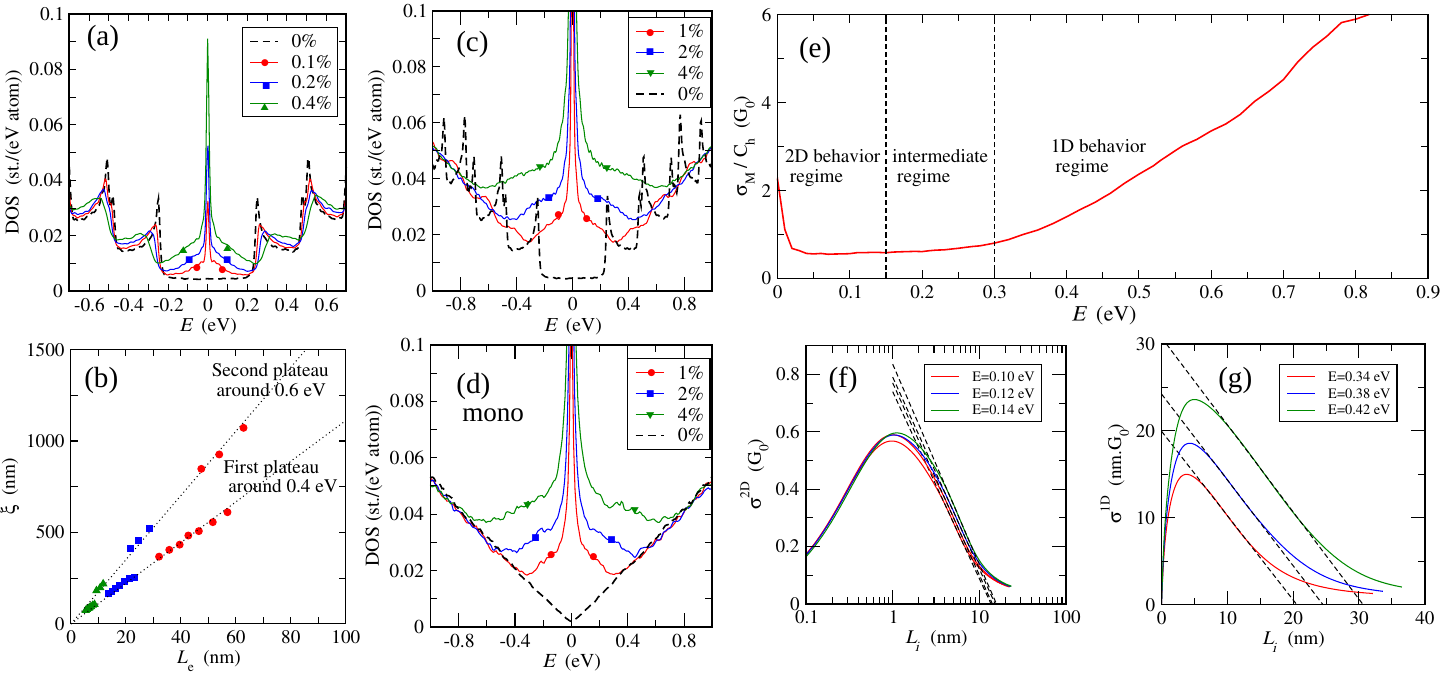}
\caption{\label{Fig_transp_nanotube}
1D / 2D regime transition in metallic carbon nanotubes.
Electronic properties of a metallic single-walled carbon nanotube $(60,0)$ with random distribution of resonant defects (vacancies):
(a) Density of states (DOS) with different small defect concentration $c = 0$, 0.1, 0.2, 0.4\% showing a 1D behavior. 
(b) Localization length $\xi$ versus elastic mean free path $L_e$ showing that Thouless relation is satisfied: 
$\xi(E_F) = (N_{\rm ch}(E_F) + 2) L_e(E_F)$ (dashed line), where $N_{\rm ch}(E_F)$ is close to the number of channels at the energy $E_F$ even if it tends to be higher. 
Points are for different energies in the corresponding plateau,
point colors indicate the concentration $c$ of defects as in panel (a).
(c) DOS for $c = 1$, 2, 4\% showing a 2D behavior, i.e.,\ very similar to (d) monolayer graphene. 
(e) Ratio $\sigma_M / C_h$  versus Fermi energy $E_F$, for $c=1$\%, where $\sigma_M$ is the  microscopic conductivity and $C_h = 60 a = 147.4\,{\rm \AA}$, the circumference of the nanotube (Sec.\,\ref{Sec_QT}).
Depending on the value of $E_F$, the conductivity at large inelastic mean free path $L_i$, $L_i \gg L_e$, crosses over to a 1D/2D localization behavior:
(f) for a small value of $E_F$, $\sigma$ is linear in $\ln L_i$ (2D law),
(g) for a large value of $E_F$, $\sigma$ is linear in $L_i$ (1D law).
(f) and (g) are drawn for $c=1$\%.
The TB model includes only first neighbors with a hopping term $t=2.7$\,eV. Dirac energy $E_D=0$.
$G_0 = {2 e^2}/{h}$.
For the numerical method see \ref{Method_Real}.
From\,\cite{Jemai19,Jemai_these}.
}
\end{figure}

Numerous studies have been carried out on the electronic transport properties of functionalized carbon nanotubes, including for instance, theoretical \cite{Latil2004,IshiiInelastic2010,Jemai19,Fan20}, 
and experimental works \cite{Jouni22} (and Refs.\ therein). 
It is obviously a 1D system, but it exhibits a 1D/2D transition in the presence of a sufficiently large number of defects and thus is of interest in this review.

The case of metallic nanotubes with resonant defects (therefore with flat bands if the system with defects is periodic) shows results related to graphene. 
The type of defects, the circumference $C_h$ of the nanotube and the Fermi energy are determinant for the localization phenomenon and the transport regime\,\cite{Jemai19}. 
In short, for a sufficiently low defect concentration $c$, the elastic mean free path $L_e$ of the charge carriers is greater $C_h$ and the transport regime is the 1D regime. 
However, for a sufficiently high $c$, it is possible to have $L_e$ and the localization length $\xi$ smaller than the circumference and the regime becomes 2D as for monolayer graphene. 
This can already be seen in the density of states (DOS), which is very close to the DOS of the free-defect nanotube at low $c$ (Fig.\,\ref{Fig_transp_nanotube}(a)) 
and very similar to monolayer graphene at high $c$ (Fig.\,\ref{Fig_transp_nanotube}(c,d)).
At low concentrations $c$, 
transport is well described by the 1D quantum transport theory except very close to the neutrality point ($E_F \approx E_D$) where even the DOS is modified by the adsorbates. 
This 1D behavior is confirmed by the linear dependence of $\xi$ on $L_e$ (Thouless relation, Fig.\,\ref{Fig_transp_nanotube}(b)), which is not achieved at high $c$.

For intermediate concentration $c$ of resonant defects, this 1D\,/\,2D regime transition exists also when $E_F$ varies as illustrated in Fig.\,\ref{Fig_transp_nanotube}(e,f,g) for $c=1$\%.
At low energies, as shown in Fig.\,\ref{Fig_transp_nanotube}(f), the scale-dependent conductivity  follows a decreasing trend (dashed lines),
$\sigma_{2D}(L_i) \approx \sigma_0 - \alpha G_0 \ln\left({L_i}/{L_e}\right)$ (Eq.\,(\ref{Eq_localization})), indicating a logarithmic scale for large $L_i$ which is caracteristic of a 2D behavior\,\cite{Lee85}. 
The slope of $\sigma_{2D}(L_i)$ on this scale is universal, with a value $\alpha \approx {1}/{\pi}$, consistent with observations in 2D monolayer graphene\,\cite{Trambly13}. 
As shown in Figs.\,\ref{Fig_transp_nanotube}(e,g), this 2D behavior occurs when the localization length $\xi$ is smaller than the nanotube circumference $(\xi < C_h)$.
The 1D regime is confirmed by the linear dependence of $\sigma$ on $L_i$ for large $L_i$,
$\sigma(Li) \approx \sigma_0-G_0 (L_i-L_e)$ (dashed lines)\,\cite{Lee85}.  
As shown in Fig.\,\ref{Fig_transp_nanotube}(e), the crossover between the 1D and 2D regimes is determined by the ratio ${\xi}/{C_h}$. Three distinct regimes can be identified:
(1) a 2D regime when $\xi < C_h$,
(2) an intermediate regime when $\xi \approx C_h$,
(3) a 1D regime when $\xi > C_h$.

The results presented in Fig.\,\ref{Fig_transp_nanotube} are calculated with the TB model including first neighbors only and similar results
with nearest-neighbor hopping or even variable range hopping even at room temperature, provided that the Fermi energy is sufficiently close from the charge neutrality point and also for sufficiently high concentration of resonant defects\,\cite{Jemai19}.

\subsection{Semiconductor 2D materials}
\label{Sec_semiCond_transp}

For the calculation of the microscopic conductivity $\sigma_M(E)$ and the diffusivity $D(E)$ we use the Kubo-Greenwood formula as explained in \ref{SecKuboG} and \ref{Method_Real}. 
Defects are included directly in the Hamiltonian of the supercell. 
The relaxation time approximation (RTA) deals with the effect of temperature on the diffusivity through an inelastic scattering time $\tau_i$, which is associated to an inelastic mean free path $L_i$  (\ref{Sec_RTA}). 
Moreover, in semiconductors, it is essential to take into account the thermodynamic average due to temperature. 
We thus define the  thermodynamic average conductivity at room temperature ($T=300$\,K) as,
\begin{eqnarray} \label{Eq_ThemCond}
\displaystyle \sigma(\mu_C,T \approx 300 {\rm K}) =   \int_{-\infty}^{+\infty} {d}E\,  \sigma_M(E) \left(- \frac{\partial f}{\partial E} \right) , \label{Eq_roomCond}
\end{eqnarray}
where  $\mu_C$ is the chemical potential, and $f(E,T=300\,{\rm K})$ the Fermi-Dirac distribution function.
The mobility $\mu$ is related to conductivity by:
\begin{eqnarray}
\sigma(\mu_C,T) = e \, |N_e(\mu_C,T)| \, \mu(\mu_c,T),
\end{eqnarray}
where $N_e$ is the number of charge carriers with respect to the undoped system without defects.
Therefore at room temperature ($T \approx 300$\,K),
\begin{eqnarray} \label{Eq_ThermMobility}
\displaystyle \mu(\mu_C,T\approx 300 {\rm K}) =  \frac{1}{e |N_e|} \int_{-\infty}^{+\infty} {d}E \, \sigma_M(E) \left(- \frac{\partial f}{\partial E} \right). \label{Eq_mobilite}
\end{eqnarray}

\subsubsection{Monolayer transition metal dichalcogenides (TMDs)}
\label{Sec_Transp_TMD}

\begin{figure}[t]
\includegraphics[width=16.4cm]{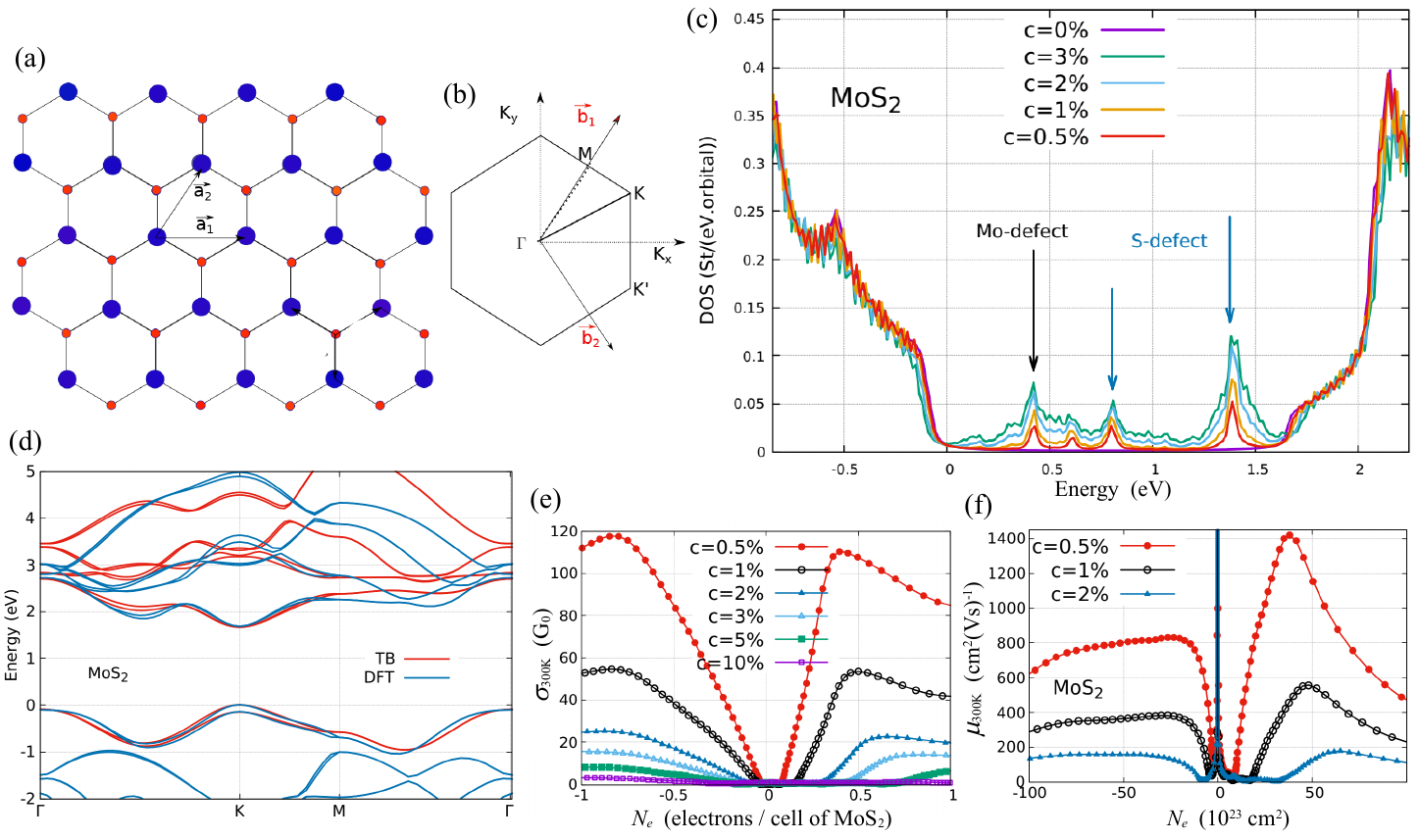}
\caption{\label{Fig_transp_MoS2}
Electronic properties of monolayer MoS$_2$.
(a) Top view of the 2H TMD atomic structure $MX_2$: Blue and red circles indicate Metal ($M=$\,Mo or W) and Chalcogen ($X=$\,S, Se or Te,) atoms, respectively.
The inner layer of $M$ atoms  is sandwiched between two layers of $X$. 
(b) First Brillouin zone. 
(c) Total density of states (DOS) with a concentration $c$ of vacancies distributed randomly in the structure. The DOS is calculated by the recursion method in a 1000$\times$1000 supercell. 
(d) TB and DFT (ABINIT\,\cite{Gonze16}) bands around the gap in MoS$_2$ without defects. 
(e),(f) Thermodynamic average conductivity conductivity $\sigma_{300\,{\rm K}}$ and mobility $\mu$ at room temperature. 
The peak of $\mu(N_e)$ is a numerical artifact, indeed
when $N_e \approx 0$ the equation (\ref{Eq_ThermMobility}) produces numerical errors because of conductivity does not go correctly
to zero (tail of Lorentzian expansion).
These calculations are performed without spin orbit coupling. 
$G_0 = {2 e^2}/{h}$.
For the numerical method see \ref{Method_Real} and Eqs.\,(\ref{Eq_roomCond}), (\ref{Eq_mobilite}).
From \cite{Venky_these}.
}
\end{figure}

There are many studies of electronic structure in TMDs (see for instance Refs.\,\cite{Huisman71,Cappelluti13,Rostami13,
Zahid13,Roldan14b,Roldan14,
Ridolfi15,SilvaGuillen16,Angeli21}), and some  works on quantum transport in these new 2D materials \cite{Yuan14,Stauber15,Krstajic16,Garcia17,
Chu19,Gali20}. 
To study quantum transport, taking into account realistic scattering
effects, we have developed Slater-Koster tight-binding (SK-TB) models \,\cite{Venky_these} for a series of monolayer TMDs ($MX_2$, where $M=$\,Mo or W, and $X=$\,S, Se or Te, Fig.\,\ref{Fig_transp_MoS2}(a))  in
agreement with our DFT calculations performed using the ABINIT software \cite{Gonze16} 
(Fig.\,\ref{Fig_transp_MoS2}(d)).

 For a good description of the electronic bands around the gap, we include the nature of the orbitals involved, $d$ orbitals of metal ($M$) and $p$ orbitals of  chalcogen ($X$).
Indeed, whereas the states around the gap at the Fermi energy $E_F$ are mainly 4$d$ (5$d$) states of 
Mo (W), 
a Hamiltonian restricted to those orbitals is not sufficient to recover states at $E_F$,
since the ligand field ($X$ atoms) splits the $d$
levels of the transition metal atoms, and thus creates a direct gap 
at the K point \cite{Huisman71}. 
Therefore, all TB models proposed in the 
literature include at least $p$ $X$ orbitals 
\cite{Cappelluti13,Rostami13,Zahid13,Ridolfi15,SilvaGuillen16}. Roughly 
speaking the valence band has mainly $d_0 = 4d_{z^2}$ ($5d_{z^2}$) Mo (W) character, 
whereas the conduction band has $d_0$ character mixed with $d_2 = 
4d_{x^2-y^2},\,4d_{xy}$ ($5d_{x^2-y^2},\,5d_{xy}$) Mo (W) character near the gap, and $d_1 = 4d_{xz},\, 
4d_{yz}$ ($5d_{xz},\, 5d_{yz}$) Mo (W) character for higher energies \cite{Cappelluti13}. 
Furthermore, it seems 
that $3p$ ($4p$) ($5p$) S (Se) (Te) orbitals, which have lower on-site energies, act as a 
perturbation of the $d$ $M$ bands. 
For this reason, several SK-TB models 
\cite{Cappelluti13,Ridolfi15,SilvaGuillen16} fit rather well to the DFT 
band structure, while they propose very different sets of parameters (on-site 
energies and SK parameters). 
Our SK-TB models\,\cite{Venky_these,Venky20} include 11 orbitals per unit cell: 5 $d$ $M$ orbitals, $d_0 = 4d_{z^2}$ ($5d_{z^2}$), 
$d_1 = 4d_{xz},\, 4d_{yz}$ ($5d_{xz},\, 5d_{yz}$), 
$d_2 = 4d_{x^2-y^2},\,4d_{xy}$ ($5d_{x^2-y^2},\,5d_{xy}$)
of 1 Mo (W) atom), and 6 
$p$ $X$ orbitals ($3p_x$ ($4p_x$) ($5p_x$), $3p_y$ ($4p_y$) ($5p_y$) and $3p_z$ ($4p_z$) ($5p_z$) of 2 S (Se) (Te) atoms.  
Note that our SK-TB model for monolayer MoS$_2$ 
is close to the model proposed in Ref.\,\cite{Ridolfi15}.

\label{Sec_transpTMD}

We consider resonant defects (vacancies) randomly distributed  on all atoms in a large supercell of MoS$_2$. 
The main results are shown Fig.\,\ref{Fig_transp_MoS2}.
Similar results for the TMD series, 
$MX_2$ ($M=$\,Mo or W, $X=$\,S, Se of Te), 
are also presented in Ref.\,\cite{Venky_these}.
The total density of states (DOS) in monolayer MoS$_2$ with vacancies is shown in Fig.\,\ref{Fig_transp_MoS2}(c) for different concentration $c$ of defects with respect to the total number of atoms. 
The vacancies lead to the appearance
of a series of peaks in the gap, which are associated to the creation of midgap states localized around the defects, whose energy and
strength depends on the specific missing atoms \cite{Yuan14,Gali20}.

Conductivity and mobility at room temperature in monolayer MoS$_2$, calculated from equations (\ref{Eq_ThemCond}), (\ref{Eq_ThermMobility}) and shown in Fig.\,\ref{Fig_transp_MoS2}(e,f), are in good agreement with already published results \cite{Yuan14,Gali20}.
Considering the TMD series\,\cite{Venky_these}, 
we found that for $n$-doped samples, $MX_2$ show similar mobilities, whereas for $p$-doped samples, the mobility of WTe$_2$ is larger than for WSe$_2$, MoTe$_2$, MoSe$_2$, or MoS$_2$. 
Numerical results show that, in general, the mobilities of TMDs are low, but they are larger for holes than for electrons, which seems to be in agreement with experimental results \cite{Zhu14,Chu14,Venky_these}. 
Further investigations are necessary, in particular to take account of the spin-orbit coupling, which is an essential and crucial characteristic for the applications of these semiconductor 2D materials. 

\subsubsection{Monolayer and multilayer black phosphorene}
\label{Sec_phosphorene}

\begin{figure}[t]
\includegraphics[width=16.4cm]{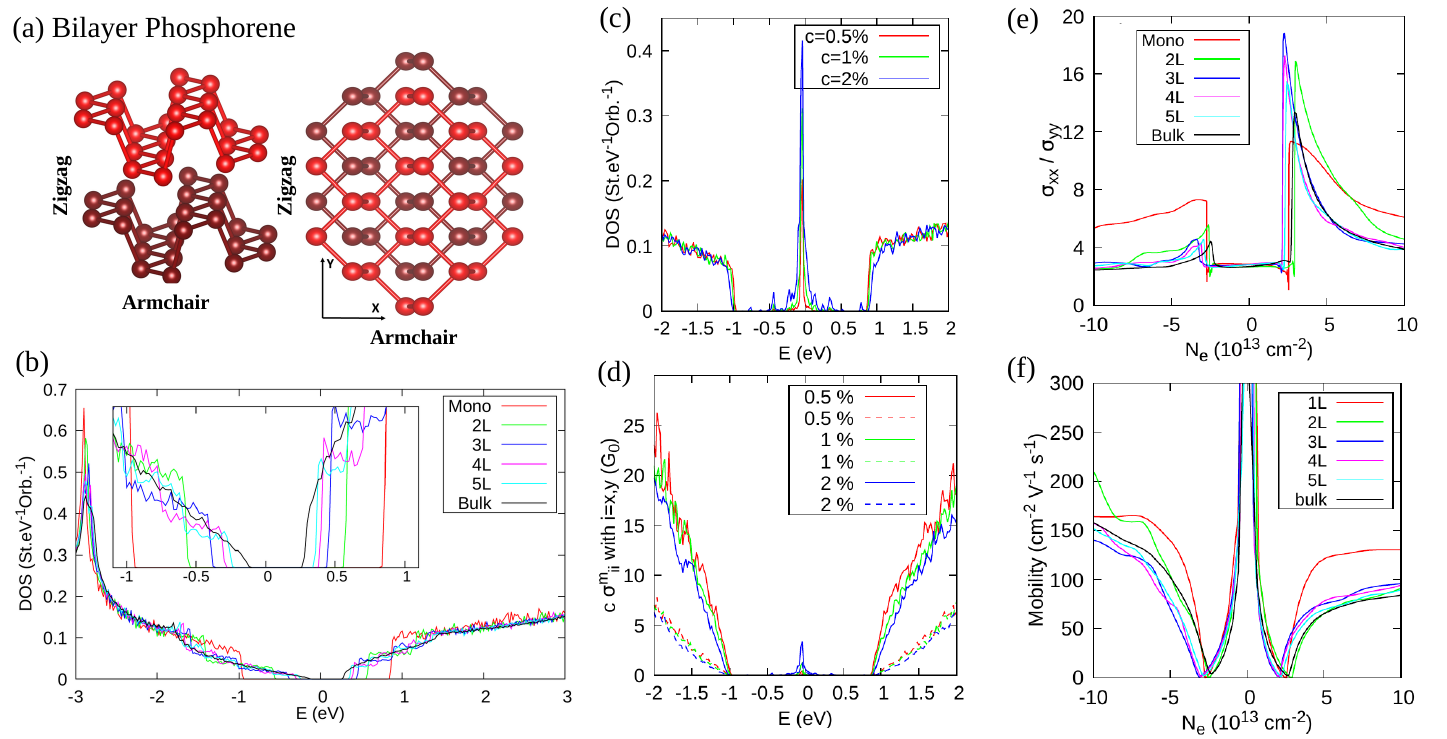}
\caption{\label{Fig_Transp_phosphorene}
Electronic properties of monolayer, multilayers ($L=1,\cdots 5$) and bulk black phosphoerene (BP) with covalent functionalization (resonant defects) simulated by a random distribution of vacant P atoms.
(a) Atomic structure of bilayer black phosphorene. 
A rectangular unit cell contains four P atoms per layer.
(b) TB DOS of defect-free
mono and multilayer phosphorene.
The TB model used was proposed by Rudenko {\it et al.} \cite{Rudenko2014,Rudenko2015}. 
(c) DOS of a monolayer with a concentration $c$ of vacancies. 
(d) Microscopic conductivity times defect concentration $c$, (bold line) $c \sigma_M^{xx}$ and (dashed line) $c \sigma_M^{yy}$ in a monolayer.
(e) Ratio of room temperature conductivity $\sigma_M^{xx}/\sigma_M^{yy}$ (calculated form equation (\ref{Eq_roomCond})) versus $N_e$ for multilayers with vacancies concentration $c = 1$\%.
$N_e$ is the density of charge carriers with respect
to the neutral case.
(f) Mobility at room temperature (calculated from equation (\ref{Eq_mobilite})) along the $x$ direction versus $N_e$ for multilayers and bulk BP and for $c = 1$\%.
$G_0 = {2 e^2}/{h}$.
For the numerical method see \ref{Method_Real} and the main text.
From\,\cite{Jouda24}. 
}
\end{figure}

Phosphorene --monolayer and multilayer black phosphorus (BP), realized experimentally in 2014\,\cite{Li2014,Liu2014,Xia2014} is a very promising 2D semiconductor due to its gap value (around $2$\,eV) which decreases as the number of planes increases, reaching $\sim 0.3$\,eV for bulk BP\,\cite{Tran2014,Qiao2014,cai2014layer,li2017direct,Gaufres19} 
(Fig.\,\ref{Fig_Transp_phosphorene}). 
This shows an increase of the transversal confinement when the number of layers decreases. 
Moreover, as the number of layers decreases, the electron-hole asymmetry becomes less pronounced. 
Indeed, for the monolayer the DOS is almost symmetric with respect to the band gap, whereas as the number of layers increases, the DOS of the valence band becomes lower than that of the conduction band\, 
(Fig.\,\ref{Fig_Transp_phosphorene}(b)).
Another characteristic of monolayer and multilayer BPs is a strong in-plane structural anisotropy, which is reflected in the band anisotropy around the gap and thus in the effective masses of electrons and holes\,\cite{Qiao2014,Carvalho2016,Xia2014,Akhtar2017}. 

Phosphorene is highly sensitive to the environment and covalent or noncovalent functionalization or disorder can modify its electronic properties\,\cite{Rudenko2014,Rudenko2015,Paez16,Mitrovic21,ryder2016,tang2017fluorinated,jellett2020prospects,van2018covalent,yan2019hydroxyl,Guzman23,Jouda24,Oubram25}. 
Quantum diffusion calculations show that resonant and nonresonant defects act similarly on the electrical conductivity \cite{Jouda24}. In fact, even at the high defect concentrations studied, the gap states of resonant defects (Fig.\,\ref{Fig_Transp_phosphorene}(c)) and the gap edge states of nonresonant defects contribute very little to the conductivity. As a result, the mobility gap remains nearly identical to the gap of the defect-free system. This suggests that electronic transport should be fairly well described by a semi-classical Bloch-Boltzmann-type approach, even when defects strongly modify the electronic structure. Particularly, the conductivity of states outside the gap is inversely proportional to the resonant defect concentration $c$ (Fig.\,\ref{Fig_Transp_phosphorene}(d)).
For sufficiently small amounts of resonant defects, the calculated electron and hole mobilities are comparable to the experimental results.
The strong anisotropy persists in the presence of the two types of defects we have studied and it is difficult to find very different behavior between resonant defects and nonresonant defects \cite{Jouda24}.
It is stronger in the case of monolayer BP and it is pretty close to that of bulk BP or multilayers. 
We also showed that the hole conductivity is greater than that of the electrons, particularly in multilayers and the bulk, and this is valid for both types of defects. 
This effect is related to the asymmetry in their DOS without defects.

\section{Conclusion}

In this review, we present some aspects on electronic structure, magnetism, and quantum transport properties of flat-band states in 2D materials and quasicrystals. 
We have focused on results obtained with tight-binding Hamiltonians that allow real materials to be simulated efficiently. 
On one hand,
in perfect (defect-free) quasicrystals, complex intermetallic phases close to quasicrystals, and quasiperiodic tilings, these states characterized by very low dispersion are present throughout the spectrum, 
and they can even lead to a sub-diffusive regime of the charge carriers. 
In twisted bilayer graphene and twisted bilayer MoS$_2$, in the absence of defects, flat-band states arising from the moir\'e pattern geometry are also found at low energies.
In these two families of materials, the flat-band states are characterized by an extremely large spatial extent, spanning an extent over hundreds or even thousands of atoms. 
On the other hand,
in systems (quasiperiodic tilings, graphene, transition metal dichalcogenides (TMDs), multilayer 2D materials) with resonant static defects,
weakly dispersive defect states associated with a relatively large extent in real space can also exist. 
For instance this  occurs  when the defects are randomly distributed over a sublattice of the atomic structure. 
In both cases --whether with or without defects--,
the usual approaches to study electronic transport, such as the semiclassical Bloch-Boltzmann method, are not applicable.
Thus one must use methods based on the fully quantum Kubo-Greenwood formula, in real space or reciprocal space, to correctly account for all quantum effects beyond the semi-classical approximation. 
Our calculations show the importance of contributions beyond the Bloch-Boltzmann approximation, which are related to the non-diagonal terms of the velocity operator in the basis of Bloch states.

Regarding quasicrystals,  quantum diffusion of charge carriers in 2D quasi-periodic tilings such as Penrose and octagonal tilings 
has numerically demonstrated the existence of sub-diffusive states, in defect-free quasiperiodic structures,  even under a pure hopping Hamiltonian.
At high diffusion times, these states behave like insulating states. 
However, it seems that in most cases, these states are not isolated and can have energies very close to super-diffusive states that behave like metallic states. 
It is now crucial to gain a deeper  understanding  of the distribution of these sub-diffusive states within the spectrum and to study in greater detail the effect of static defects and temperature on diffusion properties.

Realistic Slater-Koster tight-binding Hamiltonians allow to study twisted bilayer graphene and twisted bilayer MoS$_2$ at any rotation angle between the two layers.  
A detailed comparison with STM measurements has highlighted the crucial role of the heterostrain in real samples of magic-angle twisted bilayer graphene.
Quantum diffusion through flat-band states in the twisted bilayers, without heterostrain, with and without defects
are also discussed.
This confirms the importance of terms beyond the semi-classical Bloch-Boltzmann approximation in quantum diffusion.
However, further study needs to take account of structural effects on flat bands (such as heterostrain) and electronic correlation effects, which we have started to study using a Hubbard model in the mean-field approximation. 

Regarding the effect of local defects on electronic structure and quantum transport, our work has taken two directions. 
Firstly, a completely random distribution of defects, and secondly, selectively distributed defects on a sublattice of a bipartite lattice. 
The latter approach, which is beginning to be implemented experimentally, opens the way to finer control of electronic properties, such as the creation and control of a real gap or mobility gaps.
This kind of problem is particularly interesting in the case of Bernal bilayer graphene, which, due to its structure containing two nonequivalent atoms per cell, allows to consider many cases of selective functionalization exhibiting very different behaviors ranging from metallic to semiconducting, with in some cases highly anomalous diffusion. 
To advance in this direction, we must now include new effects in our calculations such as spin-orbit coupling in TMDs. 
A more general study of conductivity through midgap states or midgap bands seems also very promising. 

In conclusion, the present review contributes to the understanding of the electronic properties of localized states (flat bands) due to the combined effect of quantum interference and geometrical properties (here moir\'e pattern, quasiperiodicity, or bipartite lattice). 
The physics of flat bands is currently a major topic in condensed matter, either for the field of topological insulators or for remarkable electronics (correlation effect, superconductivity, ...).
It contributes to highlight the peculiarity of the localization in these systems:  
a very low 
dispersion, associated with confined electronic states with long lifetimes, i.e., a small velocity, and a spatial extent of these states over a very large number of atoms (several hundred or even several thousand). 
This type of state leads to a particular transport regime of non-interacting charge carrier conduction which we call ``small velocity regime''.

\section*{Acknowledgment}

 The authors would like to extend their warmest thanks to our collaborating colleagues, with whom many discussions and scientific projects were carried out, 
Claire Berger, Duc Nguyen-Manh, Takeo Fujiwara, Pierre Mallet, Jean-Yves Veuillen, Stephan Roche, Vincent T.\ Renard, Claude Chapelier, Lo\"ic Huder, Florie Mesple and T.\ Le Quang, Th\d{i} Thu Ph\`ung, Robert Peters. 
We would also like to thank C.\ Oguey, A.\ Sz\'all\'as, and A.\ Jagannathan, who provided us with the structures of the quasiperiodic tiling approximants studied. 
Most of the calculations were carried out at the {\it Centre de Calculs} (CDC), CY Cergy Paris Universit\'e, and using HPC resources from GENCI-IDRIS (Grant No.\ 910784).
We thank Yann Costes and Baptiste Mary, CDC, for computing assistance. 
JJK was supported by the Tunisian French Cooperation Project (CMCU 15G1306). 
GTL, AH, LM, and DM were supported by the ANR projects J2D (ANR-15-CE24-0017) and FlatMoi (ANR-21-CE30-0029).
GTL, SV, JV, AM, AH, and JJK thank CY Advanced Studies, CY Cergy Paris University, for financial support. 


\appendix

\section{Numerical methods for quantum diffusion calculations}
\label{Sec_NumMethods}

In this appendix, we briefly describe the numerical methods used to calculate electronic structure and quantum transport properties for non-interacting electrons at zero frequency.  
In the framework of the linear response, the Kubo-Greenwood formula\,\cite{Kubo57,Greenwood58}  (\ref{SecKuboG}),
allows charge carrier dc-conductivity to be calculated, taking into account the complexity of the electronic structure and all possible multiple scattering phenomena (for a review see, e.g., Refs.\,\cite{Mayou08,Fan14,Fan20}).
This approach can be used for calculations (\ref{Method_Reciprocal}) in reciprocal space for periodic systems whose size is not too large, thus allowing diagonalization of the Hamiltonian, 
and (\ref{Method_Real}) in real space with very large cell, where a static defect random  distribution can be instilled using recursion-type methods.

\subsection{Kubo-Greenwood conductivity}
\label{SecKuboG}

\subsubsection{Quantum diffusion of charge carriers and dc-conductivity}
\label{Sec_QD_and_dcCond}

In the Kubo-Greenwood approach to transport properties \cite{Kubo57,Greenwood58,Kubo91}, 
quantum diffusion, i.e.,\ the diffusivity $\mathcal{D}$, is related to the average square spreading, $\Delta X^{2}(E,t)$ of  charge carriers at each energy $E$ and each time $t$ \cite{Mayou00,Mayou08}. 
The central quantities are the {\it velocity correlation function}, $C(E,t)$, of
states of energy $E$ at time $t$,
\begin{eqnarray}
C(E,t) = \Big\langle \hat{V}_x(t)\hat{V}_x(t=0) + \hat{V}_x(t=0)\hat{V}_x(t) \Big\rangle_E
= 2\,{\rm Re}\, \Big\langle \hat{V}_x(t)\hat{V}_x(t=0) \Big\rangle_E \, ,
\label{EqAutocorVit}
\end{eqnarray}
and the {\it average square spreading} (quantum diffusion) along the $x$ direction,
\begin{equation}
\Delta X^{2}(E,t)= \left \langle \left( \hat{X}(t)- \hat{X}(t=0) \right)^{2} \right \rangle_{E},
\label{XET}
\end{equation}
where $\langle \cdots \rangle_{E}$ is the average on states with energy $E$,  
Re$\,{A}$ is the real part of ${A}$,
$\hat{V}_x(t)$ and $\hat{X}(t)$ are the Heisenberg representation of the velocity operator $\hat{V}_x$ 
and the position operator $\hat{X}$
along $x$ direction at time $t$,
\begin{eqnarray}
\hat{V}_x = \frac{{d} \hat{X}}{{d} t} =  \frac{i}{\hbar}~ \Big[ \hat{H} , \hat{X}  \Big].
\label{Eq_defOpVitesse}
\end{eqnarray}
The {\it average spreading}, $\Delta X$, is deduced from the average square spreading, $\Delta X = \sqrt{\Delta X^2}$.
It is easy to show that $C(E,t)$ is related to quantum
diffusion by the relation \cite{Kubo91,Mayou00,Mayou08},
\begin{eqnarray}
\frac{d}{{d} t} \left(\Delta X^2(E,t) \right) = \int_0^{t}C(E,t')\,{d} t'.
\label{EqX2}
\end{eqnarray}
From the Kubo-Greenwood formula, 
the dc-conductivity at zero temperature is given by the Einstein relation,
\begin{eqnarray}
\sigma(E_{F}) = e^2 n(E_{F}) \,\mathcal{D}(E_{F}),
\label{EinsteinRelation}
\end{eqnarray}
where $e$ is the electron charge, $E_{F}$ the Fermi energy,  $n$ the total density of states (DOS) (spin up + spin down), and $\mathcal{D}$ the diffusivity related 
to the average square spreading by the relation,
\begin{eqnarray}
\mathcal{D}(E_{F}) = 
\frac{1}{2}\,
\lim_{t\rightarrow \infty} \frac {{d}}{{d} t} \,  \Delta X^2(E_{F},t)
= \frac{1}{2} \,  \lim_{t\rightarrow \infty}
 \frac{\Delta X^2(E_{F},t)}{t}.
\label{EqDiffusivity}
\end{eqnarray}
It is sometimes useful to define 
a time-dependent
{\it diffusion coefficient},
\begin{equation}
D(E,t) = \frac{1}{2 t} \Delta X^2(E,t).
\label{Eq_Appendix_coefD}
\end{equation}

The above formulas are correct for a time-independent Hamiltonian. 
Thus they 
can be used when the inelastic scattering is not taken into account, typically at zero temperature, and we have used them in two distinct cases. Firstly, when the complex structure may confine electronic states (in quasicrystal and twisted bilayer graphene) without defects. 
Secondly to account for static defects (vacancies or adsorbed atoms) that are thus included in the Hamiltonian and lead to a long-time diffusive regime with a finite diffusivity.
On the other hand, we always treat inelastic scattering by the relaxation time approximation (RTA) explained in the next section. 

\subsubsection{Relaxation time approximation (RTA)}
\label{Sec_RTA}

In this approximation, the effect of inelastic scattering 
is treated in a phenomenological way, by introducing a 
inelastic relaxation time $\tau_i$,
whereas elastic scatterers are included in the Hamiltonian.
We have implemented the RTA in two different ways, numerically, which do not yield the same quantitative results, but which qualitatively produce similar behaviors.
We now present these two approaches.

(i) The simplest way of translating the effect of inelastic scatering time is to consider the diffusion coefficients $D$ (Eq.\,(\ref{Eq_Appendix_coefD})) to be constant for $t \ge \tau_i $, thus from (\ref{EqDiffusivity}), 
\begin{equation}
\mathcal{D}(E_F,\tau_i) \approx  D(E_F,t=\tau_i) = \frac{1}{2 \tau_i} \, \Delta X^2(E,t=\tau_i),
\end{equation}
where $\Delta X^2(E,t)$ can be numerically computed (Eq.\,(\ref{XET})) with or without elastic scattering defects included in the Hamiltonian. The inelastic mean free path is then,
\begin{equation}
L_i(E_F) \approx \Delta X(E_F,t=\tau_i).
\end{equation}
Although this approach is rather crude, as it does not take into account the effect of inelastic diffusion, the results are qualitatively correct. 
The above equations treat the inelastic scattering in a way that is equivalent to the standard approximation in mesoscopic physics\,\cite{Lee85}.

(ii) A better approach\, \cite{Mayou00,Mayou08,Trambly06,Ciuchi11,Lacroix_these} is to
assume that the velocity correlation function $C_{i}(E,t)$ of the system with inelastic scattering  is given by,
\begin{eqnarray}
C_{i}(E,t) ~\approx ~ C(E,t)\, {\rm e}^{-|t|/\tau_{i}} \, ,
\label{RTA_C}
\end{eqnarray}
where $C(E,t)$ is the velocity correlation of the system with elastic scattering (adatoms, vacancies, di-vacancies,...) 
but without inelastic scattering. 
Here the inelastic scattering time $\tau_i$ is the cutoff time of the weak localization effects also called dephasing time.
As shown in Refs.\,\cite{Mayou00,Mayou08,Trambly06,Ciuchi11},
the propagation given by this formalism is unaffected by inelastic scattering at short times ($t < \tau_{i}$) and diffusive at long times ($t> \tau_{i}$), as it must be.  
Using the $t=0$ initial conditions, $\Delta X^2(E,t=0)=0$ and $\frac{{d}}{{d}t} \Delta X^2(E,t=0)=0$, 
and performing two integrations by parts,
we obtain from equations
 (\ref{EqX2}), (\ref{EinsteinRelation}), (\ref{EqDiffusivity}), and (\ref{RTA_C}), \cite{Trambly13}
\begin{eqnarray}
\sigma(E_{F},\tau_{i})&=& e^{2} n(E_{F})\,\mathcal{D}(E_{F},\tau_{i}) \label{EquationEinsteinTaui} \, ,
\\
\mathcal{D}(E_{F},\tau_{i}) &=& \frac{L_{i}^{2}(E_{F},\tau_{i})}{ 2 
\tau_{i}} \, ,
\label{EquationDiffusivityTaui}
\\
L_{i}^{2}(E_{F},\tau_{i}) &=& \frac{1}{\tau_{i}} \int_0^\infty \! \Delta X^{2}(E_{F},t)\,{\rm e}^{-t/\tau_{i}} \, dt \, ,
\label{EquationLiTaui}
\end{eqnarray}
where 
$L_{i}(E_{F},\tau_{i})$ is the inelastic  mean free path
and $\mathcal{D}(E_{F},\tau_{i})$ the diffusivity. 
$\Delta X^{2}(E,t )$ is calculated for the system with the Hamiltonian given in the main text  which represents  only elastic scattering due to static defects. 

RTA deals with the effect of temperature on  diffusivity (and thus on  conductivity) through an inelastic scattering time $\tau_i$.
In this calculation, the Fermi-Dirac distribution function is taken equal to its zero temperature value. 
This is valid provided that the electronic structure varies smoothly on the thermal energy scale $k_{\rm B}T$ and is non-zero at $E_F$ (metal case).
We have checked this assumption in the case of graphene (See Supplemental Material of Ref.\,\cite{Trambly13}).
However for semiconductors, it is essential to take into account the thermodynamic average at finite temperature, as we did for the quantum transport calculation in monolayer TMDs (Sec.\,\ref{Sec_transpTMD}) and monolayer and multilayer black phophorene (Sec.\,\ref{Sec_phosphorene}). 
Therefore the conductivity and the mobility at room temperature can be calculated (see Sec.\,\ref{Sec_semiCond_transp}).

Note that as discussed in  \ref{SecRRmethod_withdefects} below, the relaxation time approximation can 
be applied to a combination of 
elastic scattering  and inelastic scattering.
Such a relaxation time approximation has been used successfully, e.g., to compute \cite{Trambly06,Trambly13} 
conductivity in approximants of quasicrystals where quantum diffusion and localization effects play an essential role \cite{Berger93,Trambly05}.

\subsubsection{Microscopic conductivity $\sigma_M$ and elastic scattering length $L_e$}

In 2D systems with static defects (resonant or non-resonant defects), 
it is known from the localization theory \cite{Lee85} that the diffusivity always tends to zero at very large time $t$; therefore, in RTA, 
$\lim_{\tau_i \to \infty} \sigma(\tau_i) = 0$, and $\sigma$ reaches a maximum value for a finite $\tau_i$, i.e.,\ for a finite $L_i$ from Eq.\,(\ref{EquationLiTaui}).
Thus, at each energy, the {\it microscopic diffusivity}, $\mathcal{D}_M$, and {\it microscopic conductivity}, $\sigma_M$, are defined as the maximum value of $\mathcal{D}(\tau_i)$ and $\sigma(\tau_i)$, respectively,
\begin{equation}
\mathcal{D}_M(E_F) = {\rm Max}\{ \mathcal{D}(E_F,\tau_i) \}_{\tau_i}
{\rm~~~and~~}
\sigma_M(E_F) = {\rm Max}\{ \sigma(E_F,\tau_i) \}_{\tau_i}.
\label{eq_dmicro_simgamicro}
\end{equation}
It is also interesting to have an estimate of the elastic mean free path $L_e$ and the inelastic mean free path $L_{iM}$ value corresponding to the diffusive regime, i.e.,\ $\sigma(L_{iM}) \approx \sigma_M$. 
We compute  the elastic mean free path $L_{e}$ along the $x$-axis, 
from the usual phenomenological formula 
by extending the short-time ballistic regime to microscopic diffusivity
\cite{Trambly13},
\begin{equation}
\label{Eq_le}
L_e(E) \approx \frac{1}{V_{0}(E)}\, {\rm Max}_{\tau_i} \left\{ \frac{L_i^{2}(E,\tau_i)}{\tau_i} \right\} = \frac{2 \mathcal{D}_M(E)}{V_{0}(E)},
\end{equation} 
where the velocity $V_0$ is the slope of $\Delta X(t)$ or $L_i(\tau_i)$ at very small $t$ or $\tau_i$, respectively
(i.e., in the ballistic regime, see Sec.\,\ref{Sec_QT}).
The elastic scattering time is thus,
\begin{equation}
\tau_e(E) \approx \frac{L_e(E)}{V_0(E)} .
\end{equation}%
It is important to note that such a definition of $L_e$ is not very accurate, and this calculation can only give an order of magnitude of the average distance  between two elastic scattering events. 
Indeed, the formula (\ref{Eq_le}) is not always valid when the electronic structure is strongly modified by static defects. 
Moreover, $V_0$ is overestimated
since the numerical calculations include not only the intraband terms but also the interband terms \cite{Mayou08}. In the case of monolayer graphene, we have shown \cite{Trambly16} that these latter increase $V_0$ by a factor of $\sqrt{2}$ which leads to an underestimation of the $L_e$.
However,
roughly speaking, $L_e$ is the value of $L_i$ 
above which the conductivity curve $\sigma(L_i)$ reaches the plateau of diffusive regime due to elastic scattering.
To better define the values of $L_i$ corresponding to the diffusive regime, we sometime define the lengths $L_{i1}$ and $L_{i2}$ such as: $\forall L_i \in [ L_{i1} ; L_{i2} ]$, 
$\sigma(L_i) > 0.9 \sigma_M$.
We also determine the value $L_{iM}$ such that $\sigma$ is a maximal, i.e.,\  $\sigma(L_{iM}) = \sigma_M$.
For instance, 
the values of $L_e$, $L_{i1}$, $L_{iM}$, and $L_{i2}$ are shown in the Supplemental Material  of Ref. \cite{Jouda21} for different concentrations of defects in Bernal bilayer graphene (BLG). 
The results show that $L_e \le L_{i1}$ with the same order of magnitude, and the ratio $L_{i2}/L_{i1}$ varies from 5-10 to very large values, depending on the type of defects and  their concentrations.

\subsubsection{Numerical calculations}
\label{Sec_NumImplementation}

To calculate numerically the conductivity  $\sigma$ and the diffusivity $\mathcal{D}$ (without RTA Eqs.\,(\ref{EinsteinRelation}), (\ref{EqDiffusivity}) or with RTA Eqs.\,(\ref{EquationEinsteinTaui}), (\ref{EquationDiffusivityTaui})), we have to calculate the density of states $n(E)$ and the average square spreading $\Delta X^2(E,t)$ (Eq.\,(\ref{XET})).  
In our studies, we used two complementary methods, which are briefly explained in the two following sections:
\begin{itemize}
\item {(Sec.~\ref{Method_Reciprocal})} For crystals without defects (except the effect of inelastic scattering), the diagonalization of the Hamiltonian in reciprocal space can be done for cells containing up to several $10^4$ orbitals. 
That permits a direct calculation of $\Delta X^2(E,t)$. 
This approach allowed us to study the quantum diffusion properties of Bloch states, considering not only the Boltzmann terms but also the non-Boltzmann terms which we had shown to be important in the small velocity regime (Sec.\,\ref{Sec_SVR}). 
This approach, which requires the structure to be periodic, does not allow randomly distributed defects to be included directly in the Hamiltonian. The effect of scattering defects, whether elastic or inelastic, can only be studied within the RTA (\ref{Sec_RTA}) by introducing a relaxation time $\tau$ (Eq.\,(\ref{eq_tau_taui_taue})) instead of $\tau_i$.

\item {(Sec. \ref{Method_Real})} With a random distribution of static defects, such as vacancies or adatoms, it is necessary to work in a very large supercell (at least several millions of orbitals per supercell), and diagonalization of the Hamiltonian, which includes elastic defects, is not possible. 
In this case, we use a recursion method in real space\,\cite{Mayou88,Mayou95,Roche97,Roche99,Triozon02}.
Inelastic defects, e.g., electron-phonon interactions, can then be treated by RTA with an inelastic relaxation time $\tau_i$ (\ref{Sec_RTA}).
\end{itemize} 

\subsection{Reciprocal-space numerical method}
\label{Method_Reciprocal}

\subsubsection{Without defects}
\label{Sec_NumMeth_withoutDef}
In pure crystals at zero temperature, once the band structure is calculated from the tight-binding Hamiltonian,
the average quadratic spreading
can be computed exactly in the basis of Bloch states \cite{Trambly06,Mayou08}.
The average square spreading is the sum of two terms,
\begin{equation}
\Delta X^2(E,t) = \Delta X_{\B}^2(E,t) + \Delta X_{\NB}^2(E,t).
\label{Eq_DeltaX2}
\end{equation}
The first term is the ballistic  Boltzmann (intraband) contribution
at the energy $E$,
\begin{equation}
\Delta X_{\B}^2(E,t) = V_{\B}^2(E) \, t^2.
\label{Eq_DeltaX2B}
\end{equation}
$V_{\B}$ is the Boltzmann velocity in the $x$ direction. 
The semi-classical theory
is equivalent to taking into account only this Boltzmann term.
The second term (interband contributions),
$\Delta X^2_{\NB}(E,t)$,  is
a non-ballistic (non-Boltzmann) contribution.
It is
due to the non-diagonal
elements in the eigenstates basis $\{| n \mathbf{k} \rangle\}$  of the velocity operator $\hat{V}_x$ (Eq.\,(\ref{Eq_defOpVitesse})). 
From the definitions (\ref{XET}) and (\ref{Eq_defOpVitesse}), we obtain \cite{Trambly06,Mayou08},
\begin{equation}
\Delta X_{\B}^2(E,t) = t^2
\left\langle\,
\sum_{\mathbf{k}}
\left| \langle n \mathbf{k} | \hat{V}_x | n \mathbf{k} \rangle \right|^2
\right\rangle_{E_n({\mathbf{k})}=E}  
\label{Eq_DeltaX2_B}
\end{equation}
and
\begin{equation}
\Delta X_{\NB}^2(E,t) = 2 \hbar^2 
\left\langle\,
\sum_{\mathbf{k}, n' (n' \ne n)}
\frac{\left| \langle n \mathbf{k} | \hat{V}_x | n' \mathbf{k} \rangle \right|^2}{\big(E_n({ \mathbf{k}})-E_{n'}({ \mathbf{k}})\big)^2}
\,\left( 1 - \cos \left[\big(E_n({ \mathbf{k}})-E_{n'}({ \mathbf{k}})\big)\frac{t}{\hbar} \right]\right) \,
\right\rangle_{E_n({\mathbf{k})}=E} ,
\label{Eq_DeltaX2_NB}
\end{equation}
where $E_n({ \mathbf{k}})$  is the energy of the eigenstate $| n  \mathbf{k} \rangle$ computed by diagonalization of the Hamiltonian in reciprocal space.
$\left\langle \cdots \right\rangle_{E_n({ \mathbf{k}})=E}$ is the average on states $| n  \mathbf{k} \rangle$ with energy equal to $E$.
The average Boltzmann velocity along the $x$ direction of the electrons at energy $E$ is thus obtained numerically from diagonal elements of $\hat{V}_x$,
\begin{equation}
V_{\B}(E) = \sqrt{
\left\langle
\left| \langle n \mathbf{k} | \hat{V}_x | n \mathbf{k} \rangle \right|^2
\right\rangle_{E_n({ \mathbf{k}})=E} } ,
\label{Eq_calcul_VB}
\end{equation}
where,
\begin{equation}
\langle n \mathbf{k} | \hat{V}_x | n \mathbf{k} \rangle =
\frac{1}{\hbar} \frac{\partial E_n({ \mathbf{k}})}{\partial k_x} ,
\label{Eq_calcul_VB_derivee}
\end{equation}
are the intraband velocities along $x$ direction.

Specifically, working in a tight-binding basis $\{|i \rangle\}$, it is possible to write down the matrix representing the operators $\hat{H}$ and $\hat{X}$, assuming diagonal matrix for $\hat{X}$.
After calculating the eigenstates of the eigenvectors of $\hat{H}$ by diagonalization, the equations (\ref{Eq_DeltaX2_B}) and (\ref{Eq_DeltaX2_NB}) allow to compute $\Delta X_{\B}^2(E,t)$ and $\Delta X_{\NB}^2(E,t)$ at any energy $E$ and time $t$.
These quantities are calculated for periodic systems where the cell size cannot exceed a few tens of thousands of orbitals for reasons of numerical calculation capacity.
These calculations are therefore made without including static defects in $\hat{H}$, unlike the $\Delta X^2(E,t)$ calculated from the real space method (Sec.\,\ref{Method_Real}) which considers a huge supercell that can contain a random distribution of static defects.

\subsubsection{Extent $L_w$ of the wave packets}

The formula seen in the previous section can be used to estimate the extent  $L_w$  of a wave packet discussed in the introduction to this article and in section \ref{Sec_SVR}. Roughly speaking, for a system without defects, $L_w$ at energy $E$ is the quadratic time average of the non-Boltzmann square spreading,
$L_w(E,t) = \sqrt{ \Delta X_{\NB}^2(E,t)}$
(Eq.\,(\ref{Eq_LW})).
In quasicrystals, quasiperiodic tilings and flat-band systems such as twisted bilayer graphene, for many energies (but not necessarily all energies), we have shown numerically that $\Delta X_{\NB}^2(E,t)$ saturates (or almost saturates) at large $t$ (see Fig.\,\ref{Fig_DX2-NB}), resulting in a finite $L_w^\infty(E)$ value which is generally large with respect to the interatomic distance. 
This new length is characteristic of the defect-free structure, and should be essential when the Boltzmann velocity is low.
Indeed, in the usual semi-classical approximation charge carriers are viewed as wave packets. Their minimal extent is $L_w^\infty(E)$ and their dynamics is  ballistic between scattering events (elastic or inelastic) only if  $L_w^\infty$ is smaller than the  mean free path due to defects (see condition (\ref{Eq_conditionSC}) and Sec.\,\ref{Sec_SVR}). 

\subsubsection{With defects}
\label{SecRRmethod_withdefects}
The effect of static disorder and/or decoherence mechanisms such as electron-electron scattering,
electron-phonon interaction 
(temperature), is not considered in the above section.
This effect can be treated in a phenomenological way by introducing an inelastic scattering time $\tau$ 
in the relaxation time approximation (RTA) (Sec.\,\ref{SecKuboG})
$\tau$ may include the effect of elastic scattering and inelastic scattering. Elastic scattering processes are represented by the time $\tau_e$ and are due to static defects like vacancies or adatoms. Inelastic scattering processes are represented by the time $\tau_i$ are due to phonon, electron-electron scattering, effect of magnetic field. A standard approximation for the relation between the different times is :
\begin{equation}
\frac{1}{\tau} = \frac{1}{\tau_e} + \frac{1}{\tau_i} \, ,
\label{eq_tau_taui_taue}
\end{equation}
$\tau$ decreases when the temperature increases and/or static defect concentration  increases.
In actual 2D materials and aluminum-based quasicrystals at room temperature, realistic values 
of $\tau_i$ are a few $10^{-13}$\,s \cite{Mayou93,Wu07}.
The conductivity $\sigma$ and diffusivity $\mathcal{D}$ can then be estimated from equations (\ref{EquationEinsteinTaui}) and (\ref{EquationDiffusivityTaui}),
where $\Delta X^{2}$ is the average square spreading in the crystal without defects.
$\mathcal{D}$ is the sum of a Boltzmann contribution $\mathcal{D}_{\rm B}$
and a non-Boltzmann contribution $\mathcal{D}_{\rm NB}$:
\begin{equation}
\mathcal{D}(E_{F},\tau) = \mathcal{D}_{\rm B}(E_{F},\tau) + \mathcal{D}_{\rm NB}(E_{F},\tau) . 
\label{Eq_Dif_B_nonB}
\end{equation} 
As explained above, the Boltzmann terms are related to the intraband contributions of the velocity correlation function. In the semi-classical approximation only these terms are taken into account. 
However, when band dispersion is very low, the  interband terms, corresponding to the non-diagonal matrix elements of the velocity operator, may become larger and dominate transport properties (Sec. \ref{Sec_SVR}). 

Note that in the presence of inelastic scattering, due for example to a non-zero temperature, 
the extent $L_w$ of the wave packets defined previously can be estimated as,
\begin{equation}
 L_w(E) \approx \sqrt{\Delta X_{\NB}^2(E,t=\tau_i)}, 
\end{equation}
or using the RTA (Sec.\,\ref{Sec_RTA}) by,
\begin{equation}
     L_w (E) \approx L_{i,NB}(E,\tau_i) ,
\end{equation}
where $L_{i,NB}(E,\tau_i)$ is calculated by the integral (\ref{EquationLiTaui}) with only $\Delta X_{\NB}$ contribution.

\subsection{Real-space numerical method}
\label{Method_Real}

This method makes it possible to calculate the quantum diffusion of charge carriers in supercells containing up to several $10^7$ orbitals. 
We can thus study the long-distance effects of non-periodic structures such as quasiperiodic structures, and/or introduce randomly distributed local structural static defects into the structure itself.  
The real-space method, which calculates $\Delta X^{2}(E,t)$  by using the polynomial expansion of the evolution operator on Chebyshev polynomials,
has been developed by D.\ Mayou, S.~N.\ Khanna, S.\ Roche, and F.\ Triozon \cite{Mayou88,Mayou95,Roche97,Roche99,Triozon02}.
A detailed presentation of the algorithm was given by F.\ Triozon in 2002 \cite{Triozon02_PhD}.
For a recent review see Refs.\,\cite{Fan14,Fan20}.
It allows very efficient numerical calculations by recursion that take into account all quantum effects.
It has been used for instance to study quantum transport in periodic approximants of quasicrystals\,\cite{Triozon02,Trambly17}, 
disordered mono and multilayer graphene \cite{Lherbier08,Lherbier11,Leconte11,Leconte11b,Lherbier12,Roche12,Roche13,Trambly13,
Missaoui17,Missaoui18,Omid20,guerrero24_tBLG_disorder_Kubo,Guerrero25} 
and chemically doped graphene \cite{Lherbier08b,Leconte10},
functionalized carbon nanotubes 
\cite{Latil2004,IshiiInelastic2010,Jemai19,Fan20},
organic semiconductors \cite{Fratini17,Missaoui19} and perovskites \cite{Lacroix20},
phosphorene\,\cite{Jouda21},
and other systems\,\cite{Fan14,Fan20}.

Static defects are included directly in the structural modeling of the system and are randomly distributed on a supercell, which can contain several millions to tens of millions of atoms. 
This corresponds to typical sizes of about one micrometer square which allows to study systems with inelastic mean free length of the order of few hundreds nanometers.
Periodic boundary conditions are applied to this supercell. 
Note that for periodic systems with a large unit cell, such as quasicrystal approximants and twisted bilayer graphene, this supercell generally contains a large number of unit cells of the system under study. 

Following the algorithm proposed in F.\ Triozon's PhD thesis (2002) \cite{Triozon02_PhD}, the average square spreading $\Delta X^{2}$ is computed for any crystal taking into account the $s$, $p$ and $d$ orbitals in a Slater-Koster tight-binding (TB) model,
\begin{equation}
\Delta X^2(E,t) = \frac{{\rm Tr} \left( [\hat{X},\hat{U}(t)]^{\dag} \delta(E-\hat{H}) [\hat{X},\hat{U}(t)] \right)}{{\rm Tr} \,\delta(E-\hat{H})},
\label{eq_DX}
\end{equation}
where $\hat{U}(t)$ is the evolution operator at time $t$, 
$\delta$ is the Dirac function and 
${\rm Tr}$ is the trace. 
Considering such a huge supercell, it is possible to evaluate the traces, ${\rm Tr} \,\hat{A}$, in the equation (\ref{eq_DX}) by the average ${\langle \hat{A} \rangle}$ on a random phase state \cite{Triozon02,Triozon02_PhD}, therefore,
\begin{equation}
{\rm Tr} \,\hat{A} = \langle \phi | \hat{A} |  \phi \rangle 
{\rm ~~~~with~~~}
|  \phi \rangle = \frac{1}{N} \sum_{i=1}^N {\rm e}^{i 2 \pi \theta_i} \,|i \rangle,
\end{equation}
where $\{|i \rangle\}$ is the tight-binding basis, $N$ the number of orbitals in a supercell, and the real numbers $\theta_i$ are random.
So the equation (\ref{eq_DX}) becomes the ratio of two 
LDOSs
on states 
$|\phi'(t)\rangle=[\hat{X},\hat{U}(t)] |  \phi \rangle$
and 
$| \phi \rangle$, respectively.
For a given step time $T$, the evolution operator is decomposed as \cite{Triozon02,Triozon02_PhD},
\begin{equation}
\hat{U}(T) =  {\rm e}^{-\frac{i}{\hbar}\hat{H}T} 
= \sum_{n=0}^{+\infty} c_n(T) \, Q_n(\hat{H})
\approx 
\sum_{n=0}^{N_{Ch}} c_n(T) \, Q_n(\hat{H}) ,
\end{equation}
where $Q_n$ is a Chebyshev polynomial of order $n$
that can be calculated by iteration and $c_n(T)$ are spectral coefficients. 
Thus $| \phi(t=mT) \rangle$ and $| \phi'(t=mT) \rangle$ calculations are done numerically by iterations for a sufficiently large $N_{Ch}$ value. 
A simple criterion is that the cutoff index $N_{Ch}$ must be large enough to maintain the norm of $|\phi\rangle$ and the DOS over time. 
This critical $N_{Ch}$ necessarily depends on the system studied, the amount of static defects contained in the supercell, and the energy precision required. 
For calculations we performed, with up to 1-2$\cdot 10^7$ orbitals per supercell, $N_{Ch}$ of a few hundred is sufficient. 
Comparisons with other methods have shown that this development on Chebyshev polynomials is numerically very efficient\,\cite{Triozon02_PhD,Fan14}. 
For details see, e.g., Refs.\,\cite{Triozon02_PhD,Missaoui_these,Lacroix_these,Venky_these}.

As explained above, the calculation of  $\Delta X^2(E,t)$  (Eq.\,(\ref{eq_DX})) is done by calculating the ratio of two DOSs.
These DOS calculations may be 
performed by the recursion method where the Hamiltonian is first written as a tridiagonal matrix in real space  of dimension $N_r$ using the Lanczos algorithm\,\cite{Lanczos50,Cullum1984,Pettifor,Dagotto1994}. 
Usually we use $N_r$ between 1000 and 3000,  and we checked that 
results do not change significantly when $N_r$ increases. 
One can then use the Lanczos coefficients for a continued-fraction expansion
of the DOS \cite{Haydock1972,Pettifor,Gagliano1987,Dagotto1994,Schenk08}. 
This recursion method
leads to a convolution of the DOS by a Lorentzian function which a small width $\epsilon$. 
The DOS is thus obtained by a Lorentzian broadening of the spectrum and 
$\epsilon$ is a kind of energy resolution of the calculation. 
Usually we choose $\epsilon$ between 1 and 10\,meV.
Alternatively, one can also diagonalize the tridiagonal Lanczos matrix of dimension $N_r\times N_r$ and apply a suitable broadening once the poles and their weights are known from the diagonalization. In the case of Lorentzian broadening, both approaches are equivalent \cite{Gagliano1987,Dagotto1994}. However,
to avoid the tail expansion of the Lorentzian function in the gap
for systems with a gap, it is preferable to diagonalize the tridiagonal Lanczos matrix 
and to evaluate the DOS
by a Gaussian broadening of the spectrum \cite{Lacroix20,Lacroix_these}.
Usually  we have used a Gaussian standard deviation of $\sigma_G =5$\,meV. 

\bibliographystyle{elsarticle-num}
\bibliography{biblio}

\end{document}